\documentclass[twocolumn,amstext,amsmath,amssymb,prc,superscriptaddress,floatfix,nofootinbib,showpacs]{revtex4-1}
\usepackage{slashed}
\usepackage{lineno}
\usepackage{graphicx}
\usepackage{color}
\usepackage{cancel}
\usepackage{ulem}
\usepackage{subfigure}

\usepackage[utf8]{inputenc} 
 
\usepackage{mathrsfs}
\usepackage{hyperref}
\hypersetup{
  colorlinks=true,        % false: boxed links; true: colored links
  linkcolor=blue,         % color of internal links
  citecolor=cyan,         % color of links to bibliography
}
\usepackage{natbib,times}
\usepackage{dcolumn}
\usepackage{xspace}
\usepackage{bbold}

\newcommand{\Tr}{\ensuremath{\operatorname{Tr}}}

\newcommand{\ua}{\ensuremath{U(1)_A}}

\def\Eq#1{Eq.~(\ref{#1})}
\def\Fig#1{Fig.~\ref{#1}}
\def\App#1{App.~\ref{#1}}
\def\Tab#1{Tab.~\ref{#1}}

\definecolor{bjcol}{rgb}{0.0,0.205,0.80}
\definecolor{mocol}{rgb}{0.90,0.1,0.70}

\newcommand{\mm}{\marginpar{\colorbox{green}{\textbf{BJ}}\\@Mario:}}

\def\roughly#1{\mathrel{\raise.3ex\hbox{$#1$\kern-.75em%
\lower1ex\hbox{$\sim$}}}}

\newcommand{\msun}{\ensuremath{\text{M}_\odot}}
 
\newcommand{\vev}[1]{\ensuremath{\left\langle #1 \right\rangle}}

\newcommand{\onefig}{0.49\textwidth}
\newcommand{\twofigs}{0.49\textwidth}

\graphicspath{
{./}
{./figures/}
}

\begin{document}

\title{Hybrid and quark star matter based on a non-perturbative
  equation of state}

\author{Konstantin Otto}
\email[E-Mail:]{konstantin.otto@physik.uni-giessen.de}
\affiliation{Institut f\"{u}r Theoretische Physik,
  Justus-Liebig-Universit\"{a}t Gie\ss en, D-35392 Gie\ss en,
  Germany\\[0.5ex]} 
\author{Micaela Oertel}
\email[E-Mail:]{micaela.oertel@obspm.fr}
\affiliation{LUTH, Observatoire de Paris, PSL Research University,
  CNRS, Universit\'e de Paris, 5 place
  Jules Janssen, 92195 Meudon, France\\[0.5ex]}%
\author{Bernd-Jochen Schaefer}
\email[E-Mail:]{bernd-jochen.schaefer@theo.physik.uni-giessen.de}
\affiliation{Institut f\"{u}r Theoretische Physik,
  Justus-Liebig-Universit\"{a}t Gie\ss en, D-35392 Gie\ss en,
  Germany\\[0.5ex]}

\pacs{
26.60.Kp   %Equations of state of neutron-star matter
11.30.Rd, % Chiral symmetries
11.10.Wx, %Finite-temperature field theory
05.10.Cc % Renormalization group methods}
}

\begin{abstract}
  With the recent dawn of the multi-messenger astronomy era a new
  window has opened to explore the constituents of matter and their
  interactions under extreme conditions. One of the pending
  challenges of modern physics is to probe the microscopic equation of
  state (EoS) of cold and dense matter via macroscopic neutron star
  observations such as their masses and radii. Still unanswered issues
  concern the detailed composition of matter in the core of neutron
  stars at high pressure and the possible presence of e.g. hyperons or
  quarks. By means of a non-perturbative functional renormalization
  group approach the influence of quantum and density fluctuations on
  the quark matter EoS in $\beta$-equilibrium is investigated within
  two- and three-flavor quark-meson model truncations and compared to
  results obtained with common mean-field approximations where
  important fluctuations are usually ignored. We find that
  they strongly impact the quark matter EoS.
  \end{abstract}

\maketitle

\section{Introduction}
\label{sec:introduction}

The recent clean mass determinations for two pulsars, PSR J1614-2230
and PSR J0348+0432 ~\cite{Arzoumanian:2017puf, *Demorest:2010bx,
  *Antoniadis:2013pzd}, confirm the existence of neutron stars (NS)
with a mass of about -- and possibly even beyond, see
\cite{Cromartie:2019kug} -- two solar masses. Since the NS mass and
radius depend strongly on the underlying equation of state (EoS), see
e.g. the reviews \cite{Lattimer:2016yq, Lattimer:2006xb,
  Oertel:2017fk}, any model for the NS interior should produce an EoS
leading to a maximum NS mass at least compatible with the above
observations. The maximum mass is thereby most sensitive to the EoS at
the highest densities. The central densities of the maximum mass
configuration can reach values of about 1 fm$^{-3}$, well above the
nuclear saturation density, $\rho_0 \sim 0.16$ fm$^{-3}$, and other
degrees of freedom than neutrons, protons and electrons might
appear. Although additional degrees of freedom soften the EoS and thus
lower the maximum mass, it has been demonstrated that the appearance
of hyperons~\cite{Chatterjee:2015pua, *Djapo2008}, mesons, or
$\Delta$-baryons \cite{Kolomeitsev:2016ptu}, is not excluded by the
observation of the two massive pulsars. Numerous studies demonstrate
in addition the possibility of a transition to quark matter at high
density in such massive stars and the formation of so-called hybrid
stars \cite{Buballa:2014jta, *Weber:2014qoa, *Orsaria:2013hna,
  *Drago:2013fsa, *Alford:2015gna, Baym2017}.  Under the hypothesis of
absolutely stable strange quark matter~\cite{witten84, *Farhi:1984qu},
even pure quark stars, also referred to as strange stars, might exist
\cite{Buballa:1998pr, *Haensel_86}.

The onset of a new degree of freedom causes not only a reduction of
the maximum mass but in general leads to smaller radii of the stars,
too. For strange quark stars, the mass-radius relation is
qualitatively different from neutron and hybrid stars.  In particular,
strange quark stars are much more compact than their neutron or hybrid
counterparts since they are mainly composed of self-bound
matter. Thus, observing a pulsar with a very small radius would be a
strong indication for a strange quark star.  However, radii are
difficult to measure and the extracted values are model dependent.
Presently, they have been essentially estimated from X-ray
observations, see \cite{Watts:2016uzu} for a recent review, and from
the tidal deformability of GW170817 measured by the LIGO-Virgo
collaboration~\cite{GW170817,Abbott:2018fj,De:2018uhw,Most:2018hfd,Capano:2019eae}.
The obtained values lie in the range of 10-14 km for a fiducial 1.4
$\msun$ star and are perfectly compatible with neutron or hybrid
stars. In the near future further precise radius determinations are to
be expected from the NICER mission and additional binary NS merger
data from LIGO-Virgo detectors.

From a theoretical point of view, it is not possible up to now to
derive the dense matter EoS from first principles over the entire
range necessary for describing compact stars, neither on the hadronic
nor on the quark side.  Therefore simplified EoS at finite densities
need to be constructed. Currently two main strategies are pursued. The
first one consists in parameterizing the EoS in the unknown density
domains, putting the least possible number of model assumptions.  The
parameters are then adjusted to existing constraints from nuclear
experiments, observations and/or theoretical calculations, including
attempts to extract the density dependence of the EoS directly from NS
mass and radius data. Examples are the EoS by \cite{Read:2008iy,
  *Ozel:2010fw, *Steiner:2012xt, *Margueron2018a, *Tews:2018fk,
  *Annala:2017llu, 1711.06748}. The second, more traditional strategy
is based on modelling dense matter. It is less flexible than the
aforementioned but has the advantage of allowing to track among others
the matter's particle content. Although some points remain open, for
instance on inhomogeneous matter in the crust, decades of considerable
effort together with constraints from experimental data and
theoretical calculations have led to reliable models for nuclear
matter up to roughly twice the saturation density. Above this density,
not only the models are less under control, but non-nucleonic degrees
of freedom might appear and the situation becomes more complicated.

Hybrid EoS thereby include both hadronic as well as quark matter and
are usually obtained by a combination of the corresponding EoS for
both sides. The quark matter part is subject to even larger
uncertainties than the hadronic one, being \textit{per} \textit{se} a
non-perturbative QCD issue. It is still under development in
particular at high baryonic densities and small temperatures
appropriate for NS. Examples are based on perturbative QCD
\cite{Kurkela:2009gj, *Xu:2015wya}, density-dependent quark-mass
models \cite{Benvenuto:1989kc, *Torres:2012xv}, NJL-models
\cite{Baldo:2002ju, *Buballa:2003qv, *Pereira:2016dfg, *Li:2018fk},
quark-meson model investigations \cite{Zacchi:2019ayh, *Zacchi:2015lwa},
Dyson-Schwinger type approaches~\cite{1506.06846, *Zhao:2015rta,
  *Isserstedt:2019pgx, *Fischer:2012vc}, holographic
\cite{Ishii:2019gta}, or quasi-particle models \cite{Peshier:1999ww,
  *Tian:2012zza, *Zhao:2015uia}. Lattice QCD calculations, which have
allowed to considerably improve our understanding of the EoS at low
baryon chemical potentials, are afflicted by the sign problem at
finite density and can thus not directly be applied to neutron
stars. For the moment some models include constraints from lattice
results at low chemical potentials, see e.g.~\cite{Schramm:2015hba} or
use QCD-like theories not subject to the sign problem, for instance
G2-QCD~\cite{Hajizadeh:2016jvj}, but much effort is still needed.

In this work an EoS for quark matter within a model based on the
underlying chiral symmetry breaking of QCD is derived and the
consequences for the structure of a non-rotating star are analyzed.
In contrast to many previous works using the mean-field approximation,
the functional renormalization group (FRG) approach is employed here
to incorporate additional non-perturbative effects via quark and meson
fluctuations. The considered quantum and
  density fluctuations are of particular importance in the vicinity
of phase transitions and are usually ignored in mean-field
approximations. These might be some reasons for the recent growing
interest in the application of the FRG method to neutron star matter
\cite{Leonhardt:2019fua, Friman:2019ncm, *Drews:2016wpi,
  *Posfay:2018sw}.

Moreover, a phase transition from hadronic
matter to quark matter can be modeled which might be of relevance for
hybrid stars. In the quark matter core of a hybrid star, strange
quarks might be suppressed due to their relatively large effective
mass~\cite{Dai:1995uj, Baldo:2002ju, Buballa:2003et, Drago:2005yj}. In
addition, large effective strange quark masses often 
destabilize hybrid stars with a strange quark matter core,
leading to gravitational collapse~\cite{Baldo:2002ju,
  Buballa:2003et}, such that we consider the two- and three-flavor
cases separately here.

With our approach it becomes feasible to investigate the influence of
quantum as well as density fluctuations on observables
for neutron stars such as the mass-radius relation in a systematic
manner.  We will consider here non-magnetized matter, since magnetic
effects on the EoS are expected to play a minor role for
pulsars~\cite{Chatterjee:2014qsa}.

The paper is organized as follows: after a brief setup of the used
effective quark-meson model for quark matter, in
Sec.~\ref{sec:constr-non-pert}, three different approximations of the
grand potential are presented which incorporate various contributions
of certain quantum, thermal and density fluctuations. The numerical
results of the phase structure are summarized in
Sec.~\ref{sec:phase-structure}.  As input for the analysis of the
mass-radius relation of a neutron star the EoS is needed. In
Sec.~\ref{sec:eos_quark_hybrid} various EoS for symmetric as well as
for $\beta$-equilibrated and charge neutral matter are calculated and
compared with a non-perturbative EoS obtained with the FRG with and
without strange quarks. In addition, a phase transition from hadronic
to quark matter is implemented and the parameter independent
consequences are studied.  The speed of sound in quark matter for
various approximations is analyzed in Sec.~\ref{sec:speed_of_sound}.
The resulting mass-radius relations for all the used approximations
are presented in Sec.~\ref{sec:star-models}.  We conclude in
Sec.~\ref{sec:summary} and summarize our findings. Parameter choices
and numerical details can be found in the Appendices
\ref{app:inputparam} and \ref{app:solutiontech}.

\section{Constructing a non-perturbative EoS}
\label{sec:constr-non-pert}

As mentioned above, we will be mainly concerned with determining the
EoS of the quark core of a hybrid star or that of a strange star. To
that end, we will consider a quark part and a leptonic part. The
latter will be treated as a relativistic non-interacting Fermi gas. In
the following, we focus on the quark part which will be calculated
within an effective two- and three-quark flavor quark-meson model
framework and we will investigate different approximations. The
quark-meson model has been widely used as an effective model of
low-energy QCD since it successfully incorporates dynamical chiral
symmetry breaking and the generation of constituent quark masses.  In
contrast to NJL-type models often employed in this context, the
quark-meson model used here does not implement a repulsive vector
interaction channel which would stiffen the EoS \cite{Weise:2018ukn}.
We confront two different mean-field approximations of the effective
potential with the results obtained within the functional
renormalization group.  This comparison enables a systematic and
parameter independent analysis of the influence of certain quantum and
density fluctuations on the EoS for vanishing temperature.

Within this first study, we neglect the possibility of diquark pairing
and do not enter the discussion of the extremely rich phase
  structure of color superconducting matter in the density range of
  neutron stars. Diquark pairing is important for transport properties, but,
 being a Fermi surface
phenomenon, has only little influence on the equation of state we are
focusing on here.

\subsection{Quark-meson Models}
\label{sec:qm-model}

The quark-meson model consists of $N_f$ flavors of 
constituent quarks $q$ and dynamical (pseudo-)scalar meson fields
$\phi_a$ encoded in the meson matrix
\begin{equation}
  \Phi := T_a \phi_a \ ,\quad a=0,\ldots,N_f^2-1
\end{equation}
with $\phi_a := \sigma_a + \mathrm{i} \pi_a$ and the generators $T_a$
of the $U(N_f)$ group transformations.  The model features besides the
usual kinetic terms for all involved fields a Yukawa-type interaction
between the quarks and mesons as well as mesonic self-interactions
through the chirally invariant potential
\begin{equation}
  U (\rho_1, \dots, \rho_{N_f}) \ ,
\end{equation}
which is in general a function of $N_f$ independent chiral invariants
\begin{equation}
  \rho_n = \mathrm{Tr} \left[ \left( \Phi^\dagger \Phi \right)^n
  \right] \quad , \quad n = 1, \dots, N_f \ . 
\end{equation}
For two light and one heavy quark flavors, $N_f=2+1$, the generators
of the corresponding $U(3)$ transformations in flavor space can be
chosen as the usual Gell-Mann matrices $\hat{\lambda}_a$, i.e.
$T_a = \hat{\lambda}_a/2$. Omitting in the effective potential the
highest chiral invariant $\rho_3$ whose associated coupling constant
is of negative mass dimension, the three-quark flavor Euclidean
Lagrangian reads
\begin{align}
  \begin{split}
\label{eq:3}
\mathcal{L}_{\text{qm}}^{(2+1)} &= \bar{q} \left(\slashed{\partial} +
  g T_a (\sigma_a + \mathrm{i} \gamma_5 \pi_a) \right) q + \mathrm{Tr}
\left( \partial_\mu \Phi^\dagger \partial_\mu \Phi \right) \\
&\hspace{2cm} + U(\rho_1, \rho_2) - c_A \xi - c_l \sigma_l - c_s
\sigma_s
  \end{split}
\end{align}
wherein additionally the lowest order axial $\ua$ symmetry breaking
term
\begin{equation}
  \xi := \det \Phi + \det \Phi^\dagger
\end{equation}
and two explicit chiral symmetry breaking terms
$-c_l \sigma_l - c_s \sigma_s$ have been added. We assume here a
perfect $SU(2)$ isospin symmetry, such that the two lightest quark
flavors \textit{up} and \textit{down} can be replaced by one index
$l=u=d$ while the \textit{strange} quark flavor is denoted by the
index $s$. The relation of the singlet-octet basis ($\sigma_0$,
$\sigma_8$) and the nonstrange-strange basis ($\sigma_l$, $\sigma_s$)
in the scalar meson sector is governed by a rotation
\begin{equation}
  \begin{pmatrix}
    \sigma_l\\
    \sigma_s
  \end{pmatrix} = \frac{1}{\sqrt{3}} \begin{pmatrix}
    \sqrt{2} & 1 \\
    1 & -\sqrt{2}
  \end{pmatrix} \begin{pmatrix}
    \sigma_0 \\
    \sigma_8
  \end{pmatrix}
\end{equation}
such that the isospin-symmetric vacuum condensate evaluates to
\begin{equation}
  \label{eq:vev}
  \langle \Phi \rangle = T_0 \sigma_0 + T_8 \sigma_8 = \mathrm{diag}
  \left( \frac{\sigma_l}{2}, \frac{\sigma_l}{2},
    \frac{\sigma_s}{\sqrt{2}} \right) \ . 
\end{equation}
For only two quark flavors, $N_f = 2$, the model simplifies to
\begin{equation}
  \mathcal{L}_{\text{qm}}^{(2)} = \bar{q} \left( \slashed{\partial} + \frac{g}{2}
    (\sigma + \mathrm{i} \gamma_5 \vec{\tau} \cdot \vec{\pi}) \right)
  q + \frac{1}{2} \left(\partial_\mu \varphi \right)^2 + U(\rho) - c
  \sigma 
\end{equation}
with the three Pauli matrices $\vec{\tau}$ as generators and only one
independent chiral invariant
\begin{equation}
  \rho \equiv \rho_1 = \frac{1}{2} (\sigma^2 + \vec{\pi}^2) 
\end{equation}
in the mesonic chiral effective potential $U$ with one scalar field
$\sigma$ and three pseudoscalar pions $\vec{\pi}$ summarized in
$\varphi^T := (\sigma, \vec{\pi})$. Implicitly a maximal
$\ua$-symmetry breaking is assumed in this representation since the
remaining chiral (pseudo)scalar multiplets, the $\eta$ and $\vec{a}$
fields, are neglected. For more details see e.g.~\cite{Mitter:2013fxa,
  *Schaefer:2008hk}.

The grand partition function $\mathcal{Z}$ in thermal equilibrium is
defined by a path integral over the quark/antiquark and meson fields
wherein the temperature is introduced via the Matsubara formalism. The
Euclidean Lagrangian generally contains $N_f$ independent quark
chemical potentials $\mu_f$ which are added to the quark-meson
Lagrangian in standard thermodynamic manner
\begin{equation}
\label{eq:13}
\mathcal{L}^{(N_f)} = \mathcal{L}^{(N_f)}_{\text{qm}} + \sum_f^{N_f} \mu_f
q_f^{\dagger}q_f\ .
\end{equation}

The three quark flavor chemical potentials are not independent.  Since
we are interested in cold neutron star matter, we assume weak
equilibrium including $\beta$-equilibrium with neutrinos freely
leaving the star,
\begin{align}
\begin{split}
\label{eq:beta_equil}
\mu_u &= \mu - \frac{2}{3} \mu_e \\
\mu_d &= \mu_s = \mu + \frac{1}{3} \mu_e~,
\end{split}
\end{align}
where $\mu$ denotes the quark chemical potential related to baryon
number, $\mu = \mu_B/3$, and $\mu_e$ the electron chemical potential
which in the present case is the negative charge chemical potential. 
In addition, electrical charge neutrality has to be fulfilled,
\begin{equation}
\label{eq:charge_neutrality}
\frac{2}{3} n_u - \frac{1}{3} n_d - \frac{1}{3} n_s - n_e = 0 \ ,
\end{equation}
such that only one independent chemical potential remains. We choose
$\mu$ as such. Note that even though the chemical potentials
introduced in Eq. \eqref{eq:beta_equil} break isospin symmetry, we
still assume only one light condensate $\sigma_l$ as an
approximation. This leads to equal masses $m_u=m_d=m_l$ even in
isospin asymmetric matter.

Finally, the logarithm of the grand partition function yields the
total grand potential which in general incorporates the thermal,
density and quantum fluctuations of the quark and meson fields
\begin{equation}
\label{eq:14}
\Omega (T, \mu) = \frac{-T \ln \mathcal{Z}}{V}\ .
\end{equation}

\subsection{Mean-field Approximation}
\label{sec:mean-field-appr}

In mean-field approximation (MFA) the grand potential for the
quark-meson model basically splits into a quark loop and a static meson
contribution
\begin{equation}
\label{eq:5}
\Omega (T,\mu) = \Omega_q + \Omega_{\text{m}}\ .
\end{equation}
For $N_f=2+1$ quark flavors the  meson contribution
$\Omega_{\text{m}}$, ignoring the $\rho_3$ invariant, reads
\begin{equation}
  \label{eq:static_meson_potential}
\Omega^{(2+1)}_\text{m} = U_{\chi} (\rho_1, \rho_2) - c_A \xi - c_l \sigma_l - c_s \sigma_s \ 
\end{equation}
with the  chirally symmetric meson potential
$U_{\chi} (\rho_1, \rho_2) = m^2 \rho_1 + \lambda_1 \rho_1^2 + \lambda_2
\rho_2$ introducing in this manner a mass parameter $m^2$ and two
quartic couplings $\lambda_i$.

For only two quark flavors the chiral potential simplifies to
\begin{equation}
\label{eq:6}
U_{\chi} (\rho) = m^2 \rho + \lambda \rho^2\ .
\end{equation}
The employed input parameters can be
found in Appendix \ref{app:inputparam}. 

The integration over the quark loop yields a UV finite and explicitly
temperature dependent contribution
\begin{align}
\begin{split}
\Omega^{(2+1)}_q &= \frac{N_c}{\pi^2} T \sum_{f=u,d,s} \int\limits_0^\infty \mathrm{d}p \, p^2 \, [ \ln ( 1- n_f(E_f,\mu_f,T) )\\
 & \hspace{2.5cm} + \ln ( 1 - n_f (E_f,-\mu_f,T) ) ]
\end{split}
\end{align}
with the usual Fermi-Dirac distribution functions
$n_f(E_f,\mu_f,T) = 1/[1+\exp((E_f-\mu_f)/T)]$, the quark energies
$E_f = \sqrt{p^2 + m_f^2}$ and corresponding light,
$m_l = g \sigma_l/2$, and strange quark masses
$m_s = g \sigma_s/\sqrt{2}$. The light and the strange
condensates $\sigma_l$ and $\sigma_s$ are the temperature and quark chemical
potential dependent minima of the full thermodynamic potential.
Again, for only two quark flavors the strange quark contribution in
this expression is simply dropped and the light quark mass is replaced by
$m_l = g \sigma/2 $ with one chiral order parameter $\sigma$.

In general, the UV divergent vacuum contribution of the quark loop to
the potential can be completely absorbed in the model parameters
because the quark-meson model is renormalizable. Ignoring this
divergent zero point part yields the standard MFA (sMFA). 
However, this vacuum contribution is automatically included in the
fully renormalized quark flow equation,
\begin{align}
  \label{eq:eMFA_flow}
\begin{split}
\Omega^{(2+1)}_{\text{rMFA},q} &= \frac{N_c}{6 \pi^2} \sum_{f=u,d,s} \int\limits_0^\Lambda \mathrm{d}k \, \frac{k^4}{E_f} \\
& \hspace{0.5cm} \left[ \tanh \left( \frac{E_f - \mu_f}{2T} \right) +
  \tanh \left( \frac{E_f + \mu_f}{2T} \right) \right] \ .
\end{split}
\end{align}
In this way important vacuum fluctuations in addition to the thermal
and density fluctuations from the quark loops are included in the
grand potential \cite{Skokov:2010sf, *Schaefer:2011ex, *Gupta:2011ez,
  *Andersen:2011pr}. In the following, we denote this approach by
renormalized MFA (rMFA).  In both mean-field approximations the
fluctuations and back-reactions on the mesonic sector are completely
left out and the same static (tree-level) meson potential is
used. However, all quark and meson fluctuations can finally be
considered by applying the functional renormalization group method.

\subsection{Functional Renormalization Group}
\label{sec:FRG}

As mentioned in the introduction a suitable framework to incorporate
quantum fluctuations in a consistent way is based on the
non-perturbative functional renormalization group in terms 
of the Wetterich equation \cite{Wetterich:1992yh}
\begin{equation}
\label{eq:1}
\partial_t \Gamma_k = \frac{1}{2} \Tr \left[ \partial_t R_k \left(
    \Gamma_k^{(2)} + R_k \right)^{-1} \right]
\end{equation}
with the RG time $t = \ln (k/\Lambda)$.  The effective average action
$\Gamma_k$ interpolates between a microscopic or bare UV action
$S_{\Lambda} = \Gamma_{k \to \Lambda}$ and the full quantum effective
action $\Gamma = \Gamma_{k \to 0}$ in the infrared and governs the
dynamics of the field expectation values after the integration of
quantum fluctuations from the UV scale $\Lambda$ down to the infrared
scale $k_{\text{IR}}$.  The infrared regulator $R_k$ specifies the
regularization of quantum fluctuations near an infrared momentum shell
with momentum $k$ and $\Gamma^{(2)}_k$ denotes the second functional
derivative of the effective average action with respect to the fields
of the given theory. The trace represents a summation and integration
over all discrete and continuous indices. The highly non-linear
Wetterich equation has a simple one-loop structure but includes higher
loop contributions in perturbation theory since the full
non-perturbative propagator enters the loop diagram. One advantage of
this approach is that it does not rely on the existence of a small
expansion parameter and thus is applicable in the non-perturbative
regime. QCD-related reviews on the functional RG approach can be found
in \cite{Berges:1996ib, *Pawlowski:2005xe, *Braun:2011pp, *Gies2006,
  *Schaefer:2006sr}.

In order to solve the functional equation numerically some truncations
are required that turn it into a finite-dimensional partial
differential equation. This truncation might induce a certain
dependence of physical observables on the regulator, but this can be
minimized by choosing optimized regulators or by implementation of RG
consistency \cite{Braun:2018svj}. In this work, a modified
three-dimensional flat regulator has been used \cite{Litim:2001up,
  *Litim:2006ag}.

The flow equation for $\Gamma_k$ must be supplemented with an initial
condition at $k=\Lambda$ which according to \Eq{eq:3} reads for three
flavors
\begin{align}
\label{eq:8}
\begin{split}
  \Gamma_{k=\Lambda} &= \int d^4x \,  \bar{q} \left(\slashed{\partial} +
  g T_a (\sigma_a + \mathrm{i} \gamma_5 \pi_a) \right) q \\
&\hspace{1,2cm} + \mathrm{Tr}
\left( \partial_\mu \Phi^\dagger \partial_\mu \Phi \right) + \Omega^{(2+1)}_{k=\Lambda}(\rho_1, \tilde\rho_2) 
  \end{split}
\end{align}
with
$\tilde{\rho}_2 := \rho_2 - \rho_1^2/3$.  
This truncation for the effective action corresponds to a
leading-order derivative expansion with standard kinetic terms for the
meson fields. Note that in this local potential approximation (LPA) no
scalar wave function renormalizations and no scale-dependence in the
Yukawa couplings between quarks and mesons are taken into account.
However, in this truncation the important dynamical back-reaction of
the mesons on the quark sector of the model is already included. For
more details see \cite{Mitter:2013fxa, *Schaefer:2008hk}.

Finally, this yields for three quark flavors the IR and UV finite flow
equation for the effective potential 
\begin{align}\label{eq:frg}
\begin{split}
\frac{\partial \Omega^{(2+1)}_k}{\partial k} &= \frac{k^4}{12 \pi^2} \Bigg
  \lbrace \sum_b \frac{1}{E_b} \coth \left( \frac{E_b}{2T} \right) -2
  N_c \sum_{f=u,d,s} \frac{1}{E_f}\\  
& \hspace{0.5cm} \left[ \tanh \left( \frac{E_f- \mu_f}{2T} \right)
    + \tanh \left( \frac{E_f + \mu_f}{2T} \right) \right] \Bigg\rbrace, 
\end{split}
\end{align}
where now the flow of the mesonic degrees of freedom is taken into
account. The meson energies $E_{b} = \sqrt{k^2+m_{b}^2}$ include the RG
scale dependent meson 
masses $m_b$ which are obtained by  diagonalizing the mass entries of the matrix
\begin{equation}
M^2_{k,ab} := \frac{\partial^2 \Omega_k}{\partial \phi_a \partial \phi_b} \ .
\end{equation} 
Details and the lengthy explicit expressions of the eigenvalues can be
found in \cite{Mitter:2013fxa, *Schaefer:2008hk}.  Evolving the system
towards the infrared yields the full thermodynamic potential evaluated
at the solution of the gap equation, i.e., the minimum of the grand
potential.

As initial UV condition for the flow \Eq{eq:frg} the meson
potential is parameterized as
\begin{equation}
  \label{eq:10}
  \Omega^{(2+1)}_{k=\Lambda} = U_{\chi,k=\Lambda} (\rho_1 ,
  \tilde{\rho}_2) - c_A \xi - c_l \sigma_l - c_s \sigma_s
\end{equation}
with the scale-dependent chiral potential
\begin{equation}
\label{eq:7}
U_{\chi, k=\Lambda} (\rho_1 , \tilde{\rho}_2) = a_{10} \rho_1 +
\frac{a_{20}}{2} \rho_1^2 + a_{01} \tilde{\rho}_2 
\end{equation}
which differs only in the second argument $\tilde{\rho}_2$ from the
corresponding chiral potential in mean-field approximation
\Eq{eq:static_meson_potential}.  Note that only the expansion
coefficients $a_{ij}$ in the potential are scale dependent while all
remaining parameters are kept constant. See \App{app:inputparam} for the
initial parameter fixing.

For two quark flavors the flow,  \Eq{eq:frg}, simplifies to
\begin{align}
  \begin{split}
    \label{eq:4}
\frac{\partial \Omega^{(2)}_k}{\partial k} &= \frac{k^4}{12 \pi^2}
\Bigg\lbrace \frac{1}{E_\sigma} \coth \left(\frac{E_\sigma}{2T}
  \right) + \frac{3}{E_\pi} \coth \left(\frac{E_\pi}{2T} \right)
\\ 
& \hspace{-0.8cm} - \frac{2 N_c}{E_q} \sum_{f=u,d} \left[ \tanh
    \left(\frac{E_q - \mu_f}{2T} \right) + \tanh \left( \frac{E_q +
        \mu_f}{2T} \right) \right] \Bigg\rbrace \ ,
\end{split}
\end{align}
with only one scalar $\sigma$ and three pion degrees of freedom.

Moreover, for $T=0$ and $N_f$ generic quark flavors the flow reduces to
\begin{equation}
\label{eq:flow_T0}
\left. \frac{\partial \Omega_k}{\partial k} \right\vert_{T=0} =
\frac{k^4}{12 \pi^2} \Big[ \sum_b \frac{1}{E_b} - 4 N_c \sum_f
  \frac{1}{E_f} \theta(E_f - \mu_f) \Big] \ . 
\end{equation}
In this limit the Fermi-Dirac distributions of the fermionic threshold
functions in \Eq{eq:frg} or in \Eq{eq:4} become a sharp Heaviside
function. Hence, for $\mu_f^2 > m^2_f$, only scales above the Fermi
sea $k^2 > k^2_{f,\text{sea}}$ with
$k^2_{f,\text{sea}} \equiv \mu_f^2-m_f^2$ contribute to the
corresponding quark loop and are integrated out yielding a finite
quark density. Hence, increasing the chemical potential suppresses
more and more the quark dynamics of the model.  The $\mu_f$
independence below the onset chemical potential for the density is
related to the so-called ``Silver Blaze property''
\cite{Cohen:2003kd}. In the following we will refer to this phenomenon
in that manner.

Note that at vanishing temperature the quark-meson model is equivalent
to its Polyakov-loop extended version, the Polyakov-quark-meson (PQM)
model \cite{Schaefer:2007pw}. In such PQM models the deconfinement
phase transition is captured statistically by including an effective
potential for the gluon background field in terms of the order
parameter fields for deconfinement, the Polyakov loops. There are
basically two major modifications by the Polyakov loops which vanish
in the zero temperature limit exactly: all known
variants of the effective Polyakov loop potential
\cite{Roessner:2006xn, Fukushima:2008wg, Herbst:2010rf,
  Herbst:2013ufa} are proportional to the temperature and the implicit
dependence of the Polyakov-loop variables on the quark loop dynamics
degenerate to a standard Fermi-Dirac distribution without a
Polyakov-loop contribution. Hence, exactly at $T=0$ the Polyakov-loop
in this context is irrelevant for the thermodynamics at
$\mu>0$. However, a phenomenological finite density generalization of
the Polyakov loop potential at zero temperature can stiffen the EoS
\cite{Ivanytskyi:2019ojt}. For a recent review see
e.g.~\cite{Fukushima:2017csk} and for recent (P)QM phase structure
investigations with the FRG see e.g. \cite{Fu:2018qsk,
  *PhysRevD.96.016009}.

\section{Phase structure}
\label{sec:phase-structure}

\begin{figure}
  \centering
  \includegraphics[width=\onefig]{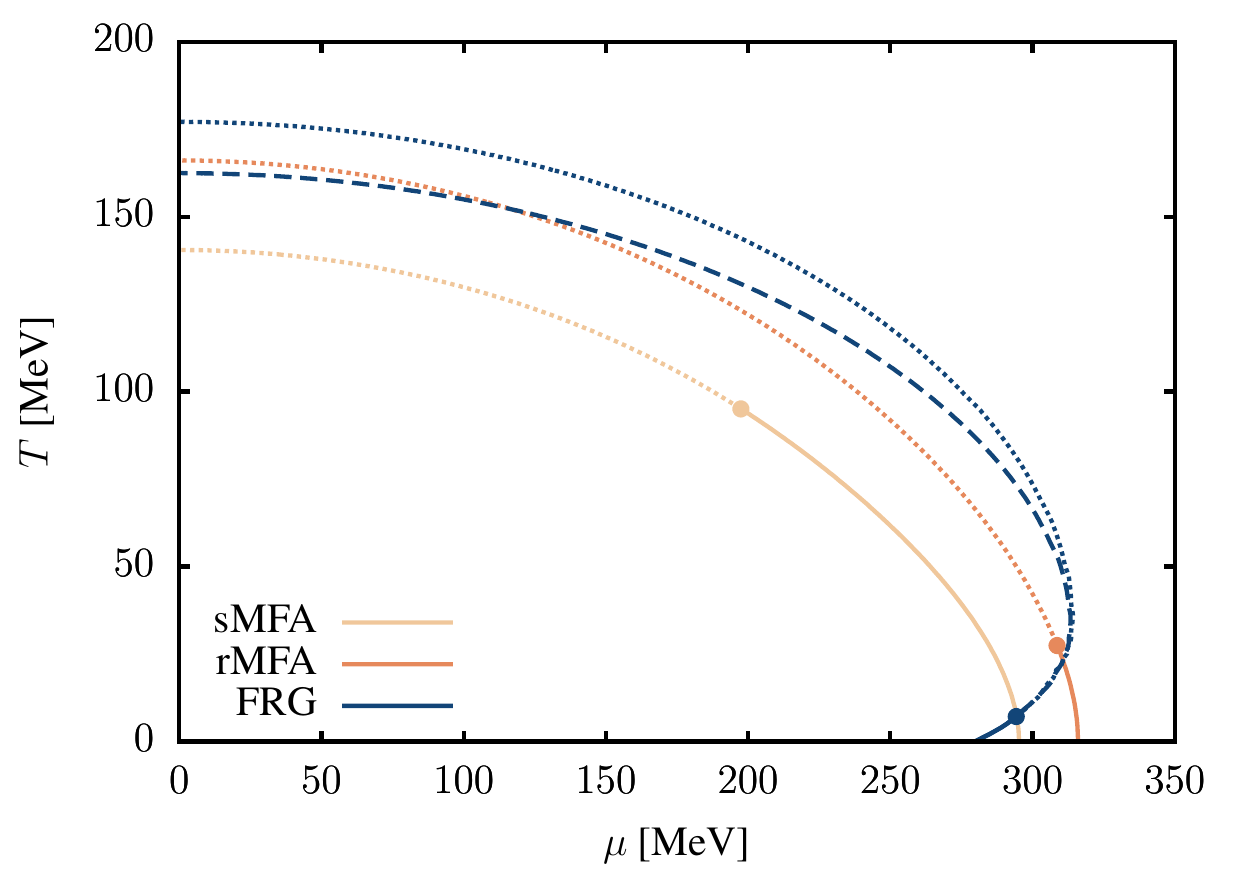}
  \caption{\label{fig:phasediag_2flav} Phase diagrams of the $N_f=2$
    quark-meson model in three different approximations (sMFA, rMFA,
    and FRG). The dots are critical endpoints. Dotted lines are
    smooth crossovers and solid lines first-order phase
    transitions. In addition, the corresponding $N_f=2+1$ flavor phase
    structure obtained with the FRG is also shown (dashed line).}
\end{figure}

\begin{figure*}
  \centering
  \subfigure[\label{fig:condensates_T_2p1} $\mu =0$.]{\includegraphics[width =
    \twofigs]{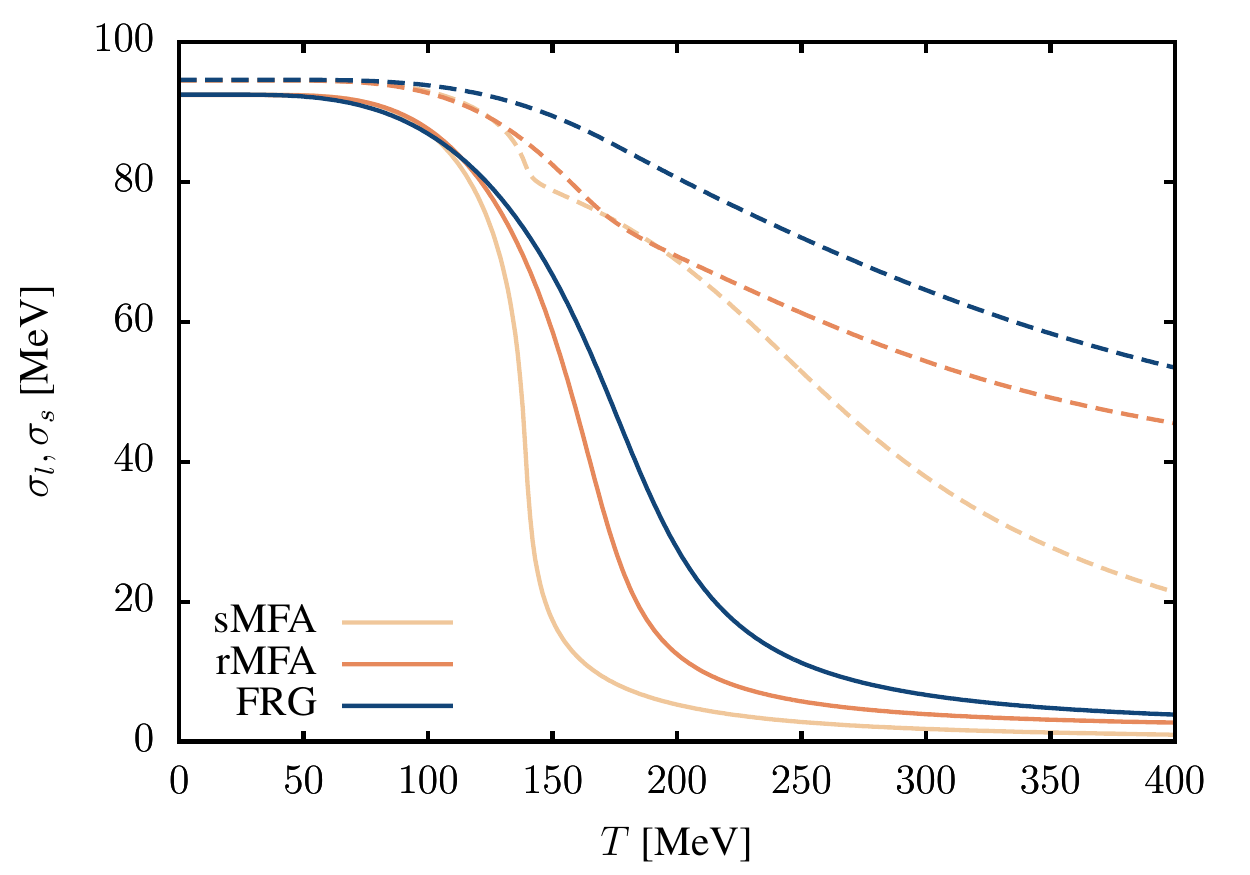}} 
\subfigure[\label{fig:condensates_mu_2p1} $T=0$.]{\includegraphics[width =
    \twofigs]{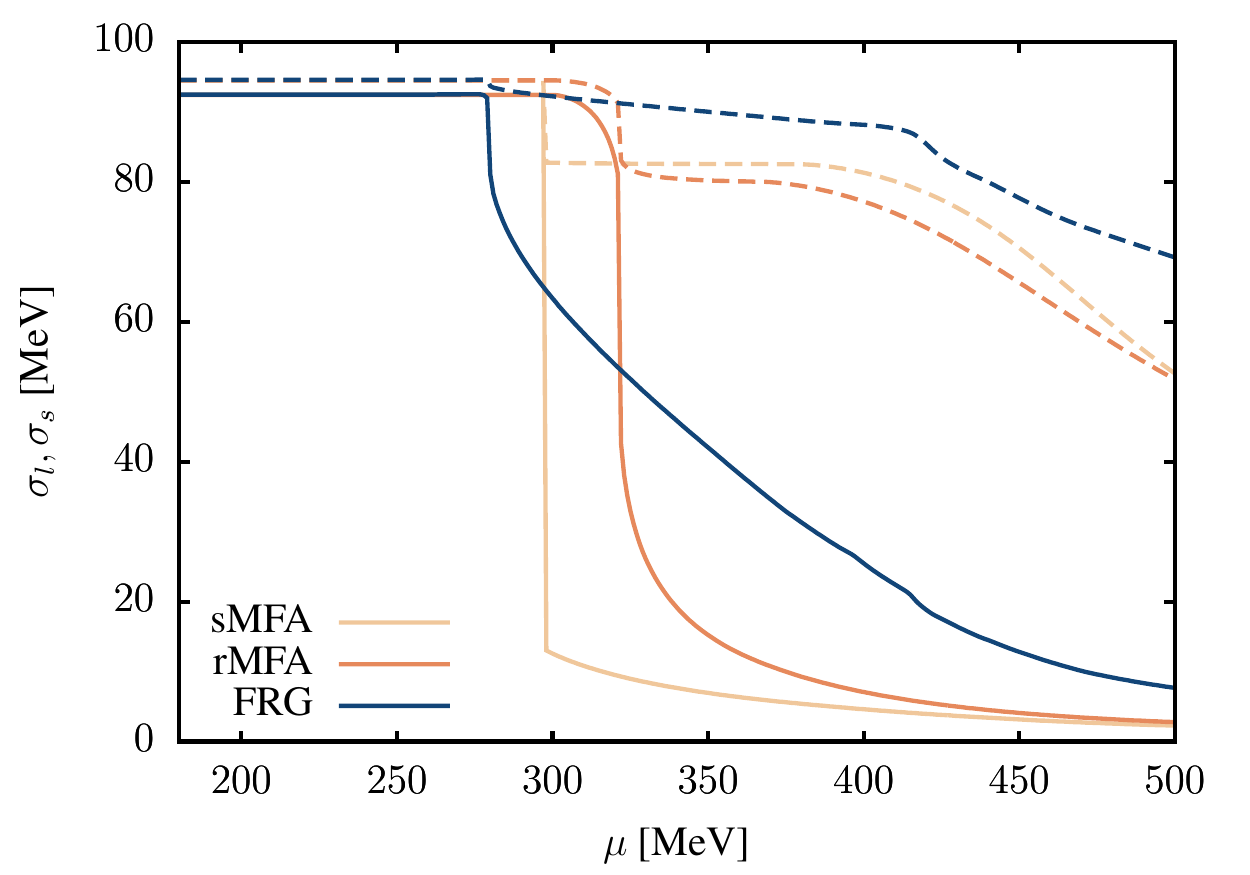}}
  \caption{Chiral condensates in the $N_f = 2+1$ flavor quark-meson
    model in three different approximations (FRG, rMFA and sMFA). The
    light condensates $\sigma_l$ (solid) are the lower lines and the
    strange condensates $\sigma_s$ (dashed) the upper lines.
  }
\end{figure*}

The bulk thermodynamics of the system is determined by the effective
potential which depends in the case of two flavors on one
(non-strange) condensate and for three or, more
  precisely, $(2+1)$ flavors on two (one non-strange and one
strange) condensates. The condensates are determined by minimizing the
total effective potential with respect to the corresponding fields,
here generically denoted by $\Phi$,
\begin{equation}
\label{eq:15}
\left. \frac{\partial \Omega}{\partial \Phi} \right|_{\Phi =
  \vev{\Phi}} = 0\ .
\end{equation}
This yields the chiral non-strange $\sigma_l$ and strange $\sigma_s$
condensates or expectation values as functions of the external
parameters, i.e. the temperature and the quark chemical potentials.
The thermodynamic pressure is just the negative of the
effective potential evaluated at the minimum and normalized in vacuum
\begin{equation}
\label{eq:12}
p = - \left( \left. \Omega- \Omega^0 \right) \right|_{\Phi = \vev{\Phi}}\  .
\end{equation}
In the FRG case, the infrared-evolved potential
$\Omega_{k=0} - \Omega_{k=0}^0$ has to be used.

By varying the external parameters the phase structure of the chiral
transition can now be analyzed. For simplicity, we consider here a
common chemical potential for all quark flavors, i.e., we set
$\mu_e = 0$ in \Eq{eq:beta_equil}.  Our numerical findings are
collected in \Fig{fig:phasediag_2flav} where the phase diagrams for
two quark flavors (dotted/solid lines) and for three flavors (dashed line)
are shown. The numerical input parameters and contingent cutoff
dependency are collected in \App{app:inputparam}; details on the
numerical implementation can be found in \App{app:solutiontech}.

Typical for the phase structure obtained with the FRG is the
back-bending behavior of the chiral phase transition line (blue lines)
for small temperatures characterized by a positive slope of the
critical chemical potential as a function of temperature. The critical
endpoints (CEPs) are denoted as dots in the figure and the crossover
regions/first-order transitions as dotted and solid lines,
respectively. The origin of this back-bending phenomenon is not yet
completely clarified. It might be related to the fact that additional
interaction channels should be considered, see \cite{Tripolt:2017zgc}
for a detailed discussion. For the chosen vacuum input parameters and
in the LPA truncation of the FRG equation, see
App.~\ref{app:inputparam}, the CEP is located at very small
temperatures, around $T \sim 10$ MeV for two and three quark
flavors. In mean-field approximations the thermodynamical behavior at
small temperatures is different and the first-order transition line
hits the chemical potential axis perpendicularly
\cite{Tripolt:2017zgc}.

Furthermore, since the inclusion of fluctuations generically
smoothens the chiral phase
transition, the crossover transition line is shifted to higher
temperatures if more fluctuations in the thermodynamic potential are
taken into account \cite{PhysRevD.96.016009}, except at very low
temperatures where the back-bending phenomenon takes over. This is
nicely demonstrated in \Fig{fig:condensates_T_2p1} where both the
light (solid lines) and the strange (dashed lines) chiral condensates
are shown as a function of temperature for vanishing quark chemical
potential.

In sMFA where only the thermal quark loop contribution is considered,
the pseudocritical crossover temperature $T_c \approx 140$ MeV at
$\mu=0$ is smallest. Already the inclusion of the vacuum quantum
fluctuations of the quarks, labeled as rMFA in the figures, lifts the
pseudocritical temperature by about 30 MeV. Interestingly, the whole
chiral phase transition is shifted constantly towards higher
temperatures by roughly this amount for all chemical potentials,
cf.~\Fig{fig:phasediag_2flav}.  As a consequence, the first-order
transition at $T=0$ is also pushed to higher chemical potentials as
visible in \Fig{fig:condensates_mu_2p1}.  This trend is continued at
least for moderate chemical potentials when additionally meson
fluctuations with the FRG are taken into account.  However, for
smaller temperatures and due to the back-bending of the transition
line, see.~\Fig{fig:phasediag_2flav}, the critical chemical potential
is pushed to smaller values in contrast to the
  previous argument. This will later be of
  relevance for the EoS.

%%%%%%%%%%%%%%%%%%%%%%%%%%%%%%%%%%%%%%%%%%%%%%%%%%%%%%%%%%%%%%%%%%%%%%%%%%%%%%%%
% figure EoS 
%%%%%%%%%%%%%%%%%%%%%%%%%%%%%%%%%%%%%%%%%%%%%%%%%%%%%%%%%%%%%%%%%%%%%%%%%%%%%%%% 
\begin{figure*}
  \centering
  \subfigure[\label{fig:eos_T0} Symmetric quark matter]{\includegraphics[width =
    \twofigs]{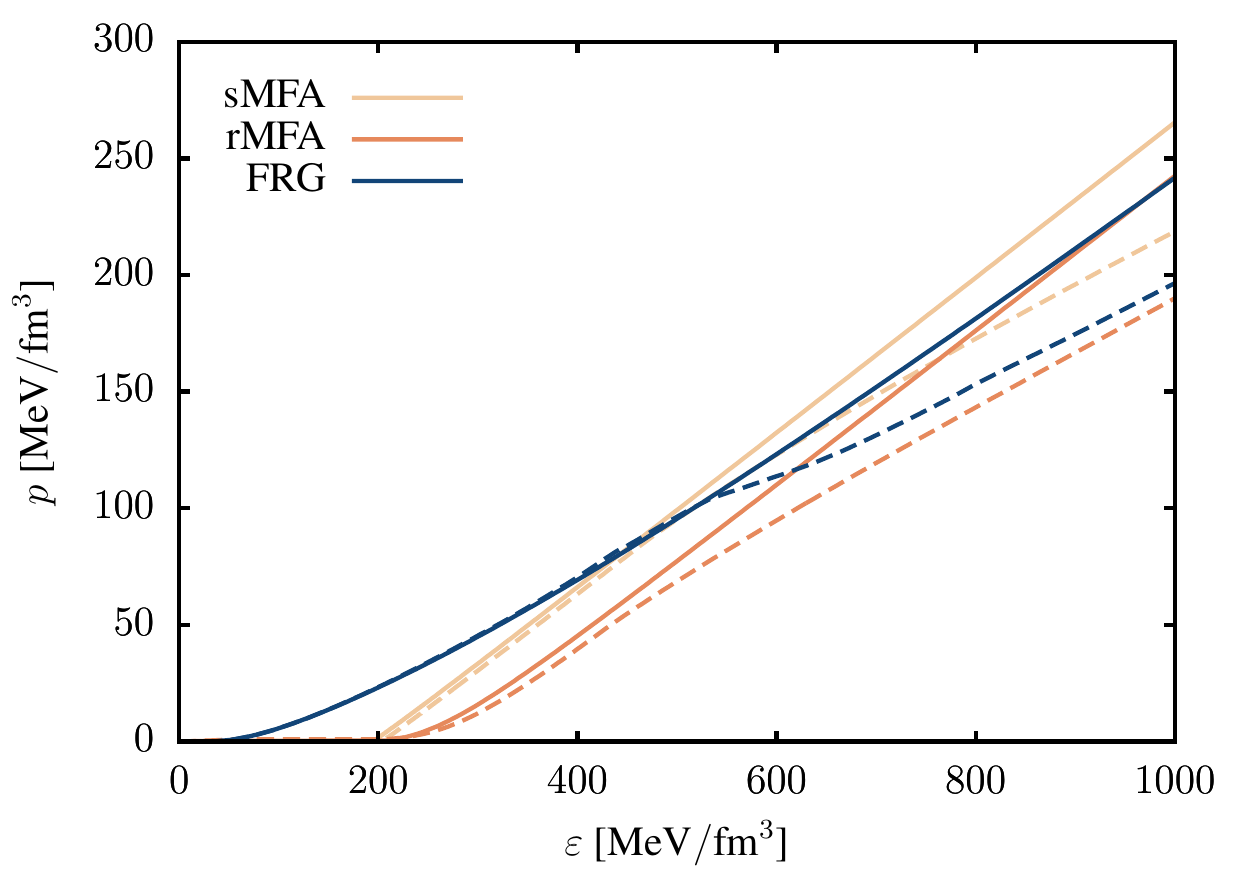}} 
\subfigure[\label{fig:eos_T0_musplit} $\beta$-stable and neutral matter]{\includegraphics[width =
    \twofigs]{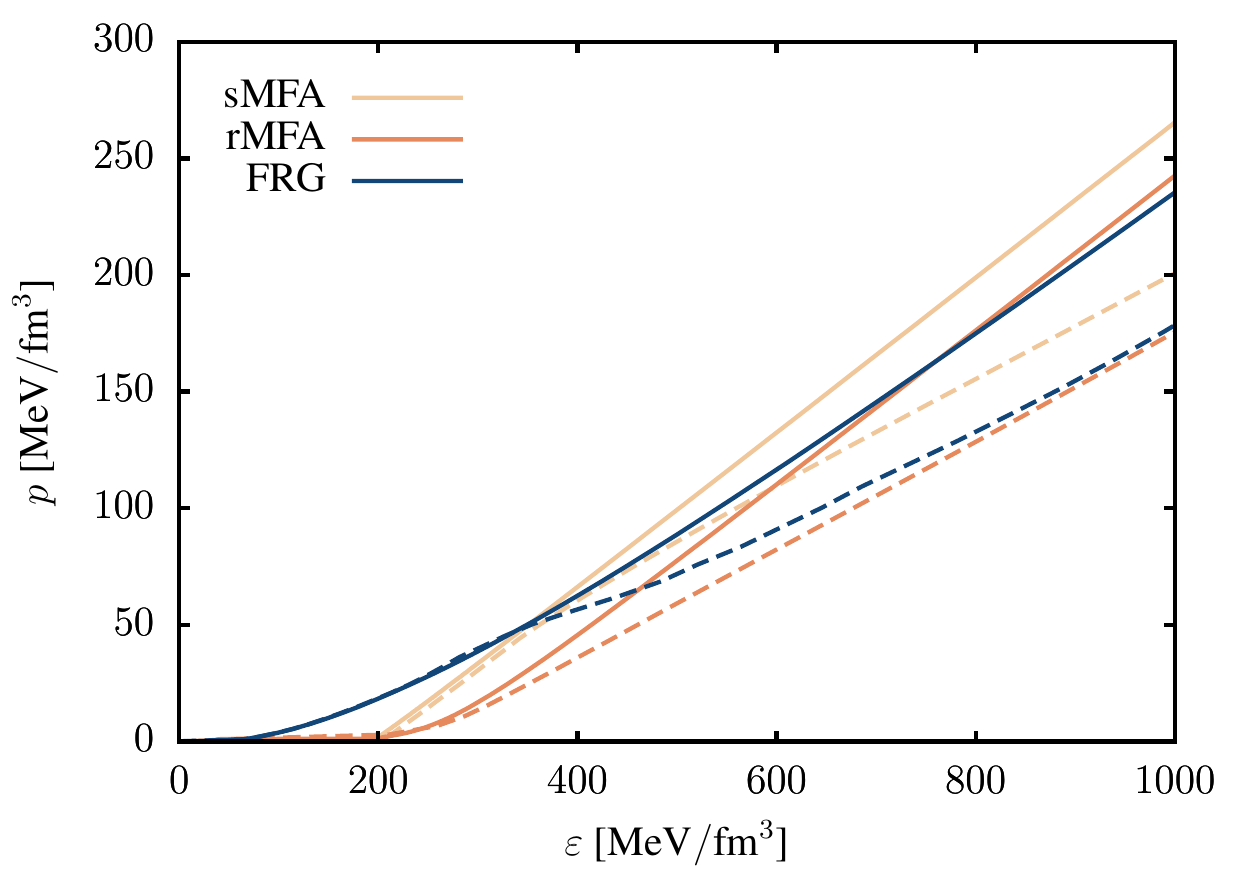}}
  \caption{Three different EoS of the $N_f = 2$
    (solid) and $N_f = 2+1$ (dashed) quark-meson model at $T=0$ for
    $m_l \approx 300$ MeV, $m_s \approx 430$ MeV and $m_\sigma = 560$
    MeV.}
\end{figure*}

All condensates exhibit for $T=0$ a first-order phase transition close
to $\mu \approx 300$ MeV corresponding to the light quark mass in the
vacuum. In sMFA, the first-order transition is strongest, in FRG
weakest. Hence, the gap in the FRG light condensate is quite small and
melts only moderately after the transition, still signaling a chirally
broken phase in this density regime of the phase diagram
\cite{Schaefer:2004en}. In rMFA the light condensate is
constant until $\mu=300$ MeV and melts down before the first-order
jump which is consistent with the Silver Blaze property. It is likely
that for a sigma mass below $560$ MeV the rMFA condensates immediately
jump at the light quark masses as well.

Just below a quark chemical potential of about $430$ MeV, the value
that coincides with the strange quark mass in the vacuum, a further
decrease is seen in all three strange condensates and a smooth chiral
phase transition takes place. 

However, when strange quarks are added to the system an opposite
behavior is found for vanishing and moderate chemical potentials and
the three-flavor crossover line is pushed down again, see the dashed line
in \Fig{fig:phasediag_2flav}. The difference to the two-flavor phase
structure shrinks for decreasing temperatures.  Below $T < 50$ MeV
almost no influence of the strange quark on the transition line is
observed where the dashed line merges with the solid two-quark flavor line.

\section{EoS for quark and hybrid stars}
\label{sec:eos_quark_hybrid}

The equations of state (EoS) for symmetric quark matter, i.e., for
equal chemical potentials, obtained in MFAs and with the FRG, are
compared to each other in \Fig{fig:eos_T0}. Solid lines are the
two-quark flavor findings and the dashed lines the corresponding
three-flavor calculations. The numerical results are almost
insensitive to the strange quark before the onset of the strange
chiral phase transition around energy densities
$\varepsilon \approx 550$ MeV/fm$^3$ but start to deviate thereafter,
see \Fig{fig:eos_T0}. In MFAs the transition is more gradually
realized and the deviation is less pronounced than within the FRG.
This can already be seen in
  Fig.~\ref{fig:condensates_mu_2p1} where the strange condensate in
  MFA decreases moderately for chemical potentials smaller than the
  strange quark mass $m_s \approx 430 \, \text{MeV}$. In the FRG curve
  there is a more rapid onset with a steeper slope around the strange
  quark mass. Note that this behavior could also be related to a
  second phase transition in the light quark sector as has been found
  in \cite{Schaefer:2004en}.

Furthermore, it is obvious that vacuum fluctuations reduce the slope
of the EoS, i.e. the sound speed, and over most of the shown density
range the EoS obtained in FRG has still a smaller slope, see
Sec.~\ref{sec:speed_of_sound}.

With the inclusion of a free relativistic electron gas and the
conditions for weak equilibrium and charge neutrality,
cf. Eqs. \eqref{eq:beta_equil} and \eqref{eq:charge_neutrality}, we
obtain slightly modified EoS. The results for $\beta$-stable and
neutral matter are presented in \Fig{fig:eos_T0_musplit}.  Differences
to the result for symmetric quark matter are visible almost
exclusively for $N_f=2+1$ where the weak equilibrium and charge
neutrality conditions render the population of strange quarks more
favorable. Hence, the onset of strangeness is pushed to smaller energy
densities, leading to a pressure reduction for a given energy density
beyond the onset.  

Note that in charge neutral and $\beta$-equilibrated matter isospin
symmetry is broken and the chemical potentials of the \textit{up} and
\textit{down} quarks split. However, our approximation with only one
chiral light condensate $\sigma_l$ for both \textit{up} and
$\textit{down}$ quark flavors yields in all cases degenerated
\textit{up-} and \textit{down-}quark masses. For
  large $\mu$ the restoration of chiral symmetry in the light quark
  sector suppresses both quark masses such that only small mass
  differences are expected there in contrast to the behavior in the
  vicinity of the chiral transition. The
impact on the EoS might be more pronounced, and the small difference
between symmetric and charge neutral matter in the two-flavor case
might be an artefact of this approximation.  For a more complete
analysis including isospin symmetry breaking the introduction of an
additional third chiral condensate is necessary which is, however,
beyond the scope of the present work.

In order to allow for a description of hybrid stars with a phase
transition from hadronic to quark matter in the interior of the star
we combine the quark matter EoS with a nuclear one. The transition is
achieved with a standard Maxwell construction\footnote{This assumes a
  high surface tension at the hadron-quark interface, see
  e.g. \cite{Glendenning:1992vb} for a discussion of this point in the
  context of a potential hadron-quark transition within hybrid stars.}
that maximizes the pressure for a given chemical potential. For the
nuclear EoS we consider some representative models compatible with
several nuclear physics constraints as well as the maximum neutron
star mass and the GW170817 tidal deformability.  Three of them are
energy density functional models, one is based on a non-relativistic
Skyrme parameterization, RG(SLy4) \cite{Gulminelli:2015csa,
  *Danielewicz:2008cm, *Chabanat:1997un}, and two are relativistic
mean field models, HS(DD2)~\cite{Typel:2009sy, *Hempel:2009mc} and
SFHo~\cite{Steiner:2012rk}. The BL EoS~\cite{Bombaci:2018ksa} is
formulated in the framework of the Brueckner-Bethe-Goldstone many-body
theory with chiral nuclear forces.

\begin{figure}[htb!]
  \centering
  \includegraphics[width=\onefig]{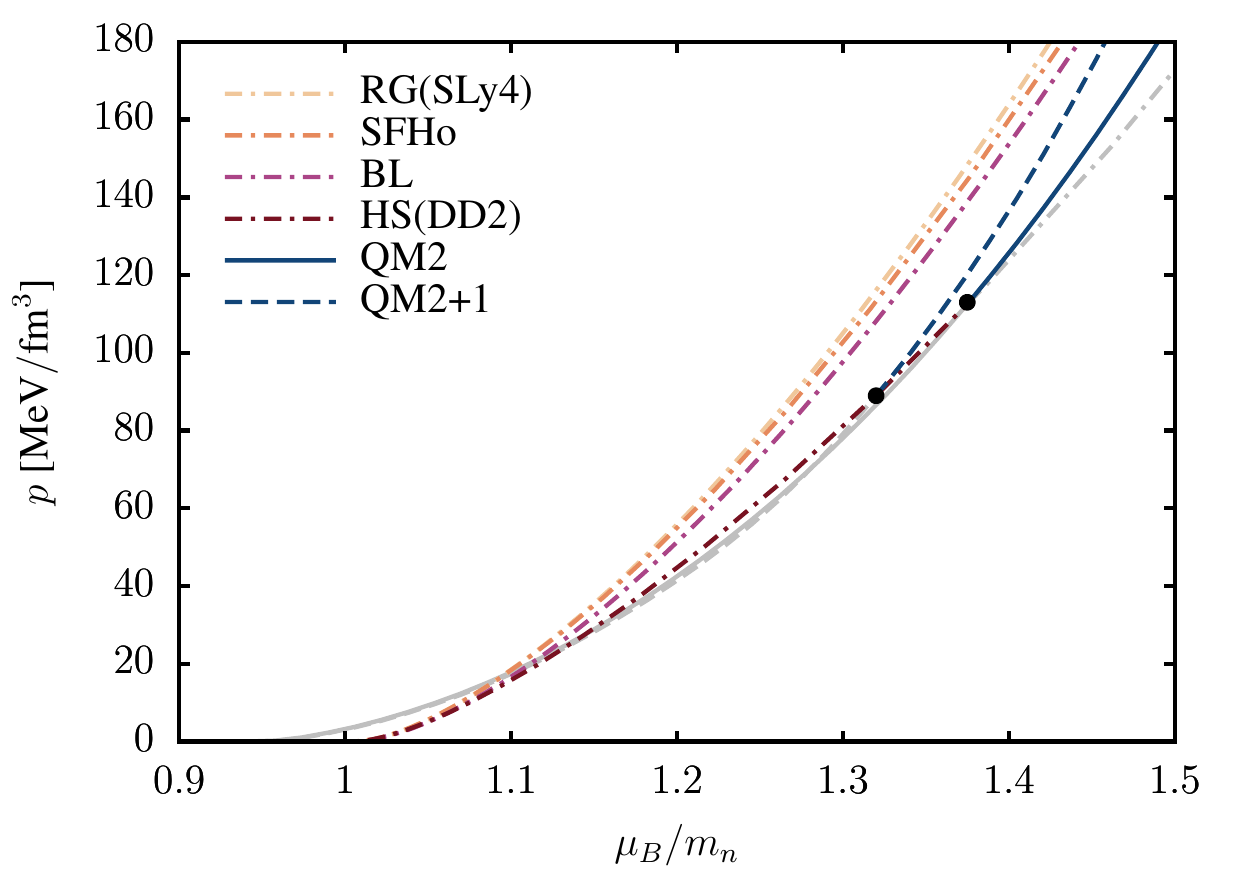}
  \caption{\label{fig:pmu_dd2} Pressure as a function of the baryon
    chemical potential in units of the neutron mass for various
    hadronic RMF models (dash-dotted) compared to the $N_f=2$ (solid)
    and $N_f=2+1$ (dashed) QM EoS in FRG ($\beta$-equilibrated and
    neutral quark matter).   Intersections (black dots) of the quark-matter
    pressure with the HS(DD2) model at higher pressure are found while 
    (unphysical) intersections at lower pressure are ignored, see text for details.}
\end{figure}

In \Fig{fig:pmu_dd2} a comparison of different nuclear EoS
(dash-dotted lines) with the $N_f=2$ (solid) and $N_f=2+1$ (dashed)
EoSs evaluated with the FRG respecting $\beta$-equilibrium and charge
neutrality is given. We do not consider the FRG results as very
realistic at low densities below $\mu_B/m_n \lesssim 1.2$, where 
a hadronic phase is expected. In addition,
the attractive meson interactions in the QM model, lead to a very high
pressure for a given chemical potential in this range. We only present
it for completeness. Disregarding the unphysical part of the QM EoS
obviously all nuclear EoS except the HS(DD2) EoS produce higher
pressure than the QM one at a given baryon chemical potential
($\mu_B = 3 \mu$) for the entire range of interest for compact
stars. Hence, no hybrid stars could exist with these model
combinations.  The pressure of the HS(DD2) EoS intersects the
two-flavor FRG pressure curve around $\mu_B/m_n \approx 1.38$
corresponding to the appropriate physical transition from nuclear to
quark matter.

\begin{figure}[htb!]
  \centering
  \includegraphics[width=\onefig]{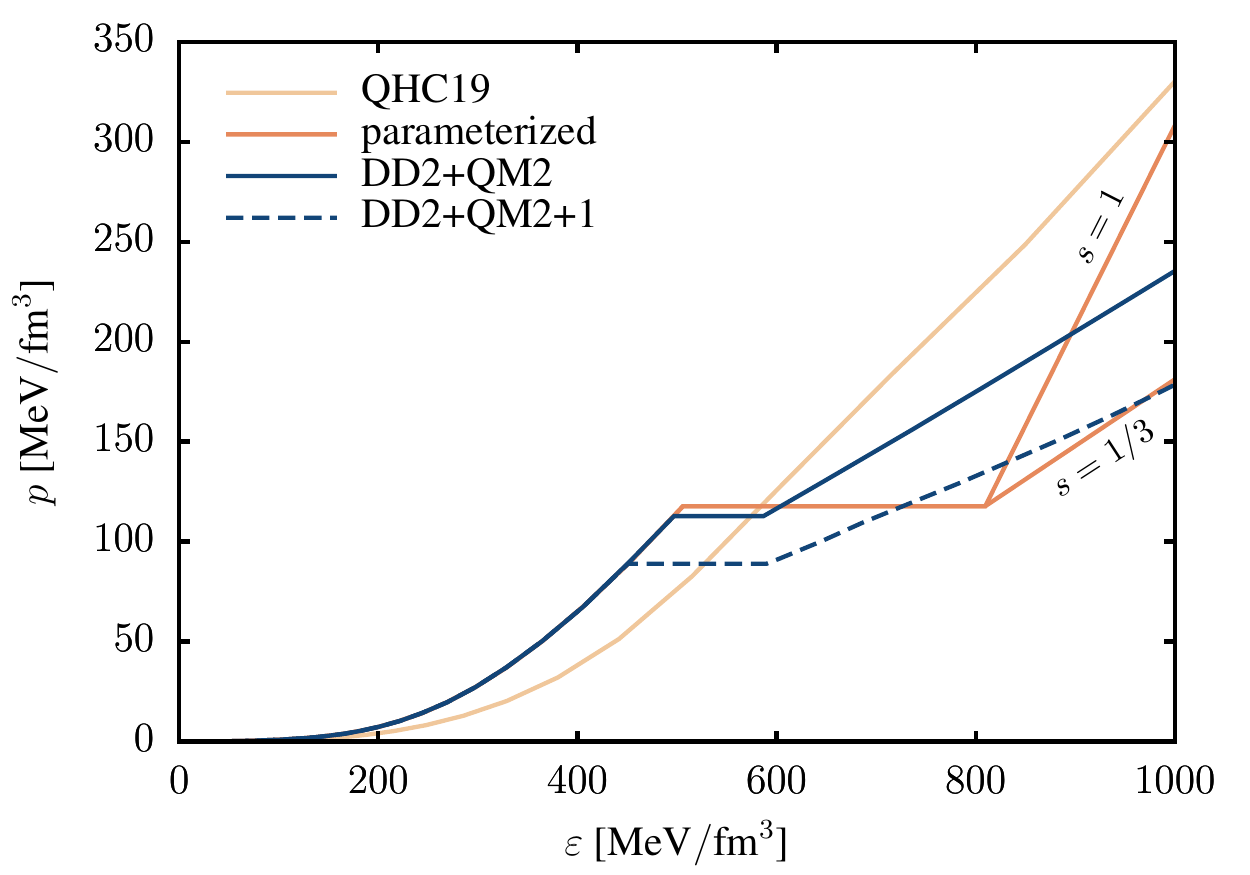}
  \caption{\label{fig:eos_hybrid} Composite EoS for the QM and DD2
    nuclear model, cf. \Fig{fig:pmu_dd2}, compared with the hadron
    quark EoS QHC19~\cite{Baym:2019iky} and a combination of the
    HS(DD2) EoS with a parameterized quark matter EoS~\cite{Alford:2017ly} 
    for $c_s^2 \equiv s = 1$ or 1/3,
    respectively. While the QHC19 model features a continuous
    quark-hadron transition, the others employ a Maxwell-constructed
    first-order transition and a discontinuity in the energy density
    $\varepsilon$.}
\end{figure}

By construction, the combination of the two-flavor QM EoS from the FRG with the
HS(DD2) EoS leads to a first-order phase transition between the
confined nuclear matter and the deconfined quark matter which is
characterized by a discontinuity in the energy density. This can be
observed in \Fig{fig:eos_hybrid} which depicts the constructed hybrid
$N_f=2$ (DD2+QM2) and $N_f=2+1$ (DD2+QM2+1) EoSs in comparison to the
hybrid EoS QHC19 \cite{Baym:2019iky} and a parameterized EoS
\cite{Alford:2017ly}.\footnote{Note that all employed nuclear EoSs, the 
  QHC19 EoS as well as our hybrid star
  EoSs with the FRG are available online in
  the CompStar Online Supernovae Equations of State (CompOSE)
  database \cite{Typel:2013rza}, see \url{https://compose.obspm.fr/}.} 
The QHC19 EoS features a smooth crossover
quark-hadron transition.  For the parameterized EoS the HS(DD2) EoS is
used for the hadronic regime and the quark matter side is parameterized
as
\begin{equation}
p(\varepsilon) = 
\begin{cases}
	p_c, & \varepsilon_c < \varepsilon < \varepsilon_c + \Delta \varepsilon \\
	p_c + s \ [\varepsilon - (\varepsilon_c + \Delta \varepsilon)], & \varepsilon > \varepsilon_c + \Delta \varepsilon \ ,
\end{cases}
\end{equation}
which describes a first-order transition at $\{\varepsilon_c,p_c\}$
with energy gap $\Delta \varepsilon$ and a constant slope
$s = \partial p/ \partial \varepsilon$, i.e. a constant quark matter
sound speed squared, thereafter. Following the idea of
Ref. \cite{Alford:2017ly}, we choose
$p_c = 1.89 \times 10^{35} \ \text{dyn cm}^{-2}$ and
$\varepsilon_c = 9.02 \times 10^{14} \ \text{g cm}^{-3}$ which
corresponds to $n_c \approx 3 n_0$ and
$\Delta \varepsilon/\varepsilon_c = 0.6$.  For the slope, we consider
two extreme parameterizations, one with $s=1/3$ corresponding to the
asymptotical QCD value, and one with $s=1$ corresponding to the
maximally allowed sound speed by causality.  This ensures the onset of
the phase transition at similar densities to those found in our hybrid
construction with two-flavor QM. The large gap in the energy density
has been chosen to produce twin stars, i.e. an additional branch of
hybrid stars with the same mass but different radius than their
nuclear counterpart \cite{Alford:2013aca}.  Such twin star
configurations are found, for example, in studies with NJL quark
matter including additional repulsive 8-quark interactions
\cite{Benic:2014jia}.  A comparison of that parameterization to the
FRG calculation in \Fig{fig:eos_hybrid} reveals that the energy gaps
in our construction are much too small to produce a disconnected
second branch but favor a single connected hybrid-nuclear
branch. In fact, our two-flavor hybrid model is well reproduced
  by values of $\Delta \varepsilon/\varepsilon_c = 0.18$ and
  $p_c/\varepsilon_c = 0.23$ and a constant $c_s^2=1/3$ on the quark
  matter side. These values lie well in between the region identified
  in \cite{Alford:2013aca} as that giving rise to a connected
  branch. Similar arguments show that the hybrid model with 2+1
  flavors on the quark side again leads to a connected branch.

%%%%%%%%%%%%%%%%%%%%%%%%%%%%%%%%%%%%%%%%%%%%%%%%%%%%%%%%%%%%%%%%%%%%%%%%%%%%%%%%
% speed of sound
%%%%%%%%%%%%%%%%%%%%%%%%%%%%%%%%%%%%%%%%%%%%%%%%%%%%%%%%%%%%%%%%%%%%%%%%%%%%%%%%
\begin{figure*}
  \centering
  \subfigure[\label{fig:speed_of_sound} Symmetric quark matter]{\includegraphics[width =
    \twofigs]{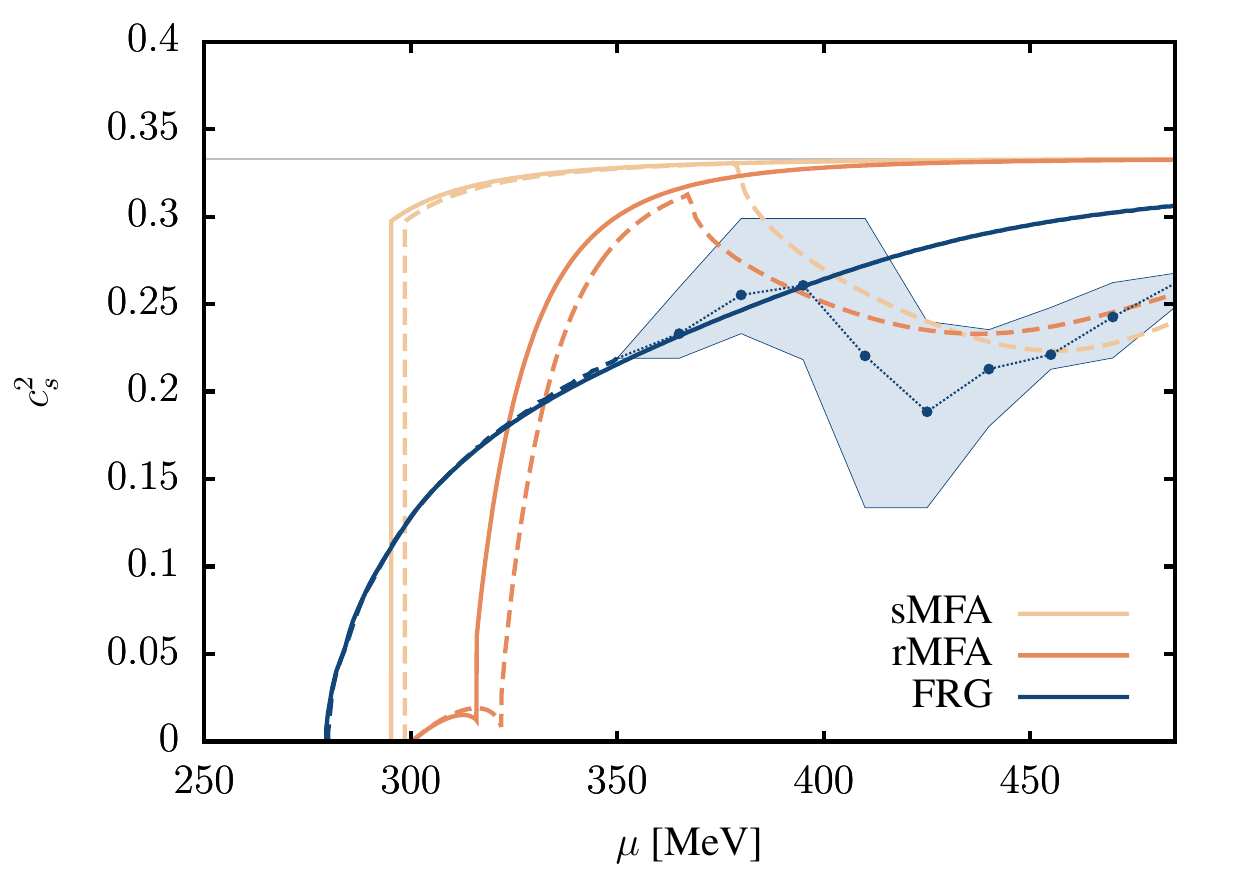}} 
\subfigure[\label{fig:speed_of_sound_musplit} $\beta$-stable and neutral matter]{\includegraphics[width =
    \twofigs]{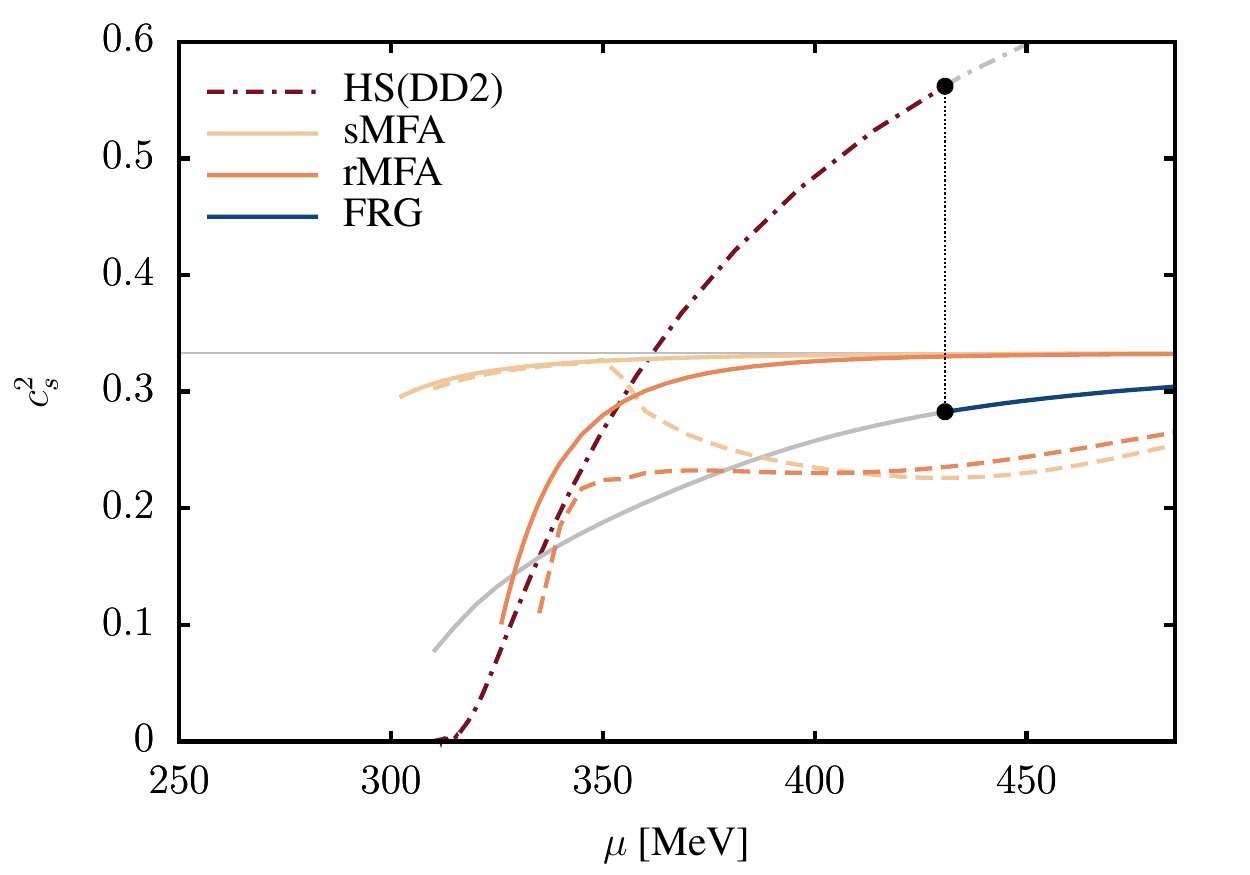}}
  \caption{Sound speed squared $c_s^2$ at zero temperature as a
    function of $\mu$ for three different quark matter EoS and the
    hadronic DD2 EoS (solid $N_f=2$, dashed $N_f=2+1$; left panel:
    flavor symmetric matter, right panel: $\beta$-equilibrated neutral
    matter). The thin horizontal lines indicate the Stefan-Boltzmann
    limit $c_s^2=1/3$. The vertical black-dotted line in the right
    panel illustrates the first-order transition from the hadronic HS(DD2)
    to the QM2 EoS, cf. \Fig{fig:eos_hybrid}.  The metastable
    phases are extrapolated in gray color. For the blue error in the
    left panel band see App.~\ref{app:solutiontech}. }
\end{figure*}

\subsection{Sound speed}
\label{sec:speed_of_sound}

More information on dense matter can be gained through a detailed
investigation of the speed of sound \cite{Tews:2018kmu}. It measures
the stiffness of the EoS for a one fluid flow by the thermodynamic
derivative of the pressure with respect to the energy density at
constant entropy and particle numbers
\begin{equation}
\label{eq:9}
c_s^2 \equiv \left. \frac{\partial p (\varepsilon)}{ \partial \varepsilon}
\right|_{S,N_i}
\end{equation}
and can be identified as the speed of propagation of sound waves. 
Causality implies an upper bound $c_s^2 \leq 1$ and thermodynamic
stability a lower bound $c_s^2 >0$. For an ideal gas composed of
point-like ultrarelativistic (massless) components the squared speed
of sound is equal to one third, $c_s^2 = 1/3$. This is common to all
systems with conformal symmetry of which an ideal massless gas is just
an example. Even for any strongly interacting system the vanishing of
the trace of the energy-momentum tensor, a feature of conformal
theories, implies that the energy density is connected to the pressure
by $\varepsilon = 3p$ hence yielding $c_s^2 = 1/3$ independently of
density, temperature, or interactions. The speed of sound is decreased
such that $c_s^2 < 1/3$ when a mass for the components is included or
when (perturbative) interactions among the components take place. In
the case of QCD at asymptotically high densities or temperatures, far
exceeding the densities in the core of compact stars, a weak coupling
expansion is valid (pQCD) such that $c_s^2$ is expected to reach the
conformal limit with increasing density from below
\cite{Bedaque:2014sqa}. This behavior is confirmed in QCD lattice
calculations at finite temperature as well as at zero and small baryon
chemical potentials \cite{Borsanyi:2012cr}.

The speed of sound has also been investigated in alternative theories
for which for example the AdS/CFT correspondence holds where
calculations in the strong coupling limit are feasible, see
e.g.~\cite{Benincasa:2006ei}. It has been conjectured that $c_s^2$ is
always bounded from above in such classes of
strongly coupled field theories by the conformal value of $1/3$
\cite{Cherman:2009tw} although recently counterexamples have been
presented \cite{Ecker:2017fyh}. For more details of this conjecture
see \cite{Bedaque:2014sqa}.

The speed of sound of the QM model in both mean-field approximations
and the FRG calculation is found generally to be always smaller than
$c_s^2=1/3$.  An alternative scenario could be the presence of a bump
in $c_s^2$ at intermediate densities before approaching the upper
bound from below asymptotically and thus implying the existence of a
maximum and a local minimum of $c_s^2$ as a function of the chemical
potential. This scenario is supported by another recent FRG analysis
including diquark condensation \cite{Leonhardt:2019fua} where a
maximum in $c_s^2 $ above $1/3$ is found. The additional inclusion of
vector interactions in the quark-meson model \cite{Zhu:2018ys} is also
expected to stiffen the EoS.

Our result for the speed of sound of quark matter with a
flavor-symmetric chemical potential is shown in
\Fig{fig:speed_of_sound}.  In the $N_f=2$ mean-field approximation
(solid lines) the speed of sound converges to the limit $c_s^2 = 1/3$
while the addition of strange quarks (dashed lines) leads to a
reduction of $c_s^2$ around scales of the strange chiral phase
transition. This behavior has already been observed in
Fig. \ref{fig:eos_T0} and is expected due to the additionally
populated strange quark states softening the EoS. In the FRG solution,
the speed of sound is generally smaller than the asymptotic mean-field
values beyond the transition which can be attributed to the quantum
fluctuations captured within the FRG approach.

Furthermore, the strength of the first-order chiral phase transition,
i.e. the gap in the order parameter, is found to correlate with the
size of the jump in the speed of sound.  Hence, the strong first-order
transition in sMFA leads to a jump of $c_s^2$ close to its asymptotic
Stefan-Boltzmann value $c_s^2=1/3$ which leads to the almost linear
behavior of the EoS even at low pressure, see \Fig{fig:eos_T0}.  The
more washed-out transition in rMFA induces an initially smaller slope
of the EoS. This becomes more significant in the FRG calculation: due
to an even smoother transition a comparably small gap
$\Delta \varepsilon$ is found in \Fig{fig:eos_T0}. In agreement with
\Fig{fig:speed_of_sound} the slope is consistently smaller than that
of the mean-field calculations.

For $N_f=2+1$, $c_s^2$ is found to be sensitive to the
numerical error caused by the employed solution method of the flow
equation which leads to visible fluctuations at high $\mu$. Therefore,
\Fig{fig:speed_of_sound} depicts for $\mu > 350 \, \text{MeV}$
averaged values in conjunction with error intervals displayed as blue
band.  For more technical details see App. \ref{app:solutiontech}.

The speed of sound for $\beta$-stable and charge-neutral quark matter
is displayed in \Fig{fig:speed_of_sound_musplit}. Note that due to the
usage of only one light condensate, the first-order transition cannot
be resolved exactly in this approximation and hence the drop of the
speed of sound to zero at low chemical potential is not shown.  Due to
the numerical uncertainties mentioned above, we postpone a careful
$N_f=2+1$ analysis to a future work. Qualitatively, we observe the
same behavior as for a flavor-symmetric chemical potential. However,
in mean-field approximation we find that the reduction of the speed of
sound due to the onset of strangeness takes place at lower chemical
potentials and more gradually than in symmetric quark matter.  This is
in agreement with the onset of strangeness already at small energy
density, see Fig.~\ref{fig:eos_T0_musplit} and the discussion in the
previous section. In \Fig{fig:speed_of_sound_musplit} we also show
the speed of sound for the HS(DD2) nuclear EoS and indicate the
transition point to quark matter in the DD2+QM2 EoS by a vertical
line.  As suggested in \cite{Ibanez:2018myp}, this discontinuity in
the speed of sound can be related to a $\delta$-function singularity
in the fundamental derivative, leading to possibly non-convex
thermodynamics.

%%%%%%%%%%%%%%%%%%%%%%%%%%%%%%%%%%%%%%%%%%%%%%%%%%%%%%%%%%%%%%%%%%%%%%%%%%%%%%%%
%  TOV results
%%%%%%%%%%%%%%%%%%%%%%%%%%%%%%%%%%%%%%%%%%%%%%%%%%%%%%%%%%%%%%%%%%%%%%%%%%%%%%%%
\begin{figure*}
  \centering
  \subfigure[\label{fig:mr_musplit} Pure quark stars]{\includegraphics[width =
    \twofigs]{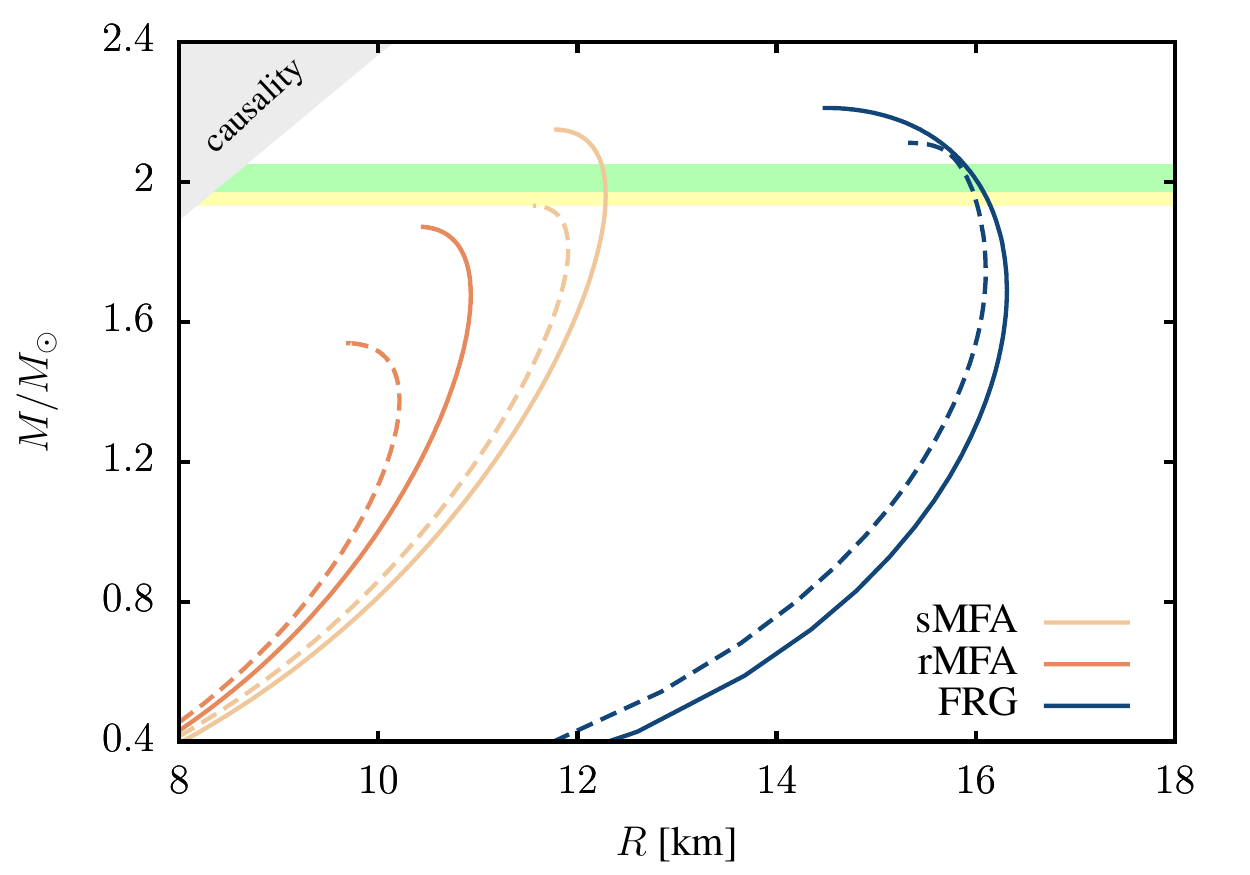}} 
\subfigure[\label{fig:mr_hybrid} Hadronic (dash-dotted) and hybrid (solid, dashed) stars]{\includegraphics[width =
    \twofigs]{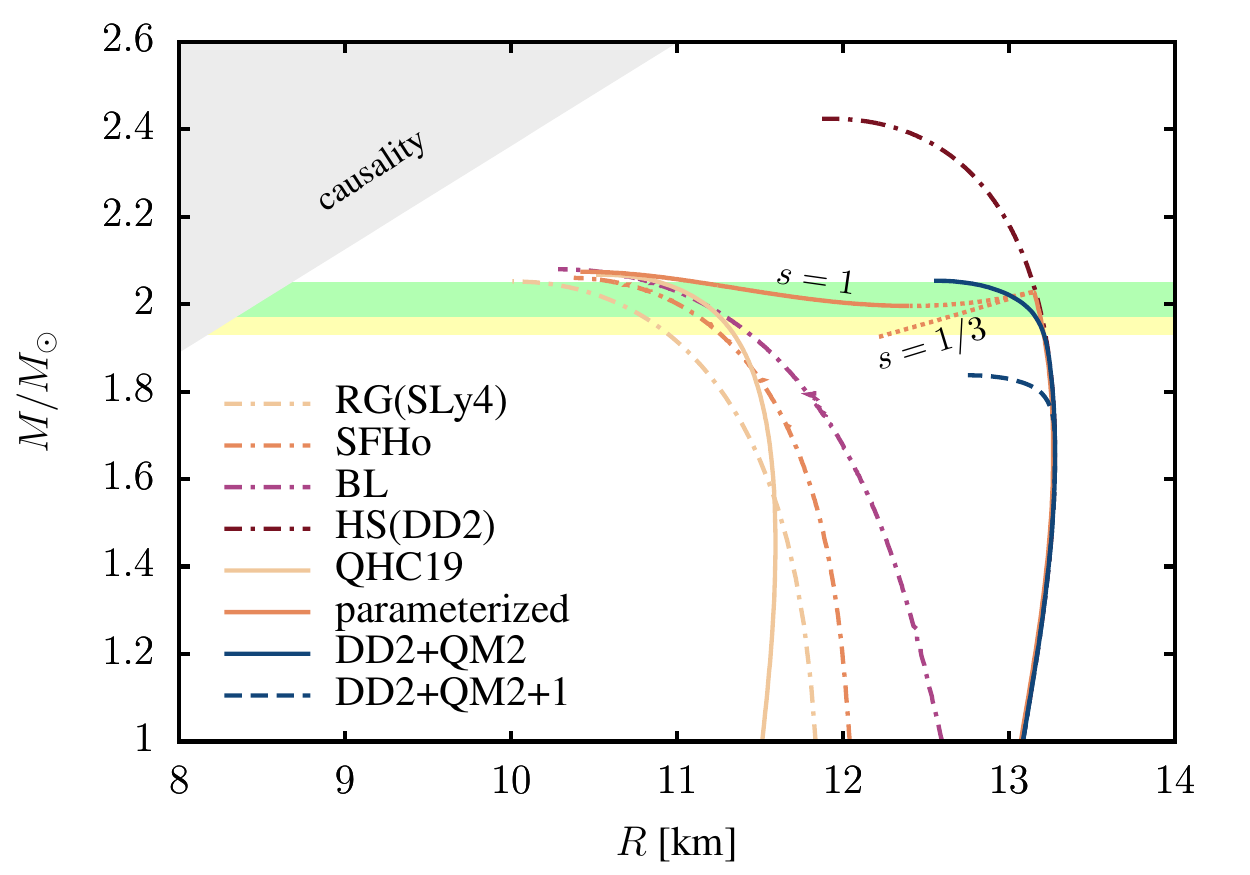}}
  \caption{Neutron star mass-radius relations for $\beta$-equilibrated
    and neutral matter. Left: pure quark stars for $N_f=2$ (solid) and
    $N_f=2+1$ (dashed). Right: purely hadronic stars (dash-dotted)
    obtained from various nuclear model EoS, cf.
    Figs. \ref{fig:pmu_dd2} and \ref{fig:eos_hybrid}.  Hybrid stars
    from the combined HS(DD2) model  with the FRG solutions for QM (solid,
    dashed). Horizontal bands: PSR J1614-2230 (yellow) and PSR
    J0348+0432 (green) mass measurements
    \cite{Demorest:2010bx,*Antoniadis:2013pzd}. See text for details.
  }
\end{figure*}

\subsection{Neutron star models}
\label{sec:star-models}

In order to determine the influence of fluctuations
  on the mass-radius relation of a neutron star, we employ hydrostatic
  equilibrium solutions for a relativistic spherically symmetric
  compact star composed of a perfect fluid, which have been derived
  from Einstein's equations by Tolman, Oppenheimer and Volkoff (TOV)
  \cite{PhysRev.55.364, *PhysRev.55.374}. In Schwarzschild
coordinates the corresponding equations determining the pressure $p$
and the enclosed gravitational mass $M$ of the star as
a function of the
radius $r$ read\footnote{In this work we employ natural units
  $c= \hbar = 1$.}
\begin{eqnarray}
  \label{eq:tovp}
  \frac{dp(r)}{dr} &=& - \frac{G}{r}\frac{\big[ p(r) + \varepsilon(r) \big] \big[
    M(r) + 4\pi r^3 p(r) \big]} {\big( r-2G M(r)\big) }\ \ \\
  \label{eq:tovm}
\frac{dM(r)}{dr} &=& 4\pi r^2 \varepsilon(r)\ ,
\end{eqnarray}
wherein $\varepsilon$ denotes the energy density and $G$ the
gravitational constant.  For a given EoS in terms of $p(\varepsilon)$
as input these equations can be integrated from the origin for a given
choice of a central pressure $p_0$, i.e.  $p(r=0) = p_0$ and
$M(r=0) =0$.  The value of the radius where the pressure vanishes,
i.e. $p(r=R)=0$, defines the surface and thus the gravitational mass $M(R)$ and radius
$R$ of the star. Varying the unknown central pressure $p_0$ yields the
mass-radius relation.

The mass-radius relations for the pure $\beta$-stable and
charge-neutral quark matter EoS are shown in \Fig{fig:mr_musplit} for
three different approximations. Please keep in mind that quark
  matter is not absolutely stable within our setup and that thus such
  pure quark stars could not exist. We, nevertheless, show the
  mass-radius relations, since they are instructive to understand the
  impact of the fluctuations. Solid lines are the solutions for a
two-flavor EoS and dashed lines the three-flavor results. In general,
all three-flavor calculations yield a smaller maximum mass than the
corresponding two-flavor results, which can be understood by the
softening of the EoS due to the additional strange degrees of
freedom. Only the two-flavor sMFA and both the two- and three-flavor
FRG results yield a maximum mass above 2$M_\odot$. Furthermore, the
inclusion of the renormalized vacuum fluctuations in the rMFA in
contrast to the sMFA leads to smaller masses and radii. The additional
consideration of mesonic fluctuations via the full FRG computation
increases the maximum mass even slightly beyond the sMFA result but
also leads to significantly larger radii. The reason is that within the present FRG setup, the density jump at the surface is much smaller than in the MFA calculations, see Fig. \ref{fig:eos_T0_musplit}, such that the star becomes much less compact and can accumulate more mass. For the sake of completeness
the causality constraint $R \leq 2.87 GM$ \cite{Lattimer:2006xb} is
also displayed in the figure.

The mass-radius relations from the combined EoSs for a hybrid star
with the $N_f=2$ or $N_f=2+1$ quark-meson matter side employing the
FRG, respectively, and a hadronic phase parameterized by the HS(DD2)
EoS are presented in \Fig{fig:mr_hybrid}, labeled again as DD2+QM2 and
DD2+QM2+1.  For $N_f=2$, the onset of quark matter leads to the
visible separation of the DD2 and the DD2+QM2 curve slightly below
$M = 2 M_\odot$ corresponding to a central baryon number density of
approximately $0.47 \, \text{fm}^{-3}$. Below this value, the hybrid
star mass-radius-relation coincides with the nuclear HS(DD2) one as it
should. For $N_f=2+1$, the DD2+QM2+1 curve exhibits a similar
behavior, but the onset of quark matter occurs at a smaller baryon
number density of approximately $0.43 \, \text{fm}^{-3}$.  Thus, since
the transition occurs well above nuclear saturation density, both
hybrid EoSs satisfy constraints from nuclear physics as implemented in
the HS(DD2) EoS. The maximum hybrid star mass of about $2.1 M_\odot$
for $N_f=2$ complies well with current observations, whereas the
$N_f=2+1$ curve does not satisfy the $2M_\odot$ limit. Since the quark
matter onset occurs only for masses slightly below $1.8 M_\odot$ and
higher, the value of the GW170817 tidal deformability obtained from
both hybrid EoS does not change significantly with respect to the
HS(DD2) value $\tilde\Lambda\approx795$ for a mass ratio of 0.8 of the
two coalescing stars. It is in slight tension with recent LIGO/Virgo
data~\cite{GW170817,Abbott:2018fj} but in agreement with the
observation that the HS(DD2) EoS leads to a relatively large radius
for intermediate mass stars. The QHC19 model leads to a radius smaller
by almost 2 km.  For comparison, the $M$-$R$ relations for four
different pure nuclear RMF models (dash-dotted lines) are also shown
in \Fig{fig:mr_hybrid}. As mentioned above, due to the small stiffness of the
FRG quark matter EoS at high densities, a hybrid
star construction with these nuclear EoS is not possible within the present setup.

As expected from \Fig{fig:eos_hybrid}, the parameterized EoS leads to
a kink in the mass-radius relation at a mass slightly above the
DD2+QM2 curve.  Contrary to the connected hadronic and hybrid branches
in the latter, the large energy gap of the $s=1$ parameterized EoS
leads to a disconnected hybrid branch and therefore twin stars at
masses of about $2M_\odot$.  For $s=1/3$, the pressure in the quark
matter phase is not sufficient to counteract the strong gravitational
pull due to the large energy density of the quark core,
cf. Ref. \cite{Alford:2017ly}, and thus does not support a stable
hybrid star branch. Hence, we can rule out the occurrence of twin
stars in our model due to the small energy gap at the phase transition
from nuclear matter to quark matter and due to the small stiffness of
the quark matter EoS.  In case of a Maxwell construction with
parameters that feature a larger energy gap, there might not even be
any stable hybrid stars with a QM model quark matter description.

In summary we found that it is feasible to construct a hybrid
nuclear-quark EoS where the quark matter part is obtained from an
effective theory within the functional renormalization group
approach. In particular, we would expect that the inclusion of
repulsive interaction channels suspected to play a significant role
might stiffen the quark matter EoS to a degree that allows for a
realistic description of combined hadronic and quark matter with other
hadronic models. Furthermore, it might deepen our understanding of the
role and possible abundance of strange matter in compact stars.

\section{Summary and conclusions}
\label{sec:summary}

The core of neutron stars contains strongly interacting matter at
extreme densities, reaching several times nuclear matter saturation
density for the most massive ones. Observations of neutron star
properties, in particular precise high mass
determinations~\cite{Arzoumanian:2017puf,*Demorest:2010bx,*Antoniadis:2013pzd}
and the GW170817 tidal deformability~\cite{GW170817, Abbott:2018fj},
thereby put constraints on the equation of state for conditions not
accessible to experiments. Although much recent progress has been
achieved to pin down the EoS, many questions remain open, in
particular on the composition of the inner core: does it contain
non-nucleonic degrees of freedom?  In addition to hyperons, nuclear
resonances or mesons, a quark matter core might appear. In this paper
we have performed a non-perturbative study of the quark matter EoS.
We have considered a two- and three flavor quark-meson model,
employing different approximations from mean-field to the functional
renormalization group. Apart from including fluctuations in a
non-perturbative way, the FRG has the potential
  power to generate the coupling parameters at low-energy scales from
  the underlying fundamental QCD at the UV, which would otherwise be
  just model parameters.

The quark-meson model fully incorporates chiral symmetry breaking and
we have performed a first study of the impact of quantum and density
fluctuations on the EoS for vanishing temperature. Since the different
approximations of the grand potential were fixed to the same input
parameters, our numerical findings are solely attributed to the impact
of the fluctuations. As anticipated from studies of the phase diagram
and confirmed by our investigations, fluctuations tend to smoothen the
chiral phase transition, see
e.g. \cite{Mitter:2013fxa}. Within the EoS, the softening due to the
appearance of strange quarks is therefore pushed to higher densities 
because the strange quarks are still heavier 
than the light quarks at higher densities. Quark stars obtained
from the FRG, including fluctuations in the quark matter EoS, have
higher maximum masses and radii compared with their mean-field
counterparts. Furthermore, we have constructed a hybrid star EoS,
combining our FRG quark matter EoS with a nuclear one via a Maxwell
construction. The results for hybrid stars with a two-flavor quark
matter core are in reasonable agreement with existing constraints. In
contrast to many studies within the mean-field NJL model, see
e.g.~\cite{Buballa:2003et}, our FRG EoS allows for gravitationally
stable hybrid stars with a three-flavor core, which however leads to a
maximum mass below the highest observed pulsar masses. We notice that
the inclusion of a repulsive vector interaction in the quark-meson
model is expected to additionally stiffen the quark matter EoS and
allow for constructing hybrid stars with nuclear EoS with smaller
radii and tidal deformabilities.

We have presented here one of the first constructions of a
non-perturbative EoS for high densities within the FRG that is a very
promising non-perturbative method for computing the EoS directly from
quark-gluon degrees of freedom, cf.~\cite{Drews:2016wpi,
  Posfay:2018sw, Leonhardt:2019fua}. Several
  approximations have been made, in particular:
  
  (i) We have employed a quark-meson truncation of the effective
  action as a low-energy effective model for QCD and solved the
  corresponding flow equation for the effective potential in lowest
  order of a derivative expansion. The quark-meson truncation
  comprises (pseudo-)scalar interaction channels which are the most
  dominant channels for chiral symmetry breaking at low temperatures
  and densities \cite{Braun:2017srn, *Mitter:2014wpa} but they might become
  insufficient at densities relevant for NSs such that in addition
  further channels like (axial-)vector ones should be taken into
  account.

(ii) For small temperatures and densities relevant for the NS interior
dynamical baryonic correlations, i.e.~the dynamical interrelation of
three-quark states and baryonic degrees of freedom with confined
quarks, become an increasingly important issue. This is obvious within
the hadronic phase, where baryons and not quarks are the relevant
degrees of freedom. Such correlations remain important at small
temperatures within a deconfined phase, too. Within the FRG framework,
the elaboration of the clustering of quarks into baryons as well as
the emergence of long-range correlations between baryons is in
principle straightforward. First successes could in parts already be
achieved in a similar QCD related context
\cite{Fu:2019hdw,*Fu:2016tey, *Braun:2014ata}. This makes the
QCD-based FRG approach towards lower densities very promising for
future investigations. It should, however, be stressed that the
computations becomes much more involved and much work is still
necessary to technically cope with this issue.

(iii) Furthermore, the phase structure in this area of the QCD phase
diagram is expected to be extremely rich with a wealth of possible
different color superconducting phases~\cite{Alford:2007xm,
  Fukushima:2010bq}. Future work should address the inclusion of
additional interaction channels like, for example, diquark-quark
channels which allow for color superconductivity and the already
mentioned repulsive (axial-)vector channels. Work along these lines is
ongoing.

\subsection*{Acknowledgments}

We are grateful to Mario Mitter for collaboration as well as providing
numerical data on the early stage of the research presented here.  We
would further like to thank D. Blaschke, M. Buballa, C.S. Fischer,
J.M. Pawlowski and
L. von Smekal for fruitful discussions.
This work was supported in part by the Helmholtz International Center
for FAIR within the LOEWE initiative of the State of Hesse,
the Deutsche Forschungsgemeinschaft (DFG) through the grant CRC-TR
211 “Strong-interaction matter under extreme conditions”, and the BMBF
under contract No. 05P18RGFCA.
K.O. acknowledges funding by the German Academic Scholarship
Foundation and the Helmholtz Graduate School for Hadron and Ion
Research (HGS-HIRe) for FAIR.  M.O. acknowledges financial support
from the action ``Physique fondamentale et ondes gravitationnelles''
of Paris Observatory.

%%%%%%%%%%%%%%%%%%%%%%%%%%%%%%%%%%%%%%%%%%%%%%%%%%%%%%%%%%%%%%%%%%%%%%%%%%%%%%%%
\appendix

\section{Input Parameters}
\label{app:inputparam}

In this appendix the input parameters for the FRG and mean-field
evaluations are summarized. 
In all $N_f=2+1$ calculations we fix the axial $\ua$ and chiral
symmetry breaking parameters to $c_A = 4807.84$ MeV,
$c_l = (120.73 \, \text{MeV})^3$ and $c_s = (336.41 \, \text{MeV})^3$
to reproduce the summed squares of the $\eta$ and $\eta'$ masses
$m_\eta^2+m_{\eta'}^2 = (1103.2 \, \text{MeV})^2$, the pion mass
$m_\pi = 138$ MeV, and the kaon mass $m_K = 496$ MeV, respectively, in
the vacuum. With the three input parameters for the chiral potential,
the sigma mass in the infrared has been set to $m_\sigma = 560$ MeV
and the vacuum minima to $\sigma_{l,0} = 92.4$ MeV and
$\sigma_{s,0} = 94.5$ MeV to yield the corresponding pion and kaon
decay constants, $f_\pi=92.4$ MeV and $f_K=113$ MeV. With a Yukawa
coupling of $g = 6.5$ the corresponding constituent quark masses are
$m_l = g \sigma_{l,0}/2 \approx 300$ MeV and
$m_s = g \sigma_{s,0}/\sqrt{2} \approx 434$ MeV. For $N_f=2$, all
strange quantities are omitted. $g$, $c_l$ and the remaining two input
parameters for the chiral potential are set to reproduce the subset
$m_l$, $m_\pi$, $m_\sigma$ and $\sigma_{k,0}$ in the same fashion as
above.

The parameter set for the FRG and rMFA flow equations have been
optimized by a global differential evolution algorithm
\cite{Storn:1997uk} with an initial UV cutoff of $\Lambda = 1$ GeV. In
the full FRG case we stopped the evolution around $k_{\text{IR}} = 80$
MeV where the condensates are already frozen and in the rMFA case we
stop at $k_{\text{IR}} = 1$ MeV. Note that all obtained numerical
results are insensitive to IR values when chosen in this region while
a UV cutoff dependence for the rMFA results can still be
seen. However, when choosing a UV cutoff larger than $\Lambda>2$ GeV for
the rMFA results any cutoff dependence disappears
\cite{Carignano:2014jla}.

The used input parameters for the mean-field potentials can be found
in \Tab{tab:starting_values_1}. The UV coefficients $a_{ij}$ for the
chiral potentials in the FRG calculations are listed in
\Tab{tab:starting_values_2}. Further details on the input parameters
can be found in Refs. \cite{Mitter:2013fxa, *Schaefer:2008hk}.

\begin{table}[ht]
\begin{center}
{\renewcommand{\arraystretch}{1.3}
\begin{tabular}{| c | c | c | c | c |}
 \hline
 $N_f$ & approx. &  $m^2 \, [\mathrm{MeV}^2]$ & $\lambda_1$ & $\lambda_2$\\
 \hline
2 & sMFA & $-(358.1)^2$ & $17.25$ & --\\
2 & rMFA & $901.09^2$ & $-5.38$ & --\\
2+1 & sMFA & $384.71^2$ & $-0.36$ & $46.48$ \\
2+1 & rMFA & $1040.94^2$ & $-2.65$ & $11.73$ \\
 \hline 
\end{tabular}}
\end{center}
\caption{Input parameters for the $N_f=2$ and $N_f=2+1$ quark-meson
  model for the sMFA and rMFA approximations. The Yukawa coupling is
  fixed at $g=6.5$ and the sigma mass at
  $m_\sigma = 560 \, \mathrm{MeV}$. For $N_f=2+1$ flavor the
  $\ua$-symmetry is explicitly broken by the parameter
  $c_A$. \label{tab:starting_values_1}}
 \end{table}

 \begin{table}[ht]
\begin{center}
{\renewcommand{\arraystretch}{1.3}
\begin{tabular}{| c | c | c | c | c | c |}
 \hline
 $N_f$ & $g$ & $m_\sigma  \, [\mathrm{MeV}]$ &  $a_{10} \, [\mathrm{MeV}^2]$ & $a_{20}$ & $a_{01}$ \\
 \hline
2 & $6.5$ & $560$ & $706.31^2$ & $21.16$ & --\\
2+1 & $6.5$ & $560$ & $515.70^2$ & $37.45$ & $47.68$ \\
 \hline 
\end{tabular}}
\end{center}
\caption{FRG input parameters similar to
  \Tab{tab:starting_values_1}. Additionally, the utilized Yukawa
  coupling $g$ and sigma mass $m_{\sigma}$ are also quoted. The
  parameter $a_{01}$ corresponds to the modified invariant
  $\tilde{\rho}_2$ by analogy with the mean-field parameter
  $\lambda_2$ in
  \Tab{tab:starting_values_1}. \label{tab:starting_values_2}}
 \end{table}

\section{Numerical Solution Techniques}
\label{app:solutiontech}

The numerical methods employed for the solution of the FRG flow
equations and for the TOV equations are outlined in this appendix.

The FRG flow equations \eqref{eq:frg} and \eqref{eq:4} are partial
differential equations (PDEs) including a partial derivative
w.r.t. the RG scale $k$ and field derivatives encoded in the mass
terms.
Several global as well as local solution schemes based on
pseudo-spectral, finite volume, Taylor expansion or Discontinuous
Galerkin ideas are known, see \cite{Grossi:2019urj,*Borchardt:2016pif}
for some recent developments. Setting up a two-dimensional grid for
the chiral potential in the variables
\begin{equation}
x = \sigma_l^2 \quad \text{and} \quad y = 2 \sigma_s^2-\sigma_l^2 \ ,
\end{equation}
cf. \cite{Mitter:2013fxa}, and interpolating the derivatives from
the discrete grid points, it is possible to reduce the PDE to a set of
coupled ordinary differential equations (ODEs).  For the
interpolation, we use cubic splines in each of the two grid
directions, respectively.  The ODE solution is obtained from an
explicit Runge-Kutta type step algorithm. For $N_f=2$, the grid
reduces to one dimension.

In the context of this work, the $T=0$ limit of the flow equation,
Eq. \eqref{eq:flow_T0}, is utilized wherein the fermionic threshold
function reduces to a Heaviside function. This also encodes the Silver
Blaze property of the theory because
\begin{equation}
E_l \geq m_l = g \sigma_{l,0}/2
\end{equation}
implies that $\theta (E_f-\mu) = 1$ for all $\mu < m_f$ and hence the
flow at the IR minimum $\sigma_{l,0}$ does not change with respect to
the vacuum flow. Here we have assumed for simplicity a
flavor-independent chemical potential. Of course, this property only
holds for $\mu < \mu_c$ where $\mu_c$ signifies the chiral first-order
transition.  In both the sMFA and FRG solutions, we observe
$\mu_c < m_l$, cf. \Fig{fig:condensates_mu_2p1}. Unfortunately,
due to the utilized grid method the Silver Blaze property is subject
to a numerical error. For any chemical potential, all grid points
located at $\sigma_l < 2\mu/g$ display a different running than in
vacuum.  Since the bosonic energies incorporate field derivatives that
are approximated from an interpolation of all grid points, the flow
experiences small modifications at the IR minimum even if $\mu <
m_l$. This effect aggravates when $\mu$ approaches $m_l$ from below,
$\mu \lesssim m_l$. It leads to fluctuations of the chiral condensate
around the vacuum IR value. Those fluctuations are small and hardly
visible in \Fig{fig:condensates_mu_2p1}.  However, they lead to
an unphysical phase of very small but non-zero pressure. Furthermore,
for a flavor-dependent chemical potential the first-order transition
is additionally distorted by the error of our approximation, see
Sec. \ref{sec:eos_quark_hybrid}.  Thus, in the numerical treatment of
the FRG EoS, data points close to the phase transition are omitted and
the EoS from the physical phase with restored chiral symmetry is
polynomially extrapolated down to $p=0$. This procedure only affects
the low-pressure outer region of the calculated pure quark stars. The
dependence of the star radius on the extrapolation error has been
checked and found to be negligible.

For $N_f=2+1$, similar numerical fluctuations as discussed above are
also found when $\mu_s$ approaches the order of the strange quark mass
$m_s$ at the current IR minimum. They are most prevalent in the
determination of the speed of sound. A strong dependence of these
fluctuations on the exact grid point configuration is observed and
gives strong evidence for the claim as a numerical
artifact. Therefore, in \Fig{fig:speed_of_sound} for $\mu > 350$ MeV
only the average derivative
$\bar{c}_s^2 := \Delta p/\Delta \varepsilon$ is shown as dots,
uniformly spaced at a distance of $15 \, \text{MeV}$. For each dot,
$\bar{c}_s^2$ has been calculated from the $(p,\varepsilon)$ tupels at
the dot to its left, itself, and the dot to its right. Furthermore,
the highest deviation of the microscopically calculated, fluctuating
speed of sound $c_s^2$ from $\bar{c}_s^2$ in the respective interval
is indicated by the border points of the shaded region such that all
$c_s^2$ data points lie within this region.

In the pure quark matter star calculations, the TOV equation is solved
with an explicit Runge-Kutta algorithm. The evolution is stopped when
the radial pressure $p(r)$ reaches a value of $10^{-5}$ relative to
the central pressure. The EoS data points are interpolated with cubic
splines, utilizing two separate splines in case of a discontinuity due
to a first-order transition such as observed in the rMFA EoS.

\bibliography{tovfrg}

\begin{thebibliography}{123}%
\makeatletter
\providecommand \@ifxundefined [1]{%
 \@ifx{#1\undefined}
}%
\providecommand \@ifnum [1]{%
 \ifnum #1\expandafter \@firstoftwo
 \else \expandafter \@secondoftwo
 \fi
}%
\providecommand \@ifx [1]{%
 \ifx #1\expandafter \@firstoftwo
 \else \expandafter \@secondoftwo
 \fi
}%
\providecommand \natexlab [1]{#1}%
\providecommand \enquote  [1]{``#1''}%
\providecommand \bibnamefont  [1]{#1}%
\providecommand \bibfnamefont [1]{#1}%
\providecommand \citenamefont [1]{#1}%
\providecommand \href@noop [0]{\@secondoftwo}%
\providecommand \href [0]{\begingroup \@sanitize@url \@href}%
\providecommand \@href[1]{\@@startlink{#1}\@@href}%
\providecommand \@@href[1]{\endgroup#1\@@endlink}%
\providecommand \@sanitize@url [0]{\catcode `\\12\catcode `\$12\catcode
  `\&12\catcode `\#12\catcode `\^12\catcode `\_12\catcode `\%12\relax}%
\providecommand \@@startlink[1]{}%
\providecommand \@@endlink[0]{}%
\providecommand \url  [0]{\begingroup\@sanitize@url \@url }%
\providecommand \@url [1]{\endgroup\@href {#1}{\urlprefix }}%
\providecommand \urlprefix  [0]{URL }%
\providecommand \Eprint [0]{\href }%
\providecommand \doibase [0]{http://dx.doi.org/}%
\providecommand \selectlanguage [0]{\@gobble}%
\providecommand \bibinfo  [0]{\@secondoftwo}%
\providecommand \bibfield  [0]{\@secondoftwo}%
\providecommand \translation [1]{[#1]}%
\providecommand \BibitemOpen [0]{}%
\providecommand \bibitemStop [0]{}%
\providecommand \bibitemNoStop [0]{.\EOS\space}%
\providecommand \EOS [0]{\spacefactor3000\relax}%
\providecommand \BibitemShut  [1]{\csname bibitem#1\endcsname}%
\let\auto@bib@innerbib\@empty
%</preamble>
\bibitem [{\citenamefont {Arzoumanian}\ et~al.(2018)\citenamefont {Arzoumanian}
  et~al.}]{Arzoumanian:2017puf}%
  \BibitemOpen
  \bibfield  {author} {\bibinfo {author} {\bibfnamefont {Z.}~\bibnamefont
  {Arzoumanian}} et~al. [\bibinfo {collaboration} {NANOGrav}],\ }\href
  {\doibase 10.3847/1538-4365/aab5b0} {\bibfield  {journal} {\bibinfo
  {journal} {Astrophys. J. Suppl.}\ }\textbf {\bibinfo {volume} {235}},\
  \bibinfo {pages} {37} (\bibinfo {year} {2018})},\ \Eprint
  {http://arxiv.org/abs/1801.01837} {arXiv:1801.01837} \BibitemShut {NoStop}%
%%CITATION = ARXIV:1801.01837;%%
\bibitem [{\citenamefont {Demorest}\ et~al.(2010)\citenamefont {Demorest},
  \citenamefont {Pennucci}, \citenamefont {Ransom}, \citenamefont {Roberts},\
  and\ \citenamefont {Hessels}}]{Demorest:2010bx}%
  \BibitemOpen
  \bibfield  {author} {\bibinfo {author} {\bibfnamefont {P.}~\bibnamefont
  {Demorest}}, \bibinfo {author} {\bibfnamefont {T.}~\bibnamefont {Pennucci}},
  \bibinfo {author} {\bibfnamefont {S.}~\bibnamefont {Ransom}}, \bibinfo
  {author} {\bibfnamefont {M.}~\bibnamefont {Roberts}}, \ and\ \bibinfo
  {author} {\bibfnamefont {J.}~\bibnamefont {Hessels}},\ }\href {\doibase
  10.1038/nature09466} {\bibfield  {journal} {\bibinfo  {journal} {Nature}\
  }\textbf {\bibinfo {volume} {467}},\ \bibinfo {pages} {1081} (\bibinfo {year}
  {2010})},\ \Eprint {http://arxiv.org/abs/1010.5788} {arXiv:1010.5788}
  \BibitemShut {NoStop}%
%%CITATION = ARXIV:1010.5788;%%
\bibitem [{\citenamefont {Antoniadis}\ et~al.(2013)\citenamefont {Antoniadis}
  et~al.}]{Antoniadis:2013pzd}%
  \BibitemOpen
  \bibfield  {author} {\bibinfo {author} {\bibfnamefont {J.}~\bibnamefont
  {Antoniadis}} et~al.,\ }\href {\doibase 10.1126/science.1233232} {\bibfield
  {journal} {\bibinfo  {journal} {Science}\ }\textbf {\bibinfo {volume}
  {340}},\ \bibinfo {pages} {6131} (\bibinfo {year} {2013})},\ \Eprint
  {http://arxiv.org/abs/1304.6875} {arXiv:1304.6875} \BibitemShut {NoStop}%
%%CITATION = ARXIV:1304.6875;%%
\bibitem [{\citenamefont {Cromartie}\ et~al.(2019)\citenamefont {Cromartie}
  et~al.}]{Cromartie:2019kug}%
  \BibitemOpen
  \bibfield  {author} {\bibinfo {author} {\bibfnamefont {H.~T.}\ \bibnamefont
  {Cromartie}} et~al.,\ }\href {\doibase 10.1038/s41550-019-0880-2} {\bibfield
  {journal} {\bibinfo  {journal} {Nature Astronomy}\ ,\ \bibinfo {pages} {439}}
  (\bibinfo {year} {2019})},\ \Eprint {http://arxiv.org/abs/1904.06759}
  {arXiv:1904.06759} \BibitemShut {NoStop}%
%%CITATION = ARXIV:1904.06759;%%
\bibitem [{\citenamefont {Lattimer}\ and\ \citenamefont
  {Prakash}(2016)}]{Lattimer:2016yq}%
  \BibitemOpen
  \bibfield  {author} {\bibinfo {author} {\bibfnamefont {J.~M.}\ \bibnamefont
  {Lattimer}}\ and\ \bibinfo {author} {\bibfnamefont {M.}~\bibnamefont
  {Prakash}},\ }\href {\doibase 10.1016/j.physrep.2015.12.005} {\bibfield
  {journal} {\bibinfo  {journal} {Phys. Rept.}\ }\textbf {\bibinfo {volume}
  {621}},\ \bibinfo {pages} {127} (\bibinfo {year} {2016})},\ \Eprint
  {http://arxiv.org/abs/1512.07820} {arXiv:1512.07820} \BibitemShut {NoStop}%
%%CITATION = ARXIV:1512.07820;%%
\bibitem [{\citenamefont {Lattimer}\ and\ \citenamefont
  {Prakash}(2007)}]{Lattimer:2006xb}%
  \BibitemOpen
  \bibfield  {author} {\bibinfo {author} {\bibfnamefont {J.~M.}\ \bibnamefont
  {Lattimer}}\ and\ \bibinfo {author} {\bibfnamefont {M.}~\bibnamefont
  {Prakash}},\ }\href {\doibase 10.1016/j.physrep.2007.02.003} {\bibfield
  {journal} {\bibinfo  {journal} {Phys. Rept.}\ }\textbf {\bibinfo {volume}
  {442}},\ \bibinfo {pages} {109} (\bibinfo {year} {2007})},\ \Eprint
  {http://arxiv.org/abs/astro-ph/0612440} {arXiv:astro-ph/0612440} \BibitemShut
  {NoStop}%
%%CITATION = ASTRO-PH/0612440;%%
\bibitem [{\citenamefont {Oertel}\ et~al.(2017)\citenamefont {Oertel},
  \citenamefont {Hempel}, \citenamefont {Kl{\"a}hn},\ and\ \citenamefont
  {Typel}}]{Oertel:2017fk}%
  \BibitemOpen
  \bibfield  {author} {\bibinfo {author} {\bibfnamefont {M.}~\bibnamefont
  {Oertel}}, \bibinfo {author} {\bibfnamefont {M.}~\bibnamefont {Hempel}},
  \bibinfo {author} {\bibfnamefont {T.}~\bibnamefont {Kl{\"a}hn}}, \ and\
  \bibinfo {author} {\bibfnamefont {S.}~\bibnamefont {Typel}},\ }\href
  {\doibase 10.1103/RevModPhys.89.015007} {\bibfield  {journal} {\bibinfo
  {journal} {Rev. Mod. Phys.}\ }\textbf {\bibinfo {volume} {89}},\ \bibinfo
  {pages} {015007} (\bibinfo {year} {2017})}\BibitemShut {NoStop}%
\bibitem [{\citenamefont {Chatterjee}\ and\ \citenamefont
  {Vida{\~n}a}(2016)}]{Chatterjee:2015pua}%
  \BibitemOpen
  \bibfield  {author} {\bibinfo {author} {\bibfnamefont {D.}~\bibnamefont
  {Chatterjee}}\ and\ \bibinfo {author} {\bibfnamefont {I.}~\bibnamefont
  {Vida{\~n}a}},\ }\href {\doibase 10.1140/epja/i2016-16029-x} {\bibfield
  {journal} {\bibinfo  {journal} {Eur. Phys. J.}\ }\textbf {\bibinfo {volume}
  {A52}},\ \bibinfo {pages} {29} (\bibinfo {year} {2016})},\ \Eprint
  {http://arxiv.org/abs/1510.06306} {arXiv:1510.06306} \BibitemShut {NoStop}%
%%CITATION = ARXIV:1510.06306;%%
\bibitem [{\citenamefont {Djapo}\ et~al.(2010)\citenamefont {Djapo},
  \citenamefont {Schaefer},\ and\ \citenamefont {Wambach}}]{Djapo2008}%
  \BibitemOpen
  \bibfield  {author} {\bibinfo {author} {\bibfnamefont {H.}~\bibnamefont
  {Djapo}}, \bibinfo {author} {\bibfnamefont {B.-J.}\ \bibnamefont {Schaefer}},
  \ and\ \bibinfo {author} {\bibfnamefont {J.}~\bibnamefont {Wambach}},\ }\href
  {\doibase 10.1103/PhysRevC.81.035803} {\bibfield  {journal} {\bibinfo
  {journal} {Phys. Rev.}\ }\textbf {\bibinfo {volume} {C81}},\ \bibinfo {pages}
  {035803} (\bibinfo {year} {2010})},\ \Eprint {http://arxiv.org/abs/0811.2939}
  {arXiv:0811.2939} \BibitemShut {NoStop}%
%%CITATION = ARXIV:0811.2939;%%
\bibitem [{\citenamefont {Kolomeitsev}\ et~al.(2017)\citenamefont
  {Kolomeitsev}, \citenamefont {Maslov},\ and\ \citenamefont
  {Voskresensky}}]{Kolomeitsev:2016ptu}%
  \BibitemOpen
  \bibfield  {author} {\bibinfo {author} {\bibfnamefont {E.~E.}\ \bibnamefont
  {Kolomeitsev}}, \bibinfo {author} {\bibfnamefont {K.~A.}\ \bibnamefont
  {Maslov}}, \ and\ \bibinfo {author} {\bibfnamefont {D.~N.}\ \bibnamefont
  {Voskresensky}},\ }\href {\doibase 10.1016/j.nuclphysa.2017.02.004}
  {\bibfield  {journal} {\bibinfo  {journal} {Nucl. Phys.}\ }\textbf {\bibinfo
  {volume} {A961}},\ \bibinfo {pages} {106} (\bibinfo {year} {2017})},\ \Eprint
  {http://arxiv.org/abs/1610.09746} {arXiv:1610.09746} \BibitemShut {NoStop}%
%%CITATION = ARXIV:1610.09746;%%
\bibitem [{\citenamefont {Buballa}\ et~al.(2014)\citenamefont {Buballa}
  et~al.}]{Buballa:2014jta}%
  \BibitemOpen
  \bibfield  {author} {\bibinfo {author} {\bibfnamefont {M.}~\bibnamefont
  {Buballa}} et~al.,\ }\href {\doibase 10.1088/0954-3899/41/12/123001}
  {\bibfield  {journal} {\bibinfo  {journal} {J. Phys.}\ }\textbf {\bibinfo
  {volume} {G41}},\ \bibinfo {pages} {123001} (\bibinfo {year} {2014})},\
  \Eprint {http://arxiv.org/abs/1402.6911} {arXiv:1402.6911} \BibitemShut
  {NoStop}%
%%CITATION = ARXIV:1402.6911;%%
\bibitem [{\citenamefont {Weber}\ et~al.(2014)\citenamefont {Weber},
  \citenamefont {Contrera}, \citenamefont {Orsaria}, \citenamefont {Spinella},\
  and\ \citenamefont {Zubairi}}]{Weber:2014qoa}%
  \BibitemOpen
  \bibfield  {author} {\bibinfo {author} {\bibfnamefont {F.}~\bibnamefont
  {Weber}}, \bibinfo {author} {\bibfnamefont {G.~A.}\ \bibnamefont {Contrera}},
  \bibinfo {author} {\bibfnamefont {M.~G.}\ \bibnamefont {Orsaria}}, \bibinfo
  {author} {\bibfnamefont {W.}~\bibnamefont {Spinella}}, \ and\ \bibinfo
  {author} {\bibfnamefont {O.}~\bibnamefont {Zubairi}},\ }\href {\doibase
  10.1142/S0217732314300225} {\bibfield  {journal} {\bibinfo  {journal} {Mod.
  Phys. Lett.}\ }\textbf {\bibinfo {volume} {A29}},\ \bibinfo {pages} {1430022}
  (\bibinfo {year} {2014})},\ \Eprint {http://arxiv.org/abs/1408.0079}
  {arXiv:1408.0079} \BibitemShut {NoStop}%
%%CITATION = ARXIV:1408.0079;%%
\bibitem [{\citenamefont {Orsaria}\ et~al.(2014)\citenamefont {Orsaria},
  \citenamefont {Rodrigues}, \citenamefont {Weber},\ and\ \citenamefont
  {Contrera}}]{Orsaria:2013hna}%
  \BibitemOpen
  \bibfield  {author} {\bibinfo {author} {\bibfnamefont {M.}~\bibnamefont
  {Orsaria}}, \bibinfo {author} {\bibfnamefont {H.}~\bibnamefont {Rodrigues}},
  \bibinfo {author} {\bibfnamefont {F.}~\bibnamefont {Weber}}, \ and\ \bibinfo
  {author} {\bibfnamefont {G.~A.}\ \bibnamefont {Contrera}},\ }\href {\doibase
  10.1103/PhysRevC.89.015806} {\bibfield  {journal} {\bibinfo  {journal} {Phys.
  Rev.}\ }\textbf {\bibinfo {volume} {C89}},\ \bibinfo {pages} {015806}
  (\bibinfo {year} {2014})},\ \Eprint {http://arxiv.org/abs/1308.1657}
  {arXiv:1308.1657} \BibitemShut {NoStop}%
%%CITATION = ARXIV:1308.1657;%%
\bibitem [{\citenamefont {Drago}\ et~al.(2014)\citenamefont {Drago},
  \citenamefont {Lavagno},\ and\ \citenamefont {Pagliara}}]{Drago:2013fsa}%
  \BibitemOpen
  \bibfield  {author} {\bibinfo {author} {\bibfnamefont {A.}~\bibnamefont
  {Drago}}, \bibinfo {author} {\bibfnamefont {A.}~\bibnamefont {Lavagno}}, \
  and\ \bibinfo {author} {\bibfnamefont {G.}~\bibnamefont {Pagliara}},\ }\href
  {\doibase 10.1103/PhysRevD.89.043014} {\bibfield  {journal} {\bibinfo
  {journal} {Phys. Rev.}\ }\textbf {\bibinfo {volume} {D89}},\ \bibinfo {pages}
  {043014} (\bibinfo {year} {2014})},\ \Eprint {http://arxiv.org/abs/1309.7263}
  {arXiv:1309.7263} \BibitemShut {NoStop}%
%%CITATION = ARXIV:1309.7263;%%
\bibitem [{\citenamefont {Alford}\ and\ \citenamefont
  {Han}(2016)}]{Alford:2015gna}%
  \BibitemOpen
  \bibfield  {author} {\bibinfo {author} {\bibfnamefont {M.~G.}\ \bibnamefont
  {Alford}}\ and\ \bibinfo {author} {\bibfnamefont {S.}~\bibnamefont {Han}},\
  }\href {\doibase 10.1140/epja/i2016-16062-9} {\bibfield  {journal} {\bibinfo
  {journal} {Eur. Phys. J.}\ }\textbf {\bibinfo {volume} {A52}},\ \bibinfo
  {pages} {62} (\bibinfo {year} {2016})},\ \Eprint
  {http://arxiv.org/abs/1508.01261} {arXiv:1508.01261} \BibitemShut {NoStop}%
%%CITATION = ARXIV:1508.01261;%%
\bibitem [{\citenamefont {Baym}\ et~al.(2018)\citenamefont {Baym},
  \citenamefont {Hatsuda}, \citenamefont {Kojo}, \citenamefont {Powell},
  \citenamefont {Song},\ and\ \citenamefont {Takatsuka}}]{Baym2017}%
  \BibitemOpen
  \bibfield  {author} {\bibinfo {author} {\bibfnamefont {G.}~\bibnamefont
  {Baym}}, \bibinfo {author} {\bibfnamefont {T.}~\bibnamefont {Hatsuda}},
  \bibinfo {author} {\bibfnamefont {T.}~\bibnamefont {Kojo}}, \bibinfo {author}
  {\bibfnamefont {P.~D.}\ \bibnamefont {Powell}}, \bibinfo {author}
  {\bibfnamefont {Y.}~\bibnamefont {Song}}, \ and\ \bibinfo {author}
  {\bibfnamefont {T.}~\bibnamefont {Takatsuka}},\ }\href {\doibase
  10.1088/1361-6633/aaae14} {\bibfield  {journal} {\bibinfo  {journal} {Rept.
  Prog. Phys.}\ }\textbf {\bibinfo {volume} {81}},\ \bibinfo {pages} {056902}
  (\bibinfo {year} {2018})},\ \Eprint {http://arxiv.org/abs/1707.04966}
  {arXiv:1707.04966} \BibitemShut {NoStop}%
%%CITATION = ARXIV:1707.04966;%%
\bibitem [{\citenamefont {{Witten}}(1984)}]{witten84}%
  \BibitemOpen
  \bibfield  {author} {\bibinfo {author} {\bibfnamefont {E.}~\bibnamefont
  {{Witten}}},\ }\href {\doibase 10.1103/PhysRevD.30.272} {\bibfield  {journal}
  {\bibinfo  {journal} {Phys. Rev. D}\ }\textbf {\bibinfo {volume} {30}},\
  \bibinfo {pages} {272} (\bibinfo {year} {1984})}\BibitemShut {NoStop}%
\bibitem [{\citenamefont {Farhi}\ and\ \citenamefont
  {Jaffe}(1984)}]{Farhi:1984qu}%
  \BibitemOpen
  \bibfield  {author} {\bibinfo {author} {\bibfnamefont {E.}~\bibnamefont
  {Farhi}}\ and\ \bibinfo {author} {\bibfnamefont {R.}~\bibnamefont {Jaffe}},\
  }\href {\doibase 10.1103/PhysRevD.30.2379} {\bibfield  {journal} {\bibinfo
  {journal} {Phys. Rev. D}\ }\textbf {\bibinfo {volume} {30}},\ \bibinfo
  {pages} {2379} (\bibinfo {year} {1984})}\BibitemShut {NoStop}%
\bibitem [{\citenamefont {Buballa}\ and\ \citenamefont
  {Oertel}(1999)}]{Buballa:1998pr}%
  \BibitemOpen
  \bibfield  {author} {\bibinfo {author} {\bibfnamefont {M.}~\bibnamefont
  {Buballa}}\ and\ \bibinfo {author} {\bibfnamefont {M.}~\bibnamefont
  {Oertel}},\ }\href {\doibase 10.1016/S0370-2693(99)00533-X} {\bibfield
  {journal} {\bibinfo  {journal} {Phys. Lett.}\ }\textbf {\bibinfo {volume}
  {B457}},\ \bibinfo {pages} {261} (\bibinfo {year} {1999})},\ \Eprint
  {http://arxiv.org/abs/hep-ph/9810529} {arXiv:hep-ph/9810529} \BibitemShut
  {NoStop}%
%%CITATION = HEP-PH/9810529;%%
\bibitem [{\citenamefont {Haensel}\ et~al.(1986)\citenamefont {Haensel},
  \citenamefont {Zdunik},\ and\ \citenamefont {Schaeffer}}]{Haensel_86}%
  \BibitemOpen
  \bibfield  {author} {\bibinfo {author} {\bibfnamefont {P.}~\bibnamefont
  {Haensel}}, \bibinfo {author} {\bibfnamefont {J.}~\bibnamefont {Zdunik}}, \
  and\ \bibinfo {author} {\bibfnamefont {R.}~\bibnamefont {Schaeffer}},\
  }\href@noop {} {\bibfield  {journal} {\bibinfo  {journal}
  {Astron.Astrophys.}\ }\textbf {\bibinfo {volume} {160}},\ \bibinfo {pages}
  {121} (\bibinfo {year} {1986})}\BibitemShut {NoStop}%
%%CITATION = AAEJA,160,121;%%
\bibitem [{\citenamefont {Watts}\ et~al.(2016)\citenamefont {Watts}
  et~al.}]{Watts:2016uzu}%
  \BibitemOpen
  \bibfield  {author} {\bibinfo {author} {\bibfnamefont {A.~L.}\ \bibnamefont
  {Watts}} et~al.,\ }\href {\doibase 10.1103/RevModPhys.88.021001} {\bibfield
  {journal} {\bibinfo  {journal} {Rev. Mod. Phys.}\ }\textbf {\bibinfo {volume}
  {88}},\ \bibinfo {pages} {021001} (\bibinfo {year} {2016})},\ \Eprint
  {http://arxiv.org/abs/1602.01081} {arXiv:1602.01081} \BibitemShut {NoStop}%
%%CITATION = ARXIV:1602.01081;%%
\bibitem [{\citenamefont {Abbott}\ et~al.(2017)\citenamefont {Abbott}
  et~al.}]{GW170817}%
  \BibitemOpen
  \bibfield  {author} {\bibinfo {author} {\bibfnamefont {B.~P.}\ \bibnamefont
  {Abbott}} et~al. [\bibinfo {collaboration} {LIGO Scientific, Virgo}],\ }\href
  {\doibase 10.1103/PhysRevLett.119.161101} {\bibfield  {journal} {\bibinfo
  {journal} {Phys. Rev. Lett.}\ }\textbf {\bibinfo {volume} {119}},\ \bibinfo
  {pages} {161101} (\bibinfo {year} {2017})},\ \Eprint
  {http://arxiv.org/abs/1710.05832} {arXiv:1710.05832} \BibitemShut {NoStop}%
%%CITATION = ARXIV:1710.05832;%%
\bibitem [{\citenamefont {Abbott}\ et~al.(2018)\citenamefont {Abbott}
  et~al.}]{Abbott:2018fj}%
  \BibitemOpen
  \bibfield  {author} {\bibinfo {author} {\bibfnamefont {B.~P.}\ \bibnamefont
  {Abbott}} et~al. [\bibinfo {collaboration} {{LIGO Scientific, Virgo}}],\
  }\href {\doibase 10.1103/PhysRevLett.121.161101} {\bibfield  {journal}
  {\bibinfo  {journal} {Phys. Rev. Lett.}\ }\textbf {\bibinfo {volume} {121}},\
  \bibinfo {pages} {161101} (\bibinfo {year} {2018})},\ \Eprint
  {http://arxiv.org/abs/1805.11581} {arXiv:1805.11581} \BibitemShut {NoStop}%
%%CITATION = ARXIV:1805.11581;%%
\bibitem [{\citenamefont {De}\ et~al.(2018)\citenamefont {De}, \citenamefont
  {Finstad}, \citenamefont {Lattimer}, \citenamefont {Brown}, \citenamefont
  {Berger},\ and\ \citenamefont {Biwer}}]{De:2018uhw}%
  \BibitemOpen
  \bibfield  {author} {\bibinfo {author} {\bibfnamefont {S.}~\bibnamefont
  {De}}, \bibinfo {author} {\bibfnamefont {D.}~\bibnamefont {Finstad}},
  \bibinfo {author} {\bibfnamefont {J.~M.}\ \bibnamefont {Lattimer}}, \bibinfo
  {author} {\bibfnamefont {D.~A.}\ \bibnamefont {Brown}}, \bibinfo {author}
  {\bibfnamefont {E.}~\bibnamefont {Berger}}, \ and\ \bibinfo {author}
  {\bibfnamefont {C.~M.}\ \bibnamefont {Biwer}},\ }\href {\doibase
  10.1103/PhysRevLett.121.259902, 10.1103/PhysRevLett.121.091102} {\bibfield
  {journal} {\bibinfo  {journal} {Phys. Rev. Lett.}\ }\textbf {\bibinfo
  {volume} {121}},\ \bibinfo {pages} {091102} (\bibinfo {year} {2018})},\
  \Eprint {http://arxiv.org/abs/1804.08583} {arXiv:1804.08583} \BibitemShut
  {NoStop}%
%%CITATION = ARXIV:1804.08583;%%
\bibitem [{\citenamefont {Most}\ et~al.(2018)\citenamefont {Most},
  \citenamefont {Weih}, \citenamefont {Rezzolla},\ and\ \citenamefont
  {Schaffner-Bielich}}]{Most:2018hfd}%
  \BibitemOpen
  \bibfield  {author} {\bibinfo {author} {\bibfnamefont {E.~R.}\ \bibnamefont
  {Most}}, \bibinfo {author} {\bibfnamefont {L.~R.}\ \bibnamefont {Weih}},
  \bibinfo {author} {\bibfnamefont {L.}~\bibnamefont {Rezzolla}}, \ and\
  \bibinfo {author} {\bibfnamefont {J.}~\bibnamefont {Schaffner-Bielich}},\
  }\href {\doibase 10.1103/PhysRevLett.120.261103} {\bibfield  {journal}
  {\bibinfo  {journal} {Phys. Rev. Lett.}\ }\textbf {\bibinfo {volume} {120}},\
  \bibinfo {pages} {261103} (\bibinfo {year} {2018})},\ \Eprint
  {http://arxiv.org/abs/1803.00549} {arXiv:1803.00549} \BibitemShut {NoStop}%
%%CITATION = ARXIV:1803.00549;%%
\bibitem [{\citenamefont {Capano}\ et~al.(2019)\citenamefont {Capano},
  \citenamefont {Tews}, \citenamefont {Brown}, \citenamefont {Margalit},
  \citenamefont {De}, \citenamefont {Kumar}, \citenamefont {Brown},
  \citenamefont {Krishnan},\ and\ \citenamefont {Reddy}}]{Capano:2019eae}%
  \BibitemOpen
  \bibfield  {author} {\bibinfo {author} {\bibfnamefont {C.~D.}\ \bibnamefont
  {Capano}}, \bibinfo {author} {\bibfnamefont {I.}~\bibnamefont {Tews}},
  \bibinfo {author} {\bibfnamefont {S.~M.}\ \bibnamefont {Brown}}, \bibinfo
  {author} {\bibfnamefont {B.}~\bibnamefont {Margalit}}, \bibinfo {author}
  {\bibfnamefont {S.}~\bibnamefont {De}}, \bibinfo {author} {\bibfnamefont
  {S.}~\bibnamefont {Kumar}}, \bibinfo {author} {\bibfnamefont {D.~A.}\
  \bibnamefont {Brown}}, \bibinfo {author} {\bibfnamefont {B.}~\bibnamefont
  {Krishnan}}, \ and\ \bibinfo {author} {\bibfnamefont {S.}~\bibnamefont
  {Reddy}},\ }\href@noop {} {\ }\Eprint {http://arxiv.org/abs/1908.10352}
  {arXiv:1908.10352} \BibitemShut {NoStop}%
%%CITATION = ARXIV:1908.10352;%%
\bibitem [{\citenamefont {Read}\ et~al.(2009)\citenamefont {Read},
  \citenamefont {Lackey}, \citenamefont {Owen},\ and\ \citenamefont
  {Friedman}}]{Read:2008iy}%
  \BibitemOpen
  \bibfield  {author} {\bibinfo {author} {\bibfnamefont {J.~S.}\ \bibnamefont
  {Read}}, \bibinfo {author} {\bibfnamefont {B.~D.}\ \bibnamefont {Lackey}},
  \bibinfo {author} {\bibfnamefont {B.~J.}\ \bibnamefont {Owen}}, \ and\
  \bibinfo {author} {\bibfnamefont {J.~L.}\ \bibnamefont {Friedman}},\ }\href
  {\doibase 10.1103/PhysRevD.79.124032} {\bibfield  {journal} {\bibinfo
  {journal} {Phys. Rev.}\ }\textbf {\bibinfo {volume} {D79}},\ \bibinfo {pages}
  {124032} (\bibinfo {year} {2009})},\ \Eprint {http://arxiv.org/abs/0812.2163}
  {arXiv:0812.2163} \BibitemShut {NoStop}%
%%CITATION = ARXIV:0812.2163;%%
\bibitem [{\citenamefont {{\"O}zel}\ et~al.(2010)\citenamefont {{\"O}zel},
  \citenamefont {Baym},\ and\ \citenamefont {Guver}}]{Ozel:2010fw}%
  \BibitemOpen
  \bibfield  {author} {\bibinfo {author} {\bibfnamefont {F.}~\bibnamefont
  {{\"O}zel}}, \bibinfo {author} {\bibfnamefont {G.}~\bibnamefont {Baym}}, \
  and\ \bibinfo {author} {\bibfnamefont {T.}~\bibnamefont {Guver}},\ }\href
  {\doibase 10.1103/PhysRevD.82.101301} {\bibfield  {journal} {\bibinfo
  {journal} {Phys. Rev.}\ }\textbf {\bibinfo {volume} {D82}},\ \bibinfo {pages}
  {101301} (\bibinfo {year} {2010})},\ \Eprint {http://arxiv.org/abs/1002.3153}
  {arXiv:1002.3153} \BibitemShut {NoStop}%
%%CITATION = ARXIV:1002.3153;%%
\bibitem [{\citenamefont {Steiner}\ et~al.(2013{\natexlab{a}})\citenamefont
  {Steiner}, \citenamefont {Lattimer},\ and\ \citenamefont
  {Brown}}]{Steiner:2012xt}%
  \BibitemOpen
  \bibfield  {author} {\bibinfo {author} {\bibfnamefont {A.~W.}\ \bibnamefont
  {Steiner}}, \bibinfo {author} {\bibfnamefont {J.~M.}\ \bibnamefont
  {Lattimer}}, \ and\ \bibinfo {author} {\bibfnamefont {E.~F.}\ \bibnamefont
  {Brown}},\ }\href {\doibase 10.1088/2041-8205/765/1/L5} {\bibfield  {journal}
  {\bibinfo  {journal} {Astrophys. J.}\ }\textbf {\bibinfo {volume} {765}},\
  \bibinfo {pages} {L5} (\bibinfo {year} {2013}{\natexlab{a}})},\ \Eprint
  {http://arxiv.org/abs/1205.6871} {arXiv:1205.6871} \BibitemShut {NoStop}%
%%CITATION = ARXIV:1205.6871;%%
\bibitem [{\citenamefont {Margueron}\ et~al.(2018)\citenamefont {Margueron},
  \citenamefont {Hoffmann~Casali},\ and\ \citenamefont
  {Gulminelli}}]{Margueron2018a}%
  \BibitemOpen
  \bibfield  {author} {\bibinfo {author} {\bibfnamefont {J.}~\bibnamefont
  {Margueron}}, \bibinfo {author} {\bibfnamefont {R.}~\bibnamefont
  {Hoffmann~Casali}}, \ and\ \bibinfo {author} {\bibfnamefont {F.}~\bibnamefont
  {Gulminelli}},\ }\href {\doibase 10.1103/PhysRevC.97.025805} {\bibfield
  {journal} {\bibinfo  {journal} {Phys. Rev.}\ }\textbf {\bibinfo {volume}
  {C97}},\ \bibinfo {pages} {025805} (\bibinfo {year} {2018})},\ \Eprint
  {http://arxiv.org/abs/1708.06894} {arXiv:1708.06894} \BibitemShut {NoStop}%
%%CITATION = ARXIV:1708.06894;%%
\bibitem [{\citenamefont {Tews}\ et~al.(2018{\natexlab{a}})\citenamefont
  {Tews}, \citenamefont {Margueron},\ and\ \citenamefont
  {Reddy}}]{Tews:2018fk}%
  \BibitemOpen
  \bibfield  {author} {\bibinfo {author} {\bibfnamefont {I.}~\bibnamefont
  {Tews}}, \bibinfo {author} {\bibfnamefont {J.}~\bibnamefont {Margueron}}, \
  and\ \bibinfo {author} {\bibfnamefont {S.}~\bibnamefont {Reddy}},\ }\href
  {\doibase 10.1103/PhysRevC.98.045804} {\bibfield  {journal} {\bibinfo
  {journal} {Phys. Rev.}\ }\textbf {\bibinfo {volume} {C98}},\ \bibinfo {pages}
  {045804} (\bibinfo {year} {2018}{\natexlab{a}})},\ \Eprint
  {http://arxiv.org/abs/1804.02783} {arXiv:1804.02783} \BibitemShut {NoStop}%
%%CITATION = ARXIV:1804.02783;%%
\bibitem [{\citenamefont {Annala}\ et~al.(2018)\citenamefont {Annala},
  \citenamefont {Gorda}, \citenamefont {Kurkela},\ and\ \citenamefont
  {Vuorinen}}]{Annala:2017llu}%
  \BibitemOpen
  \bibfield  {author} {\bibinfo {author} {\bibfnamefont {E.}~\bibnamefont
  {Annala}}, \bibinfo {author} {\bibfnamefont {T.}~\bibnamefont {Gorda}},
  \bibinfo {author} {\bibfnamefont {A.}~\bibnamefont {Kurkela}}, \ and\
  \bibinfo {author} {\bibfnamefont {A.}~\bibnamefont {Vuorinen}},\ }\href
  {\doibase 10.1103/PhysRevLett.120.172703} {\bibfield  {journal} {\bibinfo
  {journal} {Phys. Rev. Lett.}\ }\textbf {\bibinfo {volume} {120}},\ \bibinfo
  {pages} {172703} (\bibinfo {year} {2018})},\ \Eprint
  {http://arxiv.org/abs/1711.02644} {arXiv:1711.02644} \BibitemShut {NoStop}%
%%CITATION = ARXIV:1711.02644;%%
\bibitem [{\citenamefont {Fujimoto}\ et~al.(2018)\citenamefont {Fujimoto},
  \citenamefont {Fukushima},\ and\ \citenamefont {Murase}}]{1711.06748}%
  \BibitemOpen
  \bibfield  {author} {\bibinfo {author} {\bibfnamefont {Y.}~\bibnamefont
  {Fujimoto}}, \bibinfo {author} {\bibfnamefont {K.}~\bibnamefont {Fukushima}},
  \ and\ \bibinfo {author} {\bibfnamefont {K.}~\bibnamefont {Murase}},\ }\href
  {\doibase 10.1103/PhysRevD.98.023019} {\bibfield  {journal} {\bibinfo
  {journal} {Phys. Rev.}\ }\textbf {\bibinfo {volume} {D98}},\ \bibinfo {pages}
  {023019} (\bibinfo {year} {2018})},\ \Eprint
  {http://arxiv.org/abs/1711.06748} {arXiv:1711.06748} \BibitemShut {NoStop}%
%%CITATION = ARXIV:1711.06748;%%
\bibitem [{\citenamefont {Kurkela}\ et~al.(2010)\citenamefont {Kurkela},
  \citenamefont {Romatschke},\ and\ \citenamefont {Vuorinen}}]{Kurkela:2009gj}%
  \BibitemOpen
  \bibfield  {author} {\bibinfo {author} {\bibfnamefont {A.}~\bibnamefont
  {Kurkela}}, \bibinfo {author} {\bibfnamefont {P.}~\bibnamefont {Romatschke}},
  \ and\ \bibinfo {author} {\bibfnamefont {A.}~\bibnamefont {Vuorinen}},\
  }\href {\doibase 10.1103/PhysRevD.81.105021} {\bibfield  {journal} {\bibinfo
  {journal} {Phys. Rev.}\ }\textbf {\bibinfo {volume} {D81}},\ \bibinfo {pages}
  {105021} (\bibinfo {year} {2010})},\ \Eprint {http://arxiv.org/abs/0912.1856}
  {arXiv:0912.1856} \BibitemShut {NoStop}%
%%CITATION = ARXIV:0912.1856;%%
\bibitem [{\citenamefont {Xu}\ et~al.(2015{\natexlab{a}})\citenamefont {Xu},
  \citenamefont {Peng}, \citenamefont {Liu}, \citenamefont {Hou},\ and\
  \citenamefont {Chen}}]{Xu:2015wya}%
  \BibitemOpen
  \bibfield  {author} {\bibinfo {author} {\bibfnamefont {J.~F.}\ \bibnamefont
  {Xu}}, \bibinfo {author} {\bibfnamefont {G.~X.}\ \bibnamefont {Peng}},
  \bibinfo {author} {\bibfnamefont {F.}~\bibnamefont {Liu}}, \bibinfo {author}
  {\bibfnamefont {D.-F.}\ \bibnamefont {Hou}}, \ and\ \bibinfo {author}
  {\bibfnamefont {L.-W.}\ \bibnamefont {Chen}},\ }\href {\doibase
  10.1103/PhysRevD.92.025025} {\bibfield  {journal} {\bibinfo  {journal} {Phys.
  Rev.}\ }\textbf {\bibinfo {volume} {D92}},\ \bibinfo {pages} {025025}
  (\bibinfo {year} {2015}{\natexlab{a}})},\ \Eprint
  {http://arxiv.org/abs/1512.08229} {arXiv:1512.08229} \BibitemShut {NoStop}%
%%CITATION = ARXIV:1512.08229;%%
\bibitem [{\citenamefont {Benvenuto}\ and\ \citenamefont
  {Lugones}(1995)}]{Benvenuto:1989kc}%
  \BibitemOpen
  \bibfield  {author} {\bibinfo {author} {\bibfnamefont {O.~G.}\ \bibnamefont
  {Benvenuto}}\ and\ \bibinfo {author} {\bibfnamefont {G.}~\bibnamefont
  {Lugones}},\ }\href {\doibase 10.1103/PhysRevD.51.1989} {\bibfield  {journal}
  {\bibinfo  {journal} {Phys. Rev.}\ }\textbf {\bibinfo {volume} {D51}},\
  \bibinfo {pages} {1989} (\bibinfo {year} {1995})}\BibitemShut {NoStop}%
%%CITATION = PHRVA,D51,1989;%%
\bibitem [{\citenamefont {Torres}\ and\ \citenamefont
  {Menezes}(2013)}]{Torres:2012xv}%
  \BibitemOpen
  \bibfield  {author} {\bibinfo {author} {\bibfnamefont {J.~R.}\ \bibnamefont
  {Torres}}\ and\ \bibinfo {author} {\bibfnamefont {D.~P.}\ \bibnamefont
  {Menezes}},\ }\href {\doibase 10.1209/0295-5075/101/42003} {\bibfield
  {journal} {\bibinfo  {journal} {Europhys. Lett.}\ }\textbf {\bibinfo {volume}
  {101}},\ \bibinfo {pages} {42003} (\bibinfo {year} {2013})},\ \Eprint
  {http://arxiv.org/abs/1210.2350} {arXiv:1210.2350} \BibitemShut {NoStop}%
%%CITATION = ARXIV:1210.2350;%%
\bibitem [{\citenamefont {Baldo}\ et~al.(2003)\citenamefont {Baldo},
  \citenamefont {Buballa}, \citenamefont {Burgio}, \citenamefont {Neumann},
  \citenamefont {Oertel},\ and\ \citenamefont {Schulze}}]{Baldo:2002ju}%
  \BibitemOpen
  \bibfield  {author} {\bibinfo {author} {\bibfnamefont {M.}~\bibnamefont
  {Baldo}}, \bibinfo {author} {\bibfnamefont {M.}~\bibnamefont {Buballa}},
  \bibinfo {author} {\bibfnamefont {F.}~\bibnamefont {Burgio}}, \bibinfo
  {author} {\bibfnamefont {F.}~\bibnamefont {Neumann}}, \bibinfo {author}
  {\bibfnamefont {M.}~\bibnamefont {Oertel}}, \ and\ \bibinfo {author}
  {\bibfnamefont {H.~J.}\ \bibnamefont {Schulze}},\ }\href {\doibase
  10.1016/S0370-2693(03)00556-2} {\bibfield  {journal} {\bibinfo  {journal}
  {Phys. Lett.}\ }\textbf {\bibinfo {volume} {B562}},\ \bibinfo {pages} {153}
  (\bibinfo {year} {2003})},\ \Eprint {http://arxiv.org/abs/nucl-th/0212096}
  {arXiv:nucl-th/0212096} \BibitemShut {NoStop}%
%%CITATION = NUCL-TH/0212096;%%
\bibitem [{\citenamefont {Buballa}(2005)}]{Buballa:2003qv}%
  \BibitemOpen
  \bibfield  {author} {\bibinfo {author} {\bibfnamefont {M.}~\bibnamefont
  {Buballa}},\ }\href {\doibase 10.1016/j.physrep.2004.11.004} {\bibfield
  {journal} {\bibinfo  {journal} {Phys. Rept.}\ }\textbf {\bibinfo {volume}
  {407}},\ \bibinfo {pages} {205} (\bibinfo {year} {2005})},\ \Eprint
  {http://arxiv.org/abs/hep-ph/0402234} {arXiv:hep-ph/0402234} \BibitemShut
  {NoStop}%
%%CITATION = HEP-PH/0402234;%%
\bibitem [{\citenamefont {Pereira}\ et~al.(2016)\citenamefont {Pereira},
  \citenamefont {Costa},\ and\ \citenamefont
  {Provid{\^e}ncia}}]{Pereira:2016dfg}%
  \BibitemOpen
  \bibfield  {author} {\bibinfo {author} {\bibfnamefont {R.~C.}\ \bibnamefont
  {Pereira}}, \bibinfo {author} {\bibfnamefont {P.}~\bibnamefont {Costa}}, \
  and\ \bibinfo {author} {\bibfnamefont {C.}~\bibnamefont {Provid{\^e}ncia}},\
  }\href {\doibase 10.1103/PhysRevD.94.094001} {\bibfield  {journal} {\bibinfo
  {journal} {Phys. Rev.}\ }\textbf {\bibinfo {volume} {D94}},\ \bibinfo {pages}
  {094001} (\bibinfo {year} {2016})},\ \Eprint
  {http://arxiv.org/abs/1610.06435} {arXiv:1610.06435} \BibitemShut {NoStop}%
%%CITATION = ARXIV:1610.06435;%%
\bibitem [{\citenamefont {Li}\ et~al.(2018)\citenamefont {Li}, \citenamefont
  {Zhang}, \citenamefont {Yan}, \citenamefont {Huang},\ and\ \citenamefont
  {Zong}}]{Li:2018fk}%
  \BibitemOpen
  \bibfield  {author} {\bibinfo {author} {\bibfnamefont {C.-M.}\ \bibnamefont
  {Li}}, \bibinfo {author} {\bibfnamefont {J.-L.}\ \bibnamefont {Zhang}},
  \bibinfo {author} {\bibfnamefont {Y.}~\bibnamefont {Yan}}, \bibinfo {author}
  {\bibfnamefont {Y.-F.}\ \bibnamefont {Huang}}, \ and\ \bibinfo {author}
  {\bibfnamefont {H.-S.}\ \bibnamefont {Zong}},\ }\href {\doibase
  10.1103/PhysRevD.97.103013} {\bibfield  {journal} {\bibinfo  {journal} {Phys.
  Rev.}\ }\textbf {\bibinfo {volume} {D97}},\ \bibinfo {pages} {103013}
  (\bibinfo {year} {2018})},\ \Eprint {http://arxiv.org/abs/1804.10785}
  {arXiv:1804.10785} \BibitemShut {NoStop}%
%%CITATION = ARXIV:1804.10785;%%
\bibitem [{\citenamefont {Zacchi}\ and\ \citenamefont
  {Schaffner-Bielich}(2019)}]{Zacchi:2019ayh}%
  \BibitemOpen
  \bibfield  {author} {\bibinfo {author} {\bibfnamefont {A.}~\bibnamefont
  {Zacchi}}\ and\ \bibinfo {author} {\bibfnamefont {J.}~\bibnamefont
  {Schaffner-Bielich}},\ }\href@noop {} {\ }\Eprint
  {http://arxiv.org/abs/1909.12071} {arXiv:1909.12071} \BibitemShut {NoStop}%
%%CITATION = ARXIV:1909.12071;%%
\bibitem [{\citenamefont {Zacchi}\ et~al.(2015)\citenamefont {Zacchi},
  \citenamefont {Stiele},\ and\ \citenamefont
  {Schaffner-Bielich}}]{Zacchi:2015lwa}%
  \BibitemOpen
  \bibfield  {author} {\bibinfo {author} {\bibfnamefont {A.}~\bibnamefont
  {Zacchi}}, \bibinfo {author} {\bibfnamefont {R.}~\bibnamefont {Stiele}}, \
  and\ \bibinfo {author} {\bibfnamefont {J.}~\bibnamefont
  {Schaffner-Bielich}},\ }\href {\doibase 10.1103/PhysRevD.92.045022}
  {\bibfield  {journal} {\bibinfo  {journal} {Phys. Rev.}\ }\textbf {\bibinfo
  {volume} {D92}},\ \bibinfo {pages} {045022} (\bibinfo {year} {2015})},\
  \Eprint {http://arxiv.org/abs/1506.01868} {arXiv:1506.01868} \BibitemShut
  {NoStop}%
%%CITATION = ARXIV:1506.01868;%%
\bibitem [{\citenamefont {Xu}\ et~al.(2015{\natexlab{b}})\citenamefont {Xu},
  \citenamefont {Yan}, \citenamefont {Cui},\ and\ \citenamefont
  {Zong}}]{1506.06846}%
  \BibitemOpen
  \bibfield  {author} {\bibinfo {author} {\bibfnamefont {S.-S.}\ \bibnamefont
  {Xu}}, \bibinfo {author} {\bibfnamefont {Y.}~\bibnamefont {Yan}}, \bibinfo
  {author} {\bibfnamefont {Z.-F.}\ \bibnamefont {Cui}}, \ and\ \bibinfo
  {author} {\bibfnamefont {H.-S.}\ \bibnamefont {Zong}},\ }\href {\doibase
  10.1142/S0217751X15502176} {\bibfield  {journal} {\bibinfo  {journal} {Int.
  J. Mod. Phys.}\ }\textbf {\bibinfo {volume} {A30}},\ \bibinfo {pages}
  {1550217} (\bibinfo {year} {2015}{\natexlab{b}})},\ \Eprint
  {http://arxiv.org/abs/1506.06846} {arXiv:1506.06846} \BibitemShut {NoStop}%
%%CITATION = ARXIV:1506.06846;%%
\bibitem [{\citenamefont {Zhao}\ et~al.(2015{\natexlab{a}})\citenamefont
  {Zhao}, \citenamefont {Xu}, \citenamefont {Yan}, \citenamefont {Luo},
  \citenamefont {Liu},\ and\ \citenamefont {Zong}}]{Zhao:2015rta}%
  \BibitemOpen
  \bibfield  {author} {\bibinfo {author} {\bibfnamefont {T.}~\bibnamefont
  {Zhao}}, \bibinfo {author} {\bibfnamefont {S.-S.}\ \bibnamefont {Xu}},
  \bibinfo {author} {\bibfnamefont {Y.}~\bibnamefont {Yan}}, \bibinfo {author}
  {\bibfnamefont {X.-L.}\ \bibnamefont {Luo}}, \bibinfo {author} {\bibfnamefont
  {X.-J.}\ \bibnamefont {Liu}}, \ and\ \bibinfo {author} {\bibfnamefont
  {H.-S.}\ \bibnamefont {Zong}},\ }\href {\doibase 10.1103/PhysRevD.92.054012}
  {\bibfield  {journal} {\bibinfo  {journal} {Phys. Rev.}\ }\textbf {\bibinfo
  {volume} {D92}},\ \bibinfo {pages} {054012} (\bibinfo {year}
  {2015}{\natexlab{a}})},\ \Eprint {http://arxiv.org/abs/1509.03377}
  {arXiv:1509.03377} \BibitemShut {NoStop}%
%%CITATION = ARXIV:1509.03377;%%
\bibitem [{\citenamefont {Isserstedt}\ et~al.(2019)\citenamefont {Isserstedt},
  \citenamefont {Buballa}, \citenamefont {Fischer},\ and\ \citenamefont
  {Gunkel}}]{Isserstedt:2019pgx}%
  \BibitemOpen
  \bibfield  {author} {\bibinfo {author} {\bibfnamefont {P.}~\bibnamefont
  {Isserstedt}}, \bibinfo {author} {\bibfnamefont {M.}~\bibnamefont {Buballa}},
  \bibinfo {author} {\bibfnamefont {C.~S.}\ \bibnamefont {Fischer}}, \ and\
  \bibinfo {author} {\bibfnamefont {P.~J.}\ \bibnamefont {Gunkel}},\ }\href
  {\doibase 10.1103/PhysRevD.100.074011} {\bibfield  {journal} {\bibinfo
  {journal} {Phys. Rev.}\ }\textbf {\bibinfo {volume} {D100}},\ \bibinfo
  {pages} {074011} (\bibinfo {year} {2019})},\ \Eprint
  {http://arxiv.org/abs/1906.11644} {arXiv:1906.11644} \BibitemShut {NoStop}%
%%CITATION = ARXIV:1906.11644;%%
\bibitem [{\citenamefont {Fischer}\ and\ \citenamefont
  {Luecker}(2013)}]{Fischer:2012vc}%
  \BibitemOpen
  \bibfield  {author} {\bibinfo {author} {\bibfnamefont {C.~S.}\ \bibnamefont
  {Fischer}}\ and\ \bibinfo {author} {\bibfnamefont {J.}~\bibnamefont
  {Luecker}},\ }\href {\doibase 10.1016/j.physletb.2012.11.054} {\bibfield
  {journal} {\bibinfo  {journal} {Phys.Lett.}\ }\textbf {\bibinfo {volume}
  {B718}},\ \bibinfo {pages} {1036} (\bibinfo {year} {2013})},\ \Eprint
  {http://arxiv.org/abs/1206.5191} {arXiv:1206.5191} \BibitemShut {NoStop}%
%%CITATION = ARXIV:1206.5191;%%
\bibitem [{\citenamefont {Ishii}\ et~al.(2019)\citenamefont {Ishii},
  \citenamefont {J{\"a}rvinen},\ and\ \citenamefont {Nijs}}]{Ishii:2019gta}%
  \BibitemOpen
  \bibfield  {author} {\bibinfo {author} {\bibfnamefont {T.}~\bibnamefont
  {Ishii}}, \bibinfo {author} {\bibfnamefont {M.}~\bibnamefont {J{\"a}rvinen}},
  \ and\ \bibinfo {author} {\bibfnamefont {G.}~\bibnamefont {Nijs}},\ }\href
  {\doibase 10.1007/JHEP07(2019)003} {\bibfield  {journal} {\bibinfo  {journal}
  {JHEP}\ }\textbf {\bibinfo {volume} {07}},\ \bibinfo {pages} {003} (\bibinfo
  {year} {2019})},\ \Eprint {http://arxiv.org/abs/1903.06169}
  {arXiv:1903.06169} \BibitemShut {NoStop}%
%%CITATION = ARXIV:1903.06169;%%
\bibitem [{\citenamefont {Peshier}\ et~al.(2000)\citenamefont {Peshier},
  \citenamefont {Kampfer},\ and\ \citenamefont {Soff}}]{Peshier:1999ww}%
  \BibitemOpen
  \bibfield  {author} {\bibinfo {author} {\bibfnamefont {A.}~\bibnamefont
  {Peshier}}, \bibinfo {author} {\bibfnamefont {B.}~\bibnamefont {Kampfer}}, \
  and\ \bibinfo {author} {\bibfnamefont {G.}~\bibnamefont {Soff}},\ }\href
  {\doibase 10.1103/PhysRevC.61.045203} {\bibfield  {journal} {\bibinfo
  {journal} {Phys. Rev.}\ }\textbf {\bibinfo {volume} {C61}},\ \bibinfo {pages}
  {045203} (\bibinfo {year} {2000})},\ \Eprint
  {http://arxiv.org/abs/hep-ph/9911474} {arXiv:hep-ph/9911474} \BibitemShut
  {NoStop}%
%%CITATION = HEP-PH/9911474;%%
\bibitem [{\citenamefont {Tian}\ et~al.(2012)\citenamefont {Tian},
  \citenamefont {Yan}, \citenamefont {Li}, \citenamefont {Luo},\ and\
  \citenamefont {Zong}}]{Tian:2012zza}%
  \BibitemOpen
  \bibfield  {author} {\bibinfo {author} {\bibfnamefont {Y.-L.}\ \bibnamefont
  {Tian}}, \bibinfo {author} {\bibfnamefont {Y.}~\bibnamefont {Yan}}, \bibinfo
  {author} {\bibfnamefont {H.}~\bibnamefont {Li}}, \bibinfo {author}
  {\bibfnamefont {X.-L.}\ \bibnamefont {Luo}}, \ and\ \bibinfo {author}
  {\bibfnamefont {H.-S.}\ \bibnamefont {Zong}},\ }\href {\doibase
  10.1103/PhysRevD.85.045009} {\bibfield  {journal} {\bibinfo  {journal} {Phys.
  Rev.}\ }\textbf {\bibinfo {volume} {D85}},\ \bibinfo {pages} {045009}
  (\bibinfo {year} {2012})}\BibitemShut {NoStop}%
%%CITATION = PHRVA,D85,045009;%%
\bibitem [{\citenamefont {Zhao}\ et~al.(2015{\natexlab{b}})\citenamefont
  {Zhao}, \citenamefont {Yan}, \citenamefont {Luo},\ and\ \citenamefont
  {Zong}}]{Zhao:2015uia}%
  \BibitemOpen
  \bibfield  {author} {\bibinfo {author} {\bibfnamefont {T.}~\bibnamefont
  {Zhao}}, \bibinfo {author} {\bibfnamefont {Y.}~\bibnamefont {Yan}}, \bibinfo
  {author} {\bibfnamefont {X.-L.}\ \bibnamefont {Luo}}, \ and\ \bibinfo
  {author} {\bibfnamefont {H.-S.}\ \bibnamefont {Zong}},\ }\href {\doibase
  10.1103/PhysRevD.91.034018} {\bibfield  {journal} {\bibinfo  {journal} {Phys.
  Rev.}\ }\textbf {\bibinfo {volume} {D91}},\ \bibinfo {pages} {034018}
  (\bibinfo {year} {2015}{\natexlab{b}})}\BibitemShut {NoStop}%
%%CITATION = PHRVA,D91,034018;%%
\bibitem [{\citenamefont {Schramm}\ et~al.(2016)\citenamefont {Schramm},
  \citenamefont {Dexheimer},\ and\ \citenamefont
  {Negreiros}}]{Schramm:2015hba}%
  \BibitemOpen
  \bibfield  {author} {\bibinfo {author} {\bibfnamefont {S.}~\bibnamefont
  {Schramm}}, \bibinfo {author} {\bibfnamefont {V.}~\bibnamefont {Dexheimer}},
  \ and\ \bibinfo {author} {\bibfnamefont {R.}~\bibnamefont {Negreiros}},\
  }\href {\doibase 10.1140/epja/i2016-16014-5} {\bibfield  {journal} {\bibinfo
  {journal} {Eur. Phys. J.}\ }\textbf {\bibinfo {volume} {A52}},\ \bibinfo
  {pages} {14} (\bibinfo {year} {2016})},\ \Eprint
  {http://arxiv.org/abs/1508.04699} {arXiv:1508.04699} \BibitemShut {NoStop}%
%%CITATION = ARXIV:1508.04699;%%
\bibitem [{\citenamefont {Hajizadeh}\ and\ \citenamefont
  {Maas}(2016)}]{Hajizadeh:2016jvj}%
  \BibitemOpen
  \bibfield  {author} {\bibinfo {author} {\bibfnamefont {O.}~\bibnamefont
  {Hajizadeh}}\ and\ \bibinfo {author} {\bibfnamefont {A.}~\bibnamefont
  {Maas}},\ }\href {\doibase 10.22323/1.256.0358} {\bibfield  {journal}
  {\bibinfo  {journal} {PoS}\ }\textbf {\bibinfo {volume} {LATTICE2016}},\
  \bibinfo {pages} {358} (\bibinfo {year} {2016})},\ \Eprint
  {http://arxiv.org/abs/1609.06979} {arXiv:1609.06979} \BibitemShut {NoStop}%
%%CITATION = ARXIV:1609.06979;%%
\bibitem [{\citenamefont {Leonhardt}\ et~al.(2019)\citenamefont {Leonhardt},
  \citenamefont {Pospiech}, \citenamefont {Schallmo}, \citenamefont {Braun},
  \citenamefont {Drischler}, \citenamefont {Hebeler},\ and\ \citenamefont
  {Schwenk}}]{Leonhardt:2019fua}%
  \BibitemOpen
  \bibfield  {author} {\bibinfo {author} {\bibfnamefont {M.}~\bibnamefont
  {Leonhardt}}, \bibinfo {author} {\bibfnamefont {M.}~\bibnamefont {Pospiech}},
  \bibinfo {author} {\bibfnamefont {B.}~\bibnamefont {Schallmo}}, \bibinfo
  {author} {\bibfnamefont {J.}~\bibnamefont {Braun}}, \bibinfo {author}
  {\bibfnamefont {C.}~\bibnamefont {Drischler}}, \bibinfo {author}
  {\bibfnamefont {K.}~\bibnamefont {Hebeler}}, \ and\ \bibinfo {author}
  {\bibfnamefont {A.}~\bibnamefont {Schwenk}},\ }\href@noop {} {\ }\Eprint
  {http://arxiv.org/abs/1907.05814} {arXiv:1907.05814} \BibitemShut {NoStop}%
%%CITATION = ARXIV:1907.05814;%%
\bibitem [{\citenamefont {Friman}\ and\ \citenamefont
  {Weise}(2019)}]{Friman:2019ncm}%
  \BibitemOpen
  \bibfield  {author} {\bibinfo {author} {\bibfnamefont {B.}~\bibnamefont
  {Friman}}\ and\ \bibinfo {author} {\bibfnamefont {W.}~\bibnamefont {Weise}},\
  }\href@noop {} {\ }\Eprint {http://arxiv.org/abs/1908.09722}
  {arXiv:1908.09722} \BibitemShut {NoStop}%
%%CITATION = ARXIV:1908.09722;%%
\bibitem [{\citenamefont {Drews}\ and\ \citenamefont
  {Weise}(2017)}]{Drews:2016wpi}%
  \BibitemOpen
  \bibfield  {author} {\bibinfo {author} {\bibfnamefont {M.}~\bibnamefont
  {Drews}}\ and\ \bibinfo {author} {\bibfnamefont {W.}~\bibnamefont {Weise}},\
  }\href {\doibase 10.1016/j.ppnp.2016.10.002} {\bibfield  {journal} {\bibinfo
  {journal} {Prog. Part. Nucl. Phys.}\ }\textbf {\bibinfo {volume} {93}},\
  \bibinfo {pages} {69} (\bibinfo {year} {2017})},\ \Eprint
  {http://arxiv.org/abs/1610.07568} {arXiv:1610.07568} \BibitemShut {NoStop}%
%%CITATION = ARXIV:1610.07568;%%
\bibitem [{\citenamefont {P{\'o}sfay}\ et~al.(2018)\citenamefont {P{\'o}sfay},
  \citenamefont {Barnaf{\"o}ldi},\ and\ \citenamefont
  {Jakov{\'a}c}}]{Posfay:2018sw}%
  \BibitemOpen
  \bibfield  {author} {\bibinfo {author} {\bibfnamefont {P.}~\bibnamefont
  {P{\'o}sfay}}, \bibinfo {author} {\bibfnamefont {G.~G.}\ \bibnamefont
  {Barnaf{\"o}ldi}}, \ and\ \bibinfo {author} {\bibfnamefont {A.}~\bibnamefont
  {Jakov{\'a}c}},\ }\href {\doibase 10.1103/PhysRevC.97.025803} {\bibfield
  {journal} {\bibinfo  {journal} {Phys. Rev.}\ }\textbf {\bibinfo {volume}
  {C97}},\ \bibinfo {pages} {025803} (\bibinfo {year} {2018})},\ \Eprint
  {http://arxiv.org/abs/1610.03674} {arXiv:1610.03674} \BibitemShut {NoStop}%
%%CITATION = ARXIV:1610.03674;%%
\bibitem [{\citenamefont {Dai}\ et~al.(1995)\citenamefont {Dai}, \citenamefont
  {Peng},\ and\ \citenamefont {Lu}}]{Dai:1995uj}%
  \BibitemOpen
  \bibfield  {author} {\bibinfo {author} {\bibfnamefont {Z.-G.}\ \bibnamefont
  {Dai}}, \bibinfo {author} {\bibfnamefont {Q.-H.}\ \bibnamefont {Peng}}, \
  and\ \bibinfo {author} {\bibfnamefont {T.}~\bibnamefont {Lu}},\ }\href
  {\doibase 10.1086/175316} {\bibfield  {journal} {\bibinfo  {journal}
  {Astrophys. J.}\ }\textbf {\bibinfo {volume} {440}},\ \bibinfo {pages} {815}
  (\bibinfo {year} {1995})}\BibitemShut {NoStop}%
%%CITATION = ASJOA,440,815;%%
\bibitem [{\citenamefont {Buballa}\ et~al.(2004)\citenamefont {Buballa},
  \citenamefont {Neumann}, \citenamefont {Oertel},\ and\ \citenamefont
  {Shovkovy}}]{Buballa:2003et}%
  \BibitemOpen
  \bibfield  {author} {\bibinfo {author} {\bibfnamefont {M.}~\bibnamefont
  {Buballa}}, \bibinfo {author} {\bibfnamefont {F.}~\bibnamefont {Neumann}},
  \bibinfo {author} {\bibfnamefont {M.}~\bibnamefont {Oertel}}, \ and\ \bibinfo
  {author} {\bibfnamefont {I.}~\bibnamefont {Shovkovy}},\ }\href {\doibase
  10.1016/j.physletb.2004.05.064} {\bibfield  {journal} {\bibinfo  {journal}
  {Phys. Lett.}\ }\textbf {\bibinfo {volume} {B595}},\ \bibinfo {pages} {36}
  (\bibinfo {year} {2004})},\ \Eprint {http://arxiv.org/abs/nucl-th/0312078}
  {arXiv:nucl-th/0312078} \BibitemShut {NoStop}%
%%CITATION = NUCL-TH/0312078;%%
\bibitem [{\citenamefont {Drago}\ et~al.(2007)\citenamefont {Drago},
  \citenamefont {Lavagno},\ and\ \citenamefont {Parenti}}]{Drago:2005yj}%
  \BibitemOpen
  \bibfield  {author} {\bibinfo {author} {\bibfnamefont {A.}~\bibnamefont
  {Drago}}, \bibinfo {author} {\bibfnamefont {A.}~\bibnamefont {Lavagno}}, \
  and\ \bibinfo {author} {\bibfnamefont {I.}~\bibnamefont {Parenti}},\ }\href
  {\doibase 10.1086/512112} {\bibfield  {journal} {\bibinfo  {journal}
  {Astrophys. J.}\ }\textbf {\bibinfo {volume} {659}},\ \bibinfo {pages} {1519}
  (\bibinfo {year} {2007})},\ \Eprint {http://arxiv.org/abs/astro-ph/0512652}
  {arXiv:astro-ph/0512652} \BibitemShut {NoStop}%
%%CITATION = ASTRO-PH/0512652;%%
\bibitem [{\citenamefont {Chatterjee}\ et~al.(2015)\citenamefont {Chatterjee},
  \citenamefont {Elghozi}, \citenamefont {Novak},\ and\ \citenamefont
  {Oertel}}]{Chatterjee:2014qsa}%
  \BibitemOpen
  \bibfield  {author} {\bibinfo {author} {\bibfnamefont {D.}~\bibnamefont
  {Chatterjee}}, \bibinfo {author} {\bibfnamefont {T.}~\bibnamefont {Elghozi}},
  \bibinfo {author} {\bibfnamefont {J.}~\bibnamefont {Novak}}, \ and\ \bibinfo
  {author} {\bibfnamefont {M.}~\bibnamefont {Oertel}},\ }\href {\doibase
  10.1093/mnras/stu2706} {\bibfield  {journal} {\bibinfo  {journal} {Mon. Not.
  Roy. Astron. Soc.}\ }\textbf {\bibinfo {volume} {447}},\ \bibinfo {pages}
  {3785} (\bibinfo {year} {2015})},\ \Eprint {http://arxiv.org/abs/1410.6332}
  {arXiv:1410.6332} \BibitemShut {NoStop}%
%%CITATION = ARXIV:1410.6332;%%
\bibitem [{\citenamefont {Weise}(2018)}]{Weise:2018ukn}%
  \BibitemOpen
  \bibfield  {author} {\bibinfo {author} {\bibfnamefont {W.}~\bibnamefont
  {Weise}},\ }\href {\doibase 10.1142/S0218301318400049} {\bibfield  {journal}
  {\bibinfo  {journal} {Int. J. Mod. Phys.}\ }\textbf {\bibinfo {volume}
  {E27}},\ \bibinfo {pages} {1840004} (\bibinfo {year} {2018})},\ \Eprint
  {http://arxiv.org/abs/1811.09682} {arXiv:1811.09682} \BibitemShut {NoStop}%
%%CITATION = ARXIV:1811.09682;%%
\bibitem [{\citenamefont {Mitter}\ and\ \citenamefont
  {Schaefer}(2014)}]{Mitter:2013fxa}%
  \BibitemOpen
  \bibfield  {author} {\bibinfo {author} {\bibfnamefont {M.}~\bibnamefont
  {Mitter}}\ and\ \bibinfo {author} {\bibfnamefont {B.-J.}\ \bibnamefont
  {Schaefer}},\ }\href {\doibase 10.1103/PhysRevD.89.054027} {\bibfield
  {journal} {\bibinfo  {journal} {Phys.Rev.}\ }\textbf {\bibinfo {volume}
  {D89}},\ \bibinfo {pages} {054027} (\bibinfo {year} {2014})},\ \Eprint
  {http://arxiv.org/abs/1308.3176} {arXiv:1308.3176} \BibitemShut {NoStop}%
%%CITATION = ARXIV:1308.3176;%%
\bibitem [{\citenamefont {Schaefer}\ and\ \citenamefont
  {Wagner}(2009)}]{Schaefer:2008hk}%
  \BibitemOpen
  \bibfield  {author} {\bibinfo {author} {\bibfnamefont {B.-J.}\ \bibnamefont
  {Schaefer}}\ and\ \bibinfo {author} {\bibfnamefont {M.}~\bibnamefont
  {Wagner}},\ }\href {\doibase 10.1103/PhysRevD.79.014018} {\bibfield
  {journal} {\bibinfo  {journal} {Phys. Rev.}\ }\textbf {\bibinfo {volume}
  {D79}},\ \bibinfo {pages} {014018} (\bibinfo {year} {2009})},\ \Eprint
  {http://arxiv.org/abs/0808.1491} {arXiv:0808.1491} \BibitemShut {NoStop}%
%%CITATION = 0808.1491;%%
\bibitem [{\citenamefont {Skokov}\ et~al.(2010)\citenamefont {Skokov},
  \citenamefont {Friman}, \citenamefont {Nakano}, \citenamefont {Redlich},\
  and\ \citenamefont {Schaefer}}]{Skokov:2010sf}%
  \BibitemOpen
  \bibfield  {author} {\bibinfo {author} {\bibfnamefont {V.}~\bibnamefont
  {Skokov}}, \bibinfo {author} {\bibfnamefont {B.}~\bibnamefont {Friman}},
  \bibinfo {author} {\bibfnamefont {E.}~\bibnamefont {Nakano}}, \bibinfo
  {author} {\bibfnamefont {K.}~\bibnamefont {Redlich}}, \ and\ \bibinfo
  {author} {\bibfnamefont {B.-J.}\ \bibnamefont {Schaefer}},\ }\href {\doibase
  10.1103/PhysRevD.82.034029} {\bibfield  {journal} {\bibinfo  {journal} {Phys.
  Rev.}\ }\textbf {\bibinfo {volume} {D82}},\ \bibinfo {pages} {034029}
  (\bibinfo {year} {2010})},\ \Eprint {http://arxiv.org/abs/1005.3166}
  {arXiv:1005.3166} \BibitemShut {NoStop}%
%%CITATION = 1005.3166;%%
\bibitem [{\citenamefont {Schaefer}\ and\ \citenamefont
  {Wagner}(2012)}]{Schaefer:2011ex}%
  \BibitemOpen
  \bibfield  {author} {\bibinfo {author} {\bibfnamefont {B.-J.}\ \bibnamefont
  {Schaefer}}\ and\ \bibinfo {author} {\bibfnamefont {M.}~\bibnamefont
  {Wagner}},\ }\href {\doibase 10.1103/PhysRevD.85.034027} {\bibfield
  {journal} {\bibinfo  {journal} {Phys.Rev.}\ }\textbf {\bibinfo {volume}
  {D85}},\ \bibinfo {pages} {034027} (\bibinfo {year} {2012})},\ \Eprint
  {http://arxiv.org/abs/1111.6871} {arXiv:1111.6871} \BibitemShut {NoStop}%
%%CITATION = ARXIV:1111.6871;%%
\bibitem [{\citenamefont {Gupta}\ and\ \citenamefont
  {Tiwari}(2012)}]{Gupta:2011ez}%
  \BibitemOpen
  \bibfield  {author} {\bibinfo {author} {\bibfnamefont {U.~S.}\ \bibnamefont
  {Gupta}}\ and\ \bibinfo {author} {\bibfnamefont {V.~K.}\ \bibnamefont
  {Tiwari}},\ }\href {\doibase 10.1103/PhysRevD.85.014010} {\bibfield
  {journal} {\bibinfo  {journal} {Phys.Rev.}\ }\textbf {\bibinfo {volume}
  {D85}},\ \bibinfo {pages} {014010} (\bibinfo {year} {2012})},\ \Eprint
  {http://arxiv.org/abs/1107.1312} {arXiv:1107.1312} \BibitemShut {NoStop}%
%%CITATION = ARXIV:1107.1312;%%
\bibitem [{\citenamefont {Khan}\ et~al.(2016)\citenamefont {Khan},
  \citenamefont {Andersen}, \citenamefont {Kyllingstad},\ and\ \citenamefont
  {Khan}}]{Andersen:2011pr}%
  \BibitemOpen
  \bibfield  {author} {\bibinfo {author} {\bibfnamefont {R.}~\bibnamefont
  {Khan}}, \bibinfo {author} {\bibfnamefont {J.~O.}\ \bibnamefont {Andersen}},
  \bibinfo {author} {\bibfnamefont {L.~T.}\ \bibnamefont {Kyllingstad}}, \ and\
  \bibinfo {author} {\bibfnamefont {M.}~\bibnamefont {Khan}},\ }\href {\doibase
  10.1142/S0217751X16500251} {\bibfield  {journal} {\bibinfo  {journal} {Int.
  J. Mod. Phys.}\ }\textbf {\bibinfo {volume} {A31}},\ \bibinfo {pages}
  {1650025} (\bibinfo {year} {2016})},\ \Eprint
  {http://arxiv.org/abs/1102.2779} {arXiv:1102.2779} \BibitemShut {NoStop}%
%%CITATION = ARXIV:1102.2779;%%
\bibitem [{\citenamefont {Wetterich}(1993)}]{Wetterich:1992yh}%
  \BibitemOpen
  \bibfield  {author} {\bibinfo {author} {\bibfnamefont {C.}~\bibnamefont
  {Wetterich}},\ }\href {\doibase 10.1016/0370-2693(93)90726-X} {\bibfield
  {journal} {\bibinfo  {journal} {Phys. Lett.}\ }\textbf {\bibinfo {volume}
  {B301}},\ \bibinfo {pages} {90} (\bibinfo {year} {1993})}\BibitemShut
  {NoStop}%
%%CITATION = PHLTA,B301,90;%%
\bibitem [{\citenamefont {Berges}\ et~al.(1997)\citenamefont {Berges},
  \citenamefont {Tetradis},\ and\ \citenamefont {Wetterich}}]{Berges:1996ib}%
  \BibitemOpen
  \bibfield  {author} {\bibinfo {author} {\bibfnamefont {J.}~\bibnamefont
  {Berges}}, \bibinfo {author} {\bibfnamefont {N.}~\bibnamefont {Tetradis}}, \
  and\ \bibinfo {author} {\bibfnamefont {C.}~\bibnamefont {Wetterich}},\ }\href
  {\doibase 10.1016/S0370-2693(96)01654-1} {\bibfield  {journal} {\bibinfo
  {journal} {Phys.Lett.}\ }\textbf {\bibinfo {volume} {B393}},\ \bibinfo
  {pages} {387} (\bibinfo {year} {1997})},\ \Eprint
  {http://arxiv.org/abs/hep-ph/9610354} {arXiv:hep-ph/9610354} \BibitemShut
  {NoStop}%
%%CITATION = HEP-PH/9610354;%%
\bibitem [{\citenamefont {Pawlowski}(2007)}]{Pawlowski:2005xe}%
  \BibitemOpen
  \bibfield  {author} {\bibinfo {author} {\bibfnamefont {J.~M.}\ \bibnamefont
  {Pawlowski}},\ }\href {\doibase 10.1016/j.aop.2007.01.007} {\bibfield
  {journal} {\bibinfo  {journal} {Annals Phys.}\ }\textbf {\bibinfo {volume}
  {322}},\ \bibinfo {pages} {2831} (\bibinfo {year} {2007})},\ \Eprint
  {http://arxiv.org/abs/hep-th/0512261} {arXiv:hep-th/0512261} \BibitemShut
  {NoStop}%
%%CITATION = HEP-TH/0512261;%%
\bibitem [{\citenamefont {Braun}(2012)}]{Braun:2011pp}%
  \BibitemOpen
  \bibfield  {author} {\bibinfo {author} {\bibfnamefont {J.}~\bibnamefont
  {Braun}},\ }\href {\doibase 10.1088/0954-3899/39/3/033001} {\bibfield
  {journal} {\bibinfo  {journal} {J.Phys.}\ }\textbf {\bibinfo {volume}
  {G39}},\ \bibinfo {pages} {033001} (\bibinfo {year} {2012})},\ \Eprint
  {http://arxiv.org/abs/1108.4449} {arXiv:1108.4449} \BibitemShut {NoStop}%
%%CITATION = ARXIV:1108.4449;%%
\bibitem [{\citenamefont {Gies}(2012)}]{Gies2006}%
  \BibitemOpen
  \bibfield  {author} {\bibinfo {author} {\bibfnamefont {H.}~\bibnamefont
  {Gies}},\ }\href
  {http://www.citebase.org/abstract?id=oai:arXiv.org:hep-ph/0611146} {\bibfield
   {journal} {\bibinfo  {journal} {Lect. Notes Phys. 852}\ ,\ \bibinfo {pages}
  {287}} (\bibinfo {year} {2012})},\ \Eprint
  {http://arxiv.org/abs/hep-ph/0611146} {arXiv:hep-ph/0611146} \BibitemShut
  {NoStop}%
%%CITATION = HEP-PH/0611146;%%
\bibitem [{\citenamefont {Schaefer}\ and\ \citenamefont
  {Wambach}(2008)}]{Schaefer:2006sr}%
  \BibitemOpen
  \bibfield  {author} {\bibinfo {author} {\bibfnamefont {B.-J.}\ \bibnamefont
  {Schaefer}}\ and\ \bibinfo {author} {\bibfnamefont {J.}~\bibnamefont
  {Wambach}},\ }\href {\doibase 10.1134/S1063779608070083} {\bibfield
  {journal} {\bibinfo  {journal} {Phys. Part. Nucl.}\ }\textbf {\bibinfo
  {volume} {39}},\ \bibinfo {pages} {1025} (\bibinfo {year} {2008})},\ \Eprint
  {http://arxiv.org/abs/hep-ph/0611191} {arXiv:hep-ph/0611191} \BibitemShut
  {NoStop}%
%%CITATION = HEP-PH/0611191;%%
\bibitem [{\citenamefont {Braun}\ et~al.(2019)\citenamefont {Braun},
  \citenamefont {Leonhardt},\ and\ \citenamefont {Pawlowski}}]{Braun:2018svj}%
  \BibitemOpen
  \bibfield  {author} {\bibinfo {author} {\bibfnamefont {J.}~\bibnamefont
  {Braun}}, \bibinfo {author} {\bibfnamefont {M.}~\bibnamefont {Leonhardt}}, \
  and\ \bibinfo {author} {\bibfnamefont {J.~M.}\ \bibnamefont {Pawlowski}},\
  }\href {\doibase 10.21468/SciPostPhys.6.5.056} {\bibfield  {journal}
  {\bibinfo  {journal} {SciPost Phys.}\ }\textbf {\bibinfo {volume} {6}},\
  \bibinfo {pages} {056} (\bibinfo {year} {2019})},\ \Eprint
  {http://arxiv.org/abs/1806.04432} {arXiv:1806.04432} \BibitemShut {NoStop}%
%%CITATION = ARXIV:1806.04432;%%
\bibitem [{\citenamefont {Litim}(2001)}]{Litim:2001up}%
  \BibitemOpen
  \bibfield  {author} {\bibinfo {author} {\bibfnamefont {D.~F.}\ \bibnamefont
  {Litim}},\ }\href {\doibase 10.1103/PhysRevD.64.105007} {\bibfield  {journal}
  {\bibinfo  {journal} {Phys. Rev.}\ }\textbf {\bibinfo {volume} {D64}},\
  \bibinfo {pages} {105007} (\bibinfo {year} {2001})},\ \Eprint
  {http://arxiv.org/abs/hep-th/0103195} {arXiv:hep-th/0103195} \BibitemShut
  {NoStop}%
%%CITATION = HEP-TH/0103195;%%
\bibitem [{\citenamefont {Litim}\ and\ \citenamefont
  {Pawlowski}(2006)}]{Litim:2006ag}%
  \BibitemOpen
  \bibfield  {author} {\bibinfo {author} {\bibfnamefont {D.~F.}\ \bibnamefont
  {Litim}}\ and\ \bibinfo {author} {\bibfnamefont {J.~M.}\ \bibnamefont
  {Pawlowski}},\ }\href
  {http://www.citebase.org/cgi-bin/citations?id=oai:arXiv.org:hep-th/0609122}
  {\bibfield  {journal} {\bibinfo  {journal} {JHEP}\ }\textbf {\bibinfo
  {volume} {11}},\ \bibinfo {pages} {026} (\bibinfo {year} {2006})},\ \Eprint
  {http://arxiv.org/abs/hep-th/0609122} {arXiv:hep-th/0609122} \BibitemShut
  {NoStop}%
%%CITATION = HEP-TH/0609122;%%
\bibitem [{\citenamefont {Cohen}(2003)}]{Cohen:2003kd}%
  \BibitemOpen
  \bibfield  {author} {\bibinfo {author} {\bibfnamefont {T.~D.}\ \bibnamefont
  {Cohen}},\ }\href {\doibase 10.1103/PhysRevLett.91.222001} {\bibfield
  {journal} {\bibinfo  {journal} {Phys. Rev. Lett.}\ }\textbf {\bibinfo
  {volume} {91}},\ \bibinfo {pages} {222001} (\bibinfo {year} {2003})},\
  \Eprint {http://arxiv.org/abs/hep-ph/0307089} {arXiv:hep-ph/0307089}
  \BibitemShut {NoStop}%
%%CITATION = HEP-PH/0307089;%%
\bibitem [{\citenamefont {Schaefer}\ et~al.(2007)\citenamefont {Schaefer},
  \citenamefont {Pawlowski},\ and\ \citenamefont {Wambach}}]{Schaefer:2007pw}%
  \BibitemOpen
  \bibfield  {author} {\bibinfo {author} {\bibfnamefont {B.-J.}\ \bibnamefont
  {Schaefer}}, \bibinfo {author} {\bibfnamefont {J.~M.}\ \bibnamefont
  {Pawlowski}}, \ and\ \bibinfo {author} {\bibfnamefont {J.}~\bibnamefont
  {Wambach}},\ }\href {\doibase 10.1103/PhysRevD.76.074023} {\bibfield
  {journal} {\bibinfo  {journal} {Phys.Rev.}\ }\textbf {\bibinfo {volume}
  {D76}},\ \bibinfo {pages} {074023} (\bibinfo {year} {2007})},\ \Eprint
  {http://arxiv.org/abs/0704.3234} {arXiv:0704.3234} \BibitemShut {NoStop}%
%%CITATION = ARXIV:0704.3234;%%
\bibitem [{\citenamefont {Roessner}\ et~al.(2007)\citenamefont {Roessner},
  \citenamefont {Ratti},\ and\ \citenamefont {Weise}}]{Roessner:2006xn}%
  \BibitemOpen
  \bibfield  {author} {\bibinfo {author} {\bibfnamefont {S.}~\bibnamefont
  {Roessner}}, \bibinfo {author} {\bibfnamefont {C.}~\bibnamefont {Ratti}}, \
  and\ \bibinfo {author} {\bibfnamefont {W.}~\bibnamefont {Weise}},\ }\href
  {\doibase 10.1103/PhysRevD.75.034007} {\bibfield  {journal} {\bibinfo
  {journal} {Phys. Rev.}\ }\textbf {\bibinfo {volume} {D75}},\ \bibinfo {pages}
  {034007} (\bibinfo {year} {2007})},\ \Eprint
  {http://arxiv.org/abs/hep-ph/0609281} {arXiv:hep-ph/0609281} \BibitemShut
  {NoStop}%
%%CITATION = HEP-PH/0609281;%%
\bibitem [{\citenamefont {Fukushima}(2008)}]{Fukushima:2008wg}%
  \BibitemOpen
  \bibfield  {author} {\bibinfo {author} {\bibfnamefont {K.}~\bibnamefont
  {Fukushima}},\ }\href {\doibase 10.1103/PhysRevD.77.114028,
  10.1103/PhysRevD.78.039902} {\bibfield  {journal} {\bibinfo  {journal} {Phys.
  Rev.}\ }\textbf {\bibinfo {volume} {D77}},\ \bibinfo {pages} {114028}
  (\bibinfo {year} {2008})},\ \Eprint {http://arxiv.org/abs/0803.3318}
  {arXiv:0803.3318} \BibitemShut {NoStop}%
%%CITATION = ARXIV:0803.3318;%%
\bibitem [{\citenamefont {Herbst}\ et~al.(2011)\citenamefont {Herbst},
  \citenamefont {Pawlowski},\ and\ \citenamefont {Schaefer}}]{Herbst:2010rf}%
  \BibitemOpen
  \bibfield  {author} {\bibinfo {author} {\bibfnamefont {T.~K.}\ \bibnamefont
  {Herbst}}, \bibinfo {author} {\bibfnamefont {J.~M.}\ \bibnamefont
  {Pawlowski}}, \ and\ \bibinfo {author} {\bibfnamefont {B.-J.}\ \bibnamefont
  {Schaefer}},\ }\href {\doibase 10.1016/j.physletb.2010.12.003} {\bibfield
  {journal} {\bibinfo  {journal} {Phys.Lett.}\ }\textbf {\bibinfo {volume}
  {B696}},\ \bibinfo {pages} {58} (\bibinfo {year} {2011})},\ \Eprint
  {http://arxiv.org/abs/1008.0081} {arXiv:1008.0081} \BibitemShut {NoStop}%
%%CITATION = ARXIV:1008.0081;%%
\bibitem [{\citenamefont {Herbst}\ et~al.(2014)\citenamefont {Herbst},
  \citenamefont {Mitter}, \citenamefont {Pawlowski}, \citenamefont {Schaefer},\
  and\ \citenamefont {Stiele}}]{Herbst:2013ufa}%
  \BibitemOpen
  \bibfield  {author} {\bibinfo {author} {\bibfnamefont {T.~K.}\ \bibnamefont
  {Herbst}}, \bibinfo {author} {\bibfnamefont {M.}~\bibnamefont {Mitter}},
  \bibinfo {author} {\bibfnamefont {J.~M.}\ \bibnamefont {Pawlowski}}, \bibinfo
  {author} {\bibfnamefont {B.-J.}\ \bibnamefont {Schaefer}}, \ and\ \bibinfo
  {author} {\bibfnamefont {R.}~\bibnamefont {Stiele}},\ }\href {\doibase
  10.1016/j.physletb.2014.02.045} {\bibfield  {journal} {\bibinfo  {journal}
  {Phys. Lett.}\ }\textbf {\bibinfo {volume} {B731}},\ \bibinfo {pages} {248}
  (\bibinfo {year} {2014})},\ \Eprint {http://arxiv.org/abs/1308.3621}
  {arXiv:1308.3621} \BibitemShut {NoStop}%
%%CITATION = ARXIV:1308.3621;%%
\bibitem [{\citenamefont {Ivanytskyi}\ et~al.(2019)\citenamefont {Ivanytskyi},
  \citenamefont {Perez-Garcia}, \citenamefont {Sagun},\ and\ \citenamefont
  {Albertus}}]{Ivanytskyi:2019ojt}%
  \BibitemOpen
  \bibfield  {author} {\bibinfo {author} {\bibfnamefont {O.}~\bibnamefont
  {Ivanytskyi}}, \bibinfo {author} {\bibfnamefont {M.~A.}\ \bibnamefont
  {Perez-Garcia}}, \bibinfo {author} {\bibfnamefont {V.}~\bibnamefont {Sagun}},
  \ and\ \bibinfo {author} {\bibfnamefont {C.}~\bibnamefont {Albertus}},\
  }\href@noop {} {\ }\Eprint {http://arxiv.org/abs/1909.07421}
  {arXiv:1909.07421} \BibitemShut {NoStop}%
%%CITATION = ARXIV:1909.07421;%%
\bibitem [{\citenamefont {Fukushima}\ and\ \citenamefont
  {Skokov}(2017)}]{Fukushima:2017csk}%
  \BibitemOpen
  \bibfield  {author} {\bibinfo {author} {\bibfnamefont {K.}~\bibnamefont
  {Fukushima}}\ and\ \bibinfo {author} {\bibfnamefont {V.}~\bibnamefont
  {Skokov}},\ }\href {\doibase 10.1016/j.ppnp.2017.05.002} {\bibfield
  {journal} {\bibinfo  {journal} {Prog. Part. Nucl. Phys.}\ }\textbf {\bibinfo
  {volume} {96}},\ \bibinfo {pages} {154} (\bibinfo {year} {2017})},\ \Eprint
  {http://arxiv.org/abs/1705.00718} {arXiv:1705.00718} \BibitemShut {NoStop}%
%%CITATION = ARXIV:1705.00718;%%
\bibitem [{\citenamefont {Fu}\ et~al.(2018)\citenamefont {Fu}, \citenamefont
  {Pawlowski},\ and\ \citenamefont {Rennecke}}]{Fu:2018qsk}%
  \BibitemOpen
  \bibfield  {author} {\bibinfo {author} {\bibfnamefont {W.-j.}\ \bibnamefont
  {Fu}}, \bibinfo {author} {\bibfnamefont {J.~M.}\ \bibnamefont {Pawlowski}}, \
  and\ \bibinfo {author} {\bibfnamefont {F.}~\bibnamefont {Rennecke}},\
  }\href@noop {} {\ }\Eprint {http://arxiv.org/abs/1808.00410}
  {arXiv:1808.00410} \BibitemShut {NoStop}%
%%CITATION = ARXIV:1808.00410;%%
\bibitem [{\citenamefont {Rennecke}\ and\ \citenamefont
  {Schaefer}(2017)}]{PhysRevD.96.016009}%
  \BibitemOpen
  \bibfield  {author} {\bibinfo {author} {\bibfnamefont {F.}~\bibnamefont
  {Rennecke}}\ and\ \bibinfo {author} {\bibfnamefont {B.-J.}\ \bibnamefont
  {Schaefer}},\ }\href {\doibase 10.1103/PhysRevD.96.016009} {\bibfield
  {journal} {\bibinfo  {journal} {Phys. Rev.}\ }\textbf {\bibinfo {volume}
  {D96}},\ \bibinfo {pages} {016009} (\bibinfo {year} {2017})},\ \Eprint
  {http://arxiv.org/abs/1610.08748} {arXiv:1610.08748} \BibitemShut {NoStop}%
%%CITATION = ARXIV:1610.08748;%%
\bibitem [{\citenamefont {Tripolt}\ et~al.(2018)\citenamefont {Tripolt},
  \citenamefont {Schaefer}, \citenamefont {von Smekal},\ and\ \citenamefont
  {Wambach}}]{Tripolt:2017zgc}%
  \BibitemOpen
  \bibfield  {author} {\bibinfo {author} {\bibfnamefont {R.-A.}\ \bibnamefont
  {Tripolt}}, \bibinfo {author} {\bibfnamefont {B.-J.}\ \bibnamefont
  {Schaefer}}, \bibinfo {author} {\bibfnamefont {L.}~\bibnamefont {von
  Smekal}}, \ and\ \bibinfo {author} {\bibfnamefont {J.}~\bibnamefont
  {Wambach}},\ }\href {\doibase 10.1103/PhysRevD.97.034022} {\bibfield
  {journal} {\bibinfo  {journal} {Phys. Rev.}\ }\textbf {\bibinfo {volume}
  {D97}},\ \bibinfo {pages} {034022} (\bibinfo {year} {2018})},\ \Eprint
  {http://arxiv.org/abs/1709.05991} {arXiv:1709.05991} \BibitemShut {NoStop}%
%%CITATION = ARXIV:1709.05991;%%
\bibitem [{\citenamefont {Schaefer}\ and\ \citenamefont
  {Wambach}(2005)}]{Schaefer:2004en}%
  \BibitemOpen
  \bibfield  {author} {\bibinfo {author} {\bibfnamefont {B.-J.}\ \bibnamefont
  {Schaefer}}\ and\ \bibinfo {author} {\bibfnamefont {J.}~\bibnamefont
  {Wambach}},\ }\href {\doibase 10.1016/j.nuclphysa.2005.04.012} {\bibfield
  {journal} {\bibinfo  {journal} {Nucl. Phys.}\ }\textbf {\bibinfo {volume}
  {A757}},\ \bibinfo {pages} {479} (\bibinfo {year} {2005})},\ \Eprint
  {http://arxiv.org/abs/nucl-th/0403039} {arXiv:nucl-th/0403039} \BibitemShut
  {NoStop}%
%%CITATION = NUCL-TH/0403039;%%
\bibitem [{\citenamefont {Glendenning}(1992)}]{Glendenning:1992vb}%
  \BibitemOpen
  \bibfield  {author} {\bibinfo {author} {\bibfnamefont {N.~K.}\ \bibnamefont
  {Glendenning}},\ }\href {\doibase 10.1103/PhysRevD.46.1274} {\bibfield
  {journal} {\bibinfo  {journal} {Phys. Rev.}\ }\textbf {\bibinfo {volume}
  {D46}},\ \bibinfo {pages} {1274} (\bibinfo {year} {1992})}\BibitemShut
  {NoStop}%
%%CITATION = PHRVA,D46,1274;%%
\bibitem [{\citenamefont {Gulminelli}\ and\ \citenamefont
  {Raduta}(2015)}]{Gulminelli:2015csa}%
  \BibitemOpen
  \bibfield  {author} {\bibinfo {author} {\bibfnamefont {F.}~\bibnamefont
  {Gulminelli}}\ and\ \bibinfo {author} {\bibfnamefont {A.~R.}\ \bibnamefont
  {Raduta}},\ }\href {\doibase 10.1103/PhysRevC.92.055803} {\bibfield
  {journal} {\bibinfo  {journal} {Phys. Rev.}\ }\textbf {\bibinfo {volume}
  {C92}},\ \bibinfo {pages} {055803} (\bibinfo {year} {2015})},\ \Eprint
  {http://arxiv.org/abs/1504.04493} {arXiv:1504.04493} \BibitemShut {NoStop}%
%%CITATION = ARXIV:1504.04493;%%
\bibitem [{\citenamefont {Danielewicz}\ and\ \citenamefont
  {Lee}(2009)}]{Danielewicz:2008cm}%
  \BibitemOpen
  \bibfield  {author} {\bibinfo {author} {\bibfnamefont {P.}~\bibnamefont
  {Danielewicz}}\ and\ \bibinfo {author} {\bibfnamefont {J.}~\bibnamefont
  {Lee}},\ }\href {\doibase 10.1016/j.nuclphysa.2008.11.007} {\bibfield
  {journal} {\bibinfo  {journal} {Nucl. Phys.}\ }\textbf {\bibinfo {volume}
  {A818}},\ \bibinfo {pages} {36} (\bibinfo {year} {2009})},\ \Eprint
  {http://arxiv.org/abs/0807.3743} {arXiv:0807.3743} \BibitemShut {NoStop}%
%%CITATION = ARXIV:0807.3743;%%
\bibitem [{\citenamefont {Chabanat}\ et~al.(1998)\citenamefont {Chabanat},
  \citenamefont {Bonche}, \citenamefont {Haensel}, \citenamefont {Meyer},\ and\
  \citenamefont {Schaeffer}}]{Chabanat:1997un}%
  \BibitemOpen
  \bibfield  {author} {\bibinfo {author} {\bibfnamefont {E.}~\bibnamefont
  {Chabanat}}, \bibinfo {author} {\bibfnamefont {P.}~\bibnamefont {Bonche}},
  \bibinfo {author} {\bibfnamefont {P.}~\bibnamefont {Haensel}}, \bibinfo
  {author} {\bibfnamefont {J.}~\bibnamefont {Meyer}}, \ and\ \bibinfo {author}
  {\bibfnamefont {R.}~\bibnamefont {Schaeffer}},\ }\href {\doibase
  10.1016/S0375-9474(98)00570-3, 10.1016/S0375-9474(98)00180-8} {\bibfield
  {journal} {\bibinfo  {journal} {Nucl. Phys.}\ }\textbf {\bibinfo {volume}
  {A635}},\ \bibinfo {pages} {231} (\bibinfo {year} {1998})}\BibitemShut
  {NoStop}%
%%CITATION = NUPHA,A635,231;%%
\bibitem [{\citenamefont {Typel}\ et~al.(2010)\citenamefont {Typel},
  \citenamefont {Ropke}, \citenamefont {Klahn}, \citenamefont {Blaschke},\ and\
  \citenamefont {Wolter}}]{Typel:2009sy}%
  \BibitemOpen
  \bibfield  {author} {\bibinfo {author} {\bibfnamefont {S.}~\bibnamefont
  {Typel}}, \bibinfo {author} {\bibfnamefont {G.}~\bibnamefont {Ropke}},
  \bibinfo {author} {\bibfnamefont {T.}~\bibnamefont {Klahn}}, \bibinfo
  {author} {\bibfnamefont {D.}~\bibnamefont {Blaschke}}, \ and\ \bibinfo
  {author} {\bibfnamefont {H.~H.}\ \bibnamefont {Wolter}},\ }\href {\doibase
  10.1103/PhysRevC.81.015803} {\bibfield  {journal} {\bibinfo  {journal} {Phys.
  Rev.}\ }\textbf {\bibinfo {volume} {C81}},\ \bibinfo {pages} {015803}
  (\bibinfo {year} {2010})},\ \Eprint {http://arxiv.org/abs/0908.2344}
  {arXiv:0908.2344} \BibitemShut {NoStop}%
%%CITATION = ARXIV:0908.2344;%%
\bibitem [{\citenamefont {Hempel}\ and\ \citenamefont
  {Schaffner-Bielich}(2010)}]{Hempel:2009mc}%
  \BibitemOpen
  \bibfield  {author} {\bibinfo {author} {\bibfnamefont {M.}~\bibnamefont
  {Hempel}}\ and\ \bibinfo {author} {\bibfnamefont {J.}~\bibnamefont
  {Schaffner-Bielich}},\ }\href {\doibase 10.1016/j.nuclphysa.2010.02.010}
  {\bibfield  {journal} {\bibinfo  {journal} {Nucl. Phys.}\ }\textbf {\bibinfo
  {volume} {A837}},\ \bibinfo {pages} {210} (\bibinfo {year} {2010})},\ \Eprint
  {http://arxiv.org/abs/0911.4073} {arXiv:0911.4073} \BibitemShut {NoStop}%
%%CITATION = ARXIV:0911.4073;%%
\bibitem [{\citenamefont {Steiner}\ et~al.(2013{\natexlab{b}})\citenamefont
  {Steiner}, \citenamefont {Hempel},\ and\ \citenamefont
  {Fischer}}]{Steiner:2012rk}%
  \BibitemOpen
  \bibfield  {author} {\bibinfo {author} {\bibfnamefont {A.~W.}\ \bibnamefont
  {Steiner}}, \bibinfo {author} {\bibfnamefont {M.}~\bibnamefont {Hempel}}, \
  and\ \bibinfo {author} {\bibfnamefont {T.}~\bibnamefont {Fischer}},\ }\href
  {\doibase 10.1088/0004-637X/774/1/17} {\bibfield  {journal} {\bibinfo
  {journal} {Astrophys. J.}\ }\textbf {\bibinfo {volume} {774}},\ \bibinfo
  {pages} {17} (\bibinfo {year} {2013}{\natexlab{b}})},\ \Eprint
  {http://arxiv.org/abs/1207.2184} {arXiv:1207.2184} \BibitemShut {NoStop}%
%%CITATION = ARXIV:1207.2184;%%
\bibitem [{\citenamefont {Bombaci}\ and\ \citenamefont
  {Logoteta}(2018)}]{Bombaci:2018ksa}%
  \BibitemOpen
  \bibfield  {author} {\bibinfo {author} {\bibfnamefont {I.}~\bibnamefont
  {Bombaci}}\ and\ \bibinfo {author} {\bibfnamefont {D.}~\bibnamefont
  {Logoteta}},\ }\href {\doibase 10.1051/0004-6361/201731604} {\bibfield
  {journal} {\bibinfo  {journal} {Astron. Astrophys.}\ }\textbf {\bibinfo
  {volume} {609}},\ \bibinfo {pages} {A128} (\bibinfo {year} {2018})},\ \Eprint
  {http://arxiv.org/abs/1805.11846} {arXiv:1805.11846} \BibitemShut {NoStop}%
%%CITATION = ARXIV:1805.11846;%%
\bibitem [{\citenamefont {Baym}\ et~al.(2019)\citenamefont {Baym},
  \citenamefont {Furusawa}, \citenamefont {Hatsuda}, \citenamefont {Kojo},\
  and\ \citenamefont {Togashi}}]{Baym:2019iky}%
  \BibitemOpen
  \bibfield  {author} {\bibinfo {author} {\bibfnamefont {G.}~\bibnamefont
  {Baym}}, \bibinfo {author} {\bibfnamefont {S.}~\bibnamefont {Furusawa}},
  \bibinfo {author} {\bibfnamefont {T.}~\bibnamefont {Hatsuda}}, \bibinfo
  {author} {\bibfnamefont {T.}~\bibnamefont {Kojo}}, \ and\ \bibinfo {author}
  {\bibfnamefont {H.}~\bibnamefont {Togashi}},\ }\href@noop {} {\ }\Eprint
  {http://arxiv.org/abs/1903.08963} {arXiv:1903.08963} \BibitemShut {NoStop}%
%%CITATION = ARXIV:1903.08963;%%
\bibitem [{\citenamefont {Alford}\ and\ \citenamefont
  {Sedrakian}(2017)}]{Alford:2017ly}%
  \BibitemOpen
  \bibfield  {author} {\bibinfo {author} {\bibfnamefont {M.}~\bibnamefont
  {Alford}}\ and\ \bibinfo {author} {\bibfnamefont {A.}~\bibnamefont
  {Sedrakian}},\ }\href {\doibase 10.1103/PhysRevLett.119.161104} {\bibfield
  {journal} {\bibinfo  {journal} {Phys. Rev. Lett.}\ }\textbf {\bibinfo
  {volume} {119}},\ \bibinfo {pages} {161104} (\bibinfo {year}
  {2017})}\BibitemShut {NoStop}%
\bibitem [{\citenamefont {Typel}\ et~al.(2015)\citenamefont {Typel},
  \citenamefont {Oertel},\ and\ \citenamefont {Kl{\"a}hn}}]{Typel:2013rza}%
  \BibitemOpen
  \bibfield  {author} {\bibinfo {author} {\bibfnamefont {S.}~\bibnamefont
  {Typel}}, \bibinfo {author} {\bibfnamefont {M.}~\bibnamefont {Oertel}}, \
  and\ \bibinfo {author} {\bibfnamefont {T.}~\bibnamefont {Kl{\"a}hn}},\ }\href
  {\doibase 10.1134/S1063779615040061} {\bibfield  {journal} {\bibinfo
  {journal} {Phys. Part. Nucl.}\ }\textbf {\bibinfo {volume} {46}},\ \bibinfo
  {pages} {633} (\bibinfo {year} {2015})},\ \Eprint
  {http://arxiv.org/abs/1307.5715} {arXiv:1307.5715} \BibitemShut {NoStop}%
%%CITATION = ARXIV:1307.5715;%%
\bibitem [{\citenamefont {Alford}\ et~al.(2013)\citenamefont {Alford},
  \citenamefont {Han},\ and\ \citenamefont {Prakash}}]{Alford:2013aca}%
  \BibitemOpen
  \bibfield  {author} {\bibinfo {author} {\bibfnamefont {M.~G.}\ \bibnamefont
  {Alford}}, \bibinfo {author} {\bibfnamefont {S.}~\bibnamefont {Han}}, \ and\
  \bibinfo {author} {\bibfnamefont {M.}~\bibnamefont {Prakash}},\ }\href
  {\doibase 10.1103/PhysRevD.88.083013} {\bibfield  {journal} {\bibinfo
  {journal} {Phys. Rev.}\ }\textbf {\bibinfo {volume} {D88}},\ \bibinfo {pages}
  {083013} (\bibinfo {year} {2013})},\ \Eprint {http://arxiv.org/abs/1302.4732}
  {arXiv:1302.4732} \BibitemShut {NoStop}%
%%CITATION = ARXIV:1302.4732;%%
\bibitem [{\citenamefont {Benic}\ et~al.(2015)\citenamefont {Benic},
  \citenamefont {Blaschke}, \citenamefont {Alvarez-Castillo}, \citenamefont
  {Fischer},\ and\ \citenamefont {Typel}}]{Benic:2014jia}%
  \BibitemOpen
  \bibfield  {author} {\bibinfo {author} {\bibfnamefont {S.}~\bibnamefont
  {Benic}}, \bibinfo {author} {\bibfnamefont {D.}~\bibnamefont {Blaschke}},
  \bibinfo {author} {\bibfnamefont {D.~E.}\ \bibnamefont {Alvarez-Castillo}},
  \bibinfo {author} {\bibfnamefont {T.}~\bibnamefont {Fischer}}, \ and\
  \bibinfo {author} {\bibfnamefont {S.}~\bibnamefont {Typel}},\ }\href
  {\doibase 10.1051/0004-6361/201425318} {\bibfield  {journal} {\bibinfo
  {journal} {Astron. Astrophys.}\ }\textbf {\bibinfo {volume} {577}},\ \bibinfo
  {pages} {A40} (\bibinfo {year} {2015})},\ \Eprint
  {http://arxiv.org/abs/1411.2856} {arXiv:1411.2856} \BibitemShut {NoStop}%
%%CITATION = ARXIV:1411.2856;%%
\bibitem [{\citenamefont {Tews}\ et~al.(2018{\natexlab{b}})\citenamefont
  {Tews}, \citenamefont {Carlson}, \citenamefont {Gandolfi},\ and\
  \citenamefont {Reddy}}]{Tews:2018kmu}%
  \BibitemOpen
  \bibfield  {author} {\bibinfo {author} {\bibfnamefont {I.}~\bibnamefont
  {Tews}}, \bibinfo {author} {\bibfnamefont {J.}~\bibnamefont {Carlson}},
  \bibinfo {author} {\bibfnamefont {S.}~\bibnamefont {Gandolfi}}, \ and\
  \bibinfo {author} {\bibfnamefont {S.}~\bibnamefont {Reddy}},\ }\href
  {\doibase 10.3847/1538-4357/aac267} {\bibfield  {journal} {\bibinfo
  {journal} {Astrophys. J.}\ }\textbf {\bibinfo {volume} {860}},\ \bibinfo
  {pages} {149} (\bibinfo {year} {2018}{\natexlab{b}})},\ \Eprint
  {http://arxiv.org/abs/1801.01923} {arXiv:1801.01923} \BibitemShut {NoStop}%
%%CITATION = ARXIV:1801.01923;%%
\bibitem [{\citenamefont {Bedaque}\ and\ \citenamefont
  {Steiner}(2015)}]{Bedaque:2014sqa}%
  \BibitemOpen
  \bibfield  {author} {\bibinfo {author} {\bibfnamefont {P.}~\bibnamefont
  {Bedaque}}\ and\ \bibinfo {author} {\bibfnamefont {A.~W.}\ \bibnamefont
  {Steiner}},\ }\href {\doibase 10.1103/PhysRevLett.114.031103} {\bibfield
  {journal} {\bibinfo  {journal} {Phys. Rev. Lett.}\ }\textbf {\bibinfo
  {volume} {114}},\ \bibinfo {pages} {031103} (\bibinfo {year} {2015})},\
  \Eprint {http://arxiv.org/abs/1408.5116} {arXiv:1408.5116} \BibitemShut
  {NoStop}%
%%CITATION = ARXIV:1408.5116;%%
\bibitem [{\citenamefont {Borsanyi}\ et~al.(2012)\citenamefont {Borsanyi},
  \citenamefont {Endrodi}, \citenamefont {Fodor}, \citenamefont {Katz},
  \citenamefont {Krieg}, \citenamefont {Ratti},\ and\ \citenamefont
  {Szabo}}]{Borsanyi:2012cr}%
  \BibitemOpen
  \bibfield  {author} {\bibinfo {author} {\bibfnamefont {S.}~\bibnamefont
  {Borsanyi}}, \bibinfo {author} {\bibfnamefont {G.}~\bibnamefont {Endrodi}},
  \bibinfo {author} {\bibfnamefont {Z.}~\bibnamefont {Fodor}}, \bibinfo
  {author} {\bibfnamefont {S.~D.}\ \bibnamefont {Katz}}, \bibinfo {author}
  {\bibfnamefont {S.}~\bibnamefont {Krieg}}, \bibinfo {author} {\bibfnamefont
  {C.}~\bibnamefont {Ratti}}, \ and\ \bibinfo {author} {\bibfnamefont {K.~K.}\
  \bibnamefont {Szabo}},\ }\href {\doibase 10.1007/JHEP08(2012)053} {\bibfield
  {journal} {\bibinfo  {journal} {JHEP}\ }\textbf {\bibinfo {volume} {08}},\
  \bibinfo {pages} {053} (\bibinfo {year} {2012})},\ \Eprint
  {http://arxiv.org/abs/1204.6710} {arXiv:1204.6710} \BibitemShut {NoStop}%
%%CITATION = ARXIV:1204.6710;%%
\bibitem [{\citenamefont {Benincasa}\ and\ \citenamefont
  {Buchel}(2006)}]{Benincasa:2006ei}%
  \BibitemOpen
  \bibfield  {author} {\bibinfo {author} {\bibfnamefont {P.}~\bibnamefont
  {Benincasa}}\ and\ \bibinfo {author} {\bibfnamefont {A.}~\bibnamefont
  {Buchel}},\ }\href {\doibase 10.1016/j.physletb.2006.07.043} {\bibfield
  {journal} {\bibinfo  {journal} {Phys. Lett.}\ }\textbf {\bibinfo {volume}
  {B640}},\ \bibinfo {pages} {108} (\bibinfo {year} {2006})},\ \Eprint
  {http://arxiv.org/abs/hep-th/0605076} {arXiv:hep-th/0605076} \BibitemShut
  {NoStop}%
%%CITATION = HEP-TH/0605076;%%
\bibitem [{\citenamefont {Cherman}\ et~al.(2009)\citenamefont {Cherman},
  \citenamefont {Cohen},\ and\ \citenamefont {Nellore}}]{Cherman:2009tw}%
  \BibitemOpen
  \bibfield  {author} {\bibinfo {author} {\bibfnamefont {A.}~\bibnamefont
  {Cherman}}, \bibinfo {author} {\bibfnamefont {T.~D.}\ \bibnamefont {Cohen}},
  \ and\ \bibinfo {author} {\bibfnamefont {A.}~\bibnamefont {Nellore}},\ }\href
  {\doibase 10.1103/PhysRevD.80.066003} {\bibfield  {journal} {\bibinfo
  {journal} {Phys. Rev.}\ }\textbf {\bibinfo {volume} {D80}},\ \bibinfo {pages}
  {066003} (\bibinfo {year} {2009})},\ \Eprint {http://arxiv.org/abs/0905.0903}
  {arXiv:0905.0903} \BibitemShut {NoStop}%
%%CITATION = ARXIV:0905.0903;%%
\bibitem [{\citenamefont {Ecker}\ et~al.(2017)\citenamefont {Ecker},
  \citenamefont {Hoyos}, \citenamefont {Jokela}, \citenamefont
  {Rodr{\'\i}guez~Fern{\'a}ndez},\ and\ \citenamefont
  {Vuorinen}}]{Ecker:2017fyh}%
  \BibitemOpen
  \bibfield  {author} {\bibinfo {author} {\bibfnamefont {C.}~\bibnamefont
  {Ecker}}, \bibinfo {author} {\bibfnamefont {C.}~\bibnamefont {Hoyos}},
  \bibinfo {author} {\bibfnamefont {N.}~\bibnamefont {Jokela}}, \bibinfo
  {author} {\bibfnamefont {D.}~\bibnamefont {Rodr{\'\i}guez~Fern{\'a}ndez}}, \
  and\ \bibinfo {author} {\bibfnamefont {A.}~\bibnamefont {Vuorinen}},\ }\href
  {\doibase 10.1007/JHEP11(2017)031} {\bibfield  {journal} {\bibinfo  {journal}
  {JHEP}\ }\textbf {\bibinfo {volume} {11}},\ \bibinfo {pages} {031} (\bibinfo
  {year} {2017})},\ \Eprint {http://arxiv.org/abs/1707.00521}
  {arXiv:1707.00521} \BibitemShut {NoStop}%
%%CITATION = ARXIV:1707.00521;%%
\bibitem [{\citenamefont {Zhu}\ et~al.(2018)\citenamefont {Zhu}, \citenamefont
  {Zhou},\ and\ \citenamefont {Li}}]{Zhu:2018ys}%
  \BibitemOpen
  \bibfield  {author} {\bibinfo {author} {\bibfnamefont {Z.-Y.}\ \bibnamefont
  {Zhu}}, \bibinfo {author} {\bibfnamefont {E.-P.}\ \bibnamefont {Zhou}}, \
  and\ \bibinfo {author} {\bibfnamefont {A.}~\bibnamefont {Li}},\ }\href
  {\doibase 10.3847/1538-4357/aacc28} {\bibfield  {journal} {\bibinfo
  {journal} {Astrophys. J.}\ }\textbf {\bibinfo {volume} {862}},\ \bibinfo
  {pages} {98} (\bibinfo {year} {2018})},\ \Eprint
  {http://arxiv.org/abs/1802.05510} {arXiv:1802.05510} \BibitemShut {NoStop}%
%%CITATION = ARXIV:1802.05510;%%
\bibitem [{\citenamefont {Aloy}\ et~al.(2019)\citenamefont {Aloy},
  \citenamefont {Ib{\'a}{\~n}ez}, \citenamefont {Sanchis-Gual}, \citenamefont
  {Obergaulinger}, \citenamefont {Font}, \citenamefont {Serna},\ and\
  \citenamefont {Marquina}}]{Ibanez:2018myp}%
  \BibitemOpen
  \bibfield  {author} {\bibinfo {author} {\bibfnamefont {M.~A.}\ \bibnamefont
  {Aloy}}, \bibinfo {author} {\bibfnamefont {J.~M.}\ \bibnamefont
  {Ib{\'a}{\~n}ez}}, \bibinfo {author} {\bibfnamefont {N.}~\bibnamefont
  {Sanchis-Gual}}, \bibinfo {author} {\bibfnamefont {M.}~\bibnamefont
  {Obergaulinger}}, \bibinfo {author} {\bibfnamefont {J.~A.}\ \bibnamefont
  {Font}}, \bibinfo {author} {\bibfnamefont {S.}~\bibnamefont {Serna}}, \ and\
  \bibinfo {author} {\bibfnamefont {A.}~\bibnamefont {Marquina}},\ }\href
  {\doibase 10.1093/mnras/stz293} {\bibfield  {journal} {\bibinfo  {journal}
  {Mon. Not. Roy. Astron. Soc.}\ }\textbf {\bibinfo {volume} {484}},\ \bibinfo
  {pages} {4980} (\bibinfo {year} {2019})},\ \Eprint
  {http://arxiv.org/abs/1806.03314} {arXiv:1806.03314} \BibitemShut {NoStop}%
%%CITATION = ARXIV:1806.03314;%%
\bibitem [{\citenamefont {Tolman}(1939)}]{PhysRev.55.364}%
  \BibitemOpen
  \bibfield  {author} {\bibinfo {author} {\bibfnamefont {R.~C.}\ \bibnamefont
  {Tolman}},\ }\href {\doibase 10.1103/PhysRev.55.364} {\bibfield  {journal}
  {\bibinfo  {journal} {Phys. Rev.}\ }\textbf {\bibinfo {volume} {55}},\
  \bibinfo {pages} {364} (\bibinfo {year} {1939})}\BibitemShut {NoStop}%
\bibitem [{\citenamefont {Oppenheimer}\ and\ \citenamefont
  {Volkoff}(1939)}]{PhysRev.55.374}%
  \BibitemOpen
  \bibfield  {author} {\bibinfo {author} {\bibfnamefont {J.~R.}\ \bibnamefont
  {Oppenheimer}}\ and\ \bibinfo {author} {\bibfnamefont {G.~M.}\ \bibnamefont
  {Volkoff}},\ }\href {\doibase 10.1103/PhysRev.55.374} {\bibfield  {journal}
  {\bibinfo  {journal} {Phys. Rev.}\ }\textbf {\bibinfo {volume} {55}},\
  \bibinfo {pages} {374} (\bibinfo {year} {1939})}\BibitemShut {NoStop}%
\bibitem [{\citenamefont {Braun}\ et~al.(2017)\citenamefont {Braun},
  \citenamefont {Leonhardt},\ and\ \citenamefont {Pospiech}}]{Braun:2017srn}%
  \BibitemOpen
  \bibfield  {author} {\bibinfo {author} {\bibfnamefont {J.}~\bibnamefont
  {Braun}}, \bibinfo {author} {\bibfnamefont {M.}~\bibnamefont {Leonhardt}}, \
  and\ \bibinfo {author} {\bibfnamefont {M.}~\bibnamefont {Pospiech}},\ }\href
  {\doibase 10.1103/PhysRevD.96.076003} {\bibfield  {journal} {\bibinfo
  {journal} {Phys. Rev.}\ }\textbf {\bibinfo {volume} {D96}},\ \bibinfo {pages}
  {076003} (\bibinfo {year} {2017})},\ \Eprint
  {http://arxiv.org/abs/1705.00074} {arXiv:1705.00074} \BibitemShut {NoStop}%
%%CITATION = ARXIV:1705.00074;%%
\bibitem [{\citenamefont {Mitter}\ et~al.(2015)\citenamefont {Mitter},
  \citenamefont {Pawlowski},\ and\ \citenamefont
  {Strodthoff}}]{Mitter:2014wpa}%
  \BibitemOpen
  \bibfield  {author} {\bibinfo {author} {\bibfnamefont {M.}~\bibnamefont
  {Mitter}}, \bibinfo {author} {\bibfnamefont {J.~M.}\ \bibnamefont
  {Pawlowski}}, \ and\ \bibinfo {author} {\bibfnamefont {N.}~\bibnamefont
  {Strodthoff}},\ }\href {\doibase 10.1103/PhysRevD.91.054035} {\bibfield
  {journal} {\bibinfo  {journal} {Phys. Rev.}\ }\textbf {\bibinfo {volume}
  {D91}},\ \bibinfo {pages} {054035} (\bibinfo {year} {2015})},\ \Eprint
  {http://arxiv.org/abs/1411.7978} {arXiv:1411.7978} \BibitemShut {NoStop}%
%%CITATION = ARXIV:1411.7978;%%
\bibitem [{\citenamefont {Fu}\ et~al.(2019)\citenamefont {Fu}, \citenamefont
  {Pawlowski},\ and\ \citenamefont {Rennecke}}]{Fu:2019hdw}%
  \BibitemOpen
  \bibfield  {author} {\bibinfo {author} {\bibfnamefont {W.-j.}\ \bibnamefont
  {Fu}}, \bibinfo {author} {\bibfnamefont {J.~M.}\ \bibnamefont {Pawlowski}}, \
  and\ \bibinfo {author} {\bibfnamefont {F.}~\bibnamefont {Rennecke}},\
  }\href@noop {} {\ }\Eprint {http://arxiv.org/abs/1909.02991}
  {arXiv:1909.02991} \BibitemShut {NoStop}%
%%CITATION = ARXIV:1909.02991;%%
\bibitem [{\citenamefont {Fu}\ et~al.(2016)\citenamefont {Fu}, \citenamefont
  {Pawlowski}, \citenamefont {Rennecke},\ and\ \citenamefont
  {Schaefer}}]{Fu:2016tey}%
  \BibitemOpen
  \bibfield  {author} {\bibinfo {author} {\bibfnamefont {W.-j.}\ \bibnamefont
  {Fu}}, \bibinfo {author} {\bibfnamefont {J.~M.}\ \bibnamefont {Pawlowski}},
  \bibinfo {author} {\bibfnamefont {F.}~\bibnamefont {Rennecke}}, \ and\
  \bibinfo {author} {\bibfnamefont {B.-J.}\ \bibnamefont {Schaefer}},\ }\href
  {\doibase 10.1103/PhysRevD.94.116020} {\bibfield  {journal} {\bibinfo
  {journal} {Phys. Rev.}\ }\textbf {\bibinfo {volume} {D94}},\ \bibinfo {pages}
  {116020} (\bibinfo {year} {2016})},\ \Eprint
  {http://arxiv.org/abs/1608.04302} {arXiv:1608.04302} \BibitemShut {NoStop}%
%%CITATION = ARXIV:1608.04302;%%
\bibitem [{\citenamefont {Braun}\ et~al.(2016)\citenamefont {Braun},
  \citenamefont {Fister}, \citenamefont {Pawlowski},\ and\ \citenamefont
  {Rennecke}}]{Braun:2014ata}%
  \BibitemOpen
  \bibfield  {author} {\bibinfo {author} {\bibfnamefont {J.}~\bibnamefont
  {Braun}}, \bibinfo {author} {\bibfnamefont {L.}~\bibnamefont {Fister}},
  \bibinfo {author} {\bibfnamefont {J.~M.}\ \bibnamefont {Pawlowski}}, \ and\
  \bibinfo {author} {\bibfnamefont {F.}~\bibnamefont {Rennecke}},\ }\href
  {\doibase 10.1103/PhysRevD.94.034016} {\bibfield  {journal} {\bibinfo
  {journal} {Phys. Rev.}\ }\textbf {\bibinfo {volume} {D94}},\ \bibinfo {pages}
  {034016} (\bibinfo {year} {2016})},\ \Eprint {http://arxiv.org/abs/1412.1045}
  {arXiv:1412.1045} \BibitemShut {NoStop}%
%%CITATION = ARXIV:1412.1045;%%
\bibitem [{\citenamefont {Alford}\ et~al.(2008)\citenamefont {Alford},
  \citenamefont {Schmitt}, \citenamefont {Rajagopal},\ and\ \citenamefont
  {Schafer}}]{Alford:2007xm}%
  \BibitemOpen
  \bibfield  {author} {\bibinfo {author} {\bibfnamefont {M.~G.}\ \bibnamefont
  {Alford}}, \bibinfo {author} {\bibfnamefont {A.}~\bibnamefont {Schmitt}},
  \bibinfo {author} {\bibfnamefont {K.}~\bibnamefont {Rajagopal}}, \ and\
  \bibinfo {author} {\bibfnamefont {T.}~\bibnamefont {Schafer}},\ }\href
  {\doibase 10.1103/RevModPhys.80.1455} {\bibfield  {journal} {\bibinfo
  {journal} {Rev.Mod.Phys.}\ }\textbf {\bibinfo {volume} {80}},\ \bibinfo
  {pages} {1455} (\bibinfo {year} {2008})},\ \Eprint
  {http://arxiv.org/abs/0709.4635} {arXiv:0709.4635} \BibitemShut {NoStop}%
%%CITATION = ARXIV:0709.4635;%%
\bibitem [{\citenamefont {Fukushima}\ and\ \citenamefont
  {Hatsuda}(2011)}]{Fukushima:2010bq}%
  \BibitemOpen
  \bibfield  {author} {\bibinfo {author} {\bibfnamefont {K.}~\bibnamefont
  {Fukushima}}\ and\ \bibinfo {author} {\bibfnamefont {T.}~\bibnamefont
  {Hatsuda}},\ }\href {\doibase 10.1088/0034-4885/74/1/014001} {\bibfield
  {journal} {\bibinfo  {journal} {Rept. Prog. Phys.}\ }\textbf {\bibinfo
  {volume} {74}},\ \bibinfo {pages} {014001} (\bibinfo {year} {2011})},\
  \Eprint {http://arxiv.org/abs/1005.4814} {arXiv:1005.4814} \BibitemShut
  {NoStop}%
%%CITATION = 1005.4814;%%
\bibitem [{\citenamefont {Storn}\ and\ \citenamefont
  {Price}(1997)}]{Storn:1997uk}%
  \BibitemOpen
  \bibfield  {author} {\bibinfo {author} {\bibfnamefont {R.}~\bibnamefont
  {Storn}}\ and\ \bibinfo {author} {\bibfnamefont {K.}~\bibnamefont {Price}},\
  }\href {\doibase 10.1023/A:1008202821328} {\bibfield  {journal} {\bibinfo
  {journal} {Journal of Global Optimization}\ }\textbf {\bibinfo {volume}
  {11}},\ \bibinfo {pages} {341} (\bibinfo {year} {1997})}\BibitemShut
  {NoStop}%
\bibitem [{\citenamefont {Carignano}\ et~al.(2014)\citenamefont {Carignano},
  \citenamefont {Buballa},\ and\ \citenamefont {Schaefer}}]{Carignano:2014jla}%
  \BibitemOpen
  \bibfield  {author} {\bibinfo {author} {\bibfnamefont {S.}~\bibnamefont
  {Carignano}}, \bibinfo {author} {\bibfnamefont {M.}~\bibnamefont {Buballa}},
  \ and\ \bibinfo {author} {\bibfnamefont {B.-J.}\ \bibnamefont {Schaefer}},\
  }\href {\doibase 10.1103/PhysRevD.90.014033} {\bibfield  {journal} {\bibinfo
  {journal} {Phys. Rev.}\ }\textbf {\bibinfo {volume} {D90}},\ \bibinfo {pages}
  {014033} (\bibinfo {year} {2014})},\ \Eprint {http://arxiv.org/abs/1404.0057}
  {arXiv:1404.0057} \BibitemShut {NoStop}%
%%CITATION = ARXIV:1404.0057;%%
\bibitem [{\citenamefont {Grossi}\ and\ \citenamefont
  {Wink}(2019)}]{Grossi:2019urj}%
  \BibitemOpen
  \bibfield  {author} {\bibinfo {author} {\bibfnamefont {E.}~\bibnamefont
  {Grossi}}\ and\ \bibinfo {author} {\bibfnamefont {N.}~\bibnamefont {Wink}},\
  }\href@noop {} {\ }\Eprint {http://arxiv.org/abs/1903.09503}
  {arXiv:1903.09503} \BibitemShut {NoStop}%
%%CITATION = ARXIV:1903.09503;%%
\bibitem [{\citenamefont {Borchardt}\ and\ \citenamefont
  {Knorr}(2016)}]{Borchardt:2016pif}%
  \BibitemOpen
  \bibfield  {author} {\bibinfo {author} {\bibfnamefont {J.}~\bibnamefont
  {Borchardt}}\ and\ \bibinfo {author} {\bibfnamefont {B.}~\bibnamefont
  {Knorr}},\ }\href {\doibase 10.1103/PhysRevD.94.025027} {\bibfield  {journal}
  {\bibinfo  {journal} {Phys. Rev.}\ }\textbf {\bibinfo {volume} {D94}},\
  \bibinfo {pages} {025027} (\bibinfo {year} {2016})},\ \Eprint
  {http://arxiv.org/abs/1603.06726} {arXiv:1603.06726} \BibitemShut {NoStop}%
%%CITATION = ARXIV:1603.06726;%%
\end{thebibliography}%

\end{document}